\documentclass[preprint,12pt]{elsarticle}

\usepackage{amssymb}
\usepackage{amsmath}

\journal{Reliability Engineering and System Safety}

\begin{document}

\begin{frontmatter}



\title{Uncertainty Quantification on State-Based Conflict
Detection and Resolution Algorithms}



\author{Muhammad Fazlur Rahman\corref{cor1}}
\ead{m.f.rahman@tudelft.nl}
\cortext[cor1]{Corresponding author.}

\author{Joost Ellerbroek}
\author{Jacco Hoekstra}

\address{Faculty of Aerospace Engineering, Delft University of Technology, 
Kluyverweg 1, 2629 HS Delft, The Netherlands}

\begin{abstract}
This study investigates how navigation uncertainty affects conflict detection and resolution (CD\&R) for uncrewed aircraft in U-space. Position and velocity errors are modelled as zero-mean Gaussian noise consistent with ADS-L accuracy, and propagated through conflict metrics using Monte Carlo and analytical approximations. Under uncertainty, state-based detection becomes probabilistic. The probability of detection depends on both the level of uncertainty and the encounter geometry, and falls below 50\% when the nominal intrusion time equals the look-ahead. Operationally, detection is re-evaluated over time as the encounter develops, yielding multiple observations with varying probabilities. Two resolution algorithms are compared: Modified Voltage Potential (MVP) and Velocity Obstacle (VO). MVP proves more robust under uncertainty because it explicitly maximises distance at the closest point of approach (CPA). By maximising CPA distance, MVP maintains an outward push and avoids reversal behaviour during the manoeuvre, whereas VO performance degrades at low relative speeds and shallow angles. BlueSky simulations confirm these effects: MVP achieves higher intrusion-prevention rates and larger post-resolution miss distances across conflict scenarios, with its advantage most pronounced at low relative velocity The findings highlight the importance of maximising CPA distance as a conflict resolution strategy. Moreover, the look-ahead horizon and protected zone can be tuned to achieve a desired target level of safety.

\end{abstract}




\end{frontmatter}



\begin{abstract}
This study investigates how navigation uncertainty affects conflict detection and resolution (CD\&R) for uncrewed aircraft in U-space. Position and velocity errors are modelled as zero-mean Gaussian noise consistent with ADS-L accuracy, and propagated through conflict metrics using Monte Carlo and analytical approximations. Under uncertainty, state-based detection becomes probabilistic. The probability of detection depends on both the level of uncertainty and the encounter geometry, and falls below 50\% when the nominal intrusion time equals the look-ahead. Operationally, detection is re-evaluated over time as the encounter develops, yielding multiple observations with varying probabilities. Two resolution algorithms are compared: Modified Voltage Potential (MVP) and Velocity Obstacle (VO). MVP proves more robust under uncertainty because it explicitly maximises distance at the closest point of approach (CPA). By maximising CPA distance, MVP maintains an outward push and avoids reversal behaviour during the manoeuvre, whereas VO performance degrades at low relative speeds and shallow angles. BlueSky simulations confirm these effects: MVP achieves higher intrusion-prevention rates and larger post-resolution miss distances across conflict scenarios, with its advantage most pronounced at low relative velocity The findings highlight the importance of maximising CPA distance as a conflict resolution strategy. Moreover, the look-ahead horizon and protected zone can be tuned to achieve a desired target level of safety.

\end{abstract}
\section{Introduction}

The number of Uncrewed Aerial Systems (UASs) operating in European airspace is expected to increase markedly in the coming decade. Estimates indicate that several hundred thousand drones may be active by 2030 \cite{european_commission_drone_2022}. The safe integration of this traffic alongside existing aviation requires dedicated operational concepts, such as the U-space framework \cite{sesar_joint_undertaking_u-space_2023}. Within U-space, the tactical separation layer is intended to address conflicts not resolved during pre-flight strategic planning, by using real-time state information to detect and resolve conflicts between aircraft on short time horizons \cite{international_civil_aviation_organization_global_2005}.

This tactical conflict resolution, typically relies on observations of the current aircraft state information such as position and velocity to be shared among the airspace users. These tactical attributes are provided by communication, navigation, and surveillance (CNS) systems. One of the means to communicate this state information to other airspace users is the Automatic Dependent Surveillance–Light (ADS-L) system, proposed for U-space operations \cite{european_union_aviation_safety_agency_easa_technical_2022}. These state data are subject to uncertainty where position and velocity estimates contain errors that could degrade the performance of tactical conflict detection and resolution \cite{rahman_effect_2024}. While the same problems exist in crewed aircraft \cite{langejan_effect_2016, khan_surveillance_2011}, the scales of the separation minima and manoeuvring speeds are different. Crewed aircraft apply minimum separation standard of about 5 nautical miles, whereas UAS operations typically require only 50 to 200 meters \cite{weibel_establishing_2011, ribeiro_review_2020, alejo_multi-uav_2009, weinert_near_2022, jenie_selective_2015}. The much smaller separation distances make navigation uncertainty proportionally more significant for UAS.

In the presence of navigation errors, conflict detection and resolution become probabilistic. Analysis of these stochastic effects enables a systematic characterisation of how errors in state information propagate through detection and resolution logic, and how this propagation influences sensitivity to thresholds such as look-ahead time and protected-zone radius. Such models would allow a derivation of requirements on aspects like look-ahead time and separation minima from a target level of safety (TLS).

The aim of this paper is to derive such models by analysing the impact of navigation uncertainties, represented as errors in position and velocity, on state-based conflict detection and decentralized resolution algorithms in U-space. Analytical solutions and first-order approximations are developed to describe how such errors propagate into detection variables. These results are combined with Monte Carlo simulation to quantify detection probabilities under different uncertainty conditions. The analysis is extended to resolution algorithms, where uncertainty in the input state is propagated through candidate manoeuvres to assess the distribution of post-resolution separation. Large-scale traffic simulations are conducted to connect these microscopic effects to macroscopic safety indicators.

The remainder of the paper is organized as follows. Section~\ref{sec:autonomous} introduces the tactical conflict detection and resolution algorithms. Section~\ref{sec:nav_uncertainty} presents the uncertainty models and their propagation within state-based detection and resolution. Section~\ref{sec:methodology} describes the evaluation methodology. Section~\ref{sec:results} reports and discusses the results, and Section~\ref{sec:conclusion} concludes the paper.
\section{Tactical Separation Management}
\label{sec:autonomous}

Suppose two aircraft are in conflict and require separation. This conflict is defined as a predicted loss of separation within a specified look-ahead time, rather than an actual breach of the minimum separation standard. When the separation minima between two aircraft are actually violated, it results in a loss of separation (LoS). To detect conflicts before they escalate into LoS, the future positions of the aircraft must be estimated. This is typically done by extrapolating their current velocity vectors, although more advanced methods may employ intent information shared between the conflicting aircraft \cite{yang_using_1998}.

In today's civil aviation, the air traffic controller is responsible for resolving such conflicts through a centralized approach. The controller maintains a global view of the traffic situation and issues coordinated manoeuvres to ensure safe separation. The main advantage of this approach is its ability to optimize decisions globally, taking into account the entire airspace picture, which allows for efficient conflict resolution and adherence to traffic flow constraints. However, centralized systems depend heavily on infrastructure and reliable communication, and their scalability becomes a significant challenge in environments with dense or rapidly changing traffic, such as those expected in future U-space operations. This challenge is intensified by the highly combinatorial nature of multi aircraft conflict resolution, where the number of possible configurations grows quadratically with the size of the group \cite{granger_traffic_2003}.

On the other hand, in a decentralized approach  \cite{hoekstra_designing_2002}, the decision-making for conflict resolution is assigned to each pilot. Such a decentralized approach can be applied to UAS by assigning the conflict resolution task to the autonomous system. The benefit of this method is that it relies on multiple agents to resolve the conflict, in contrast to the centralized control, thus removing the single point of failure and improving robustness. However, the downside is that this approach typically has limited global situational awareness, and because  conflicts are solved  locally, resolutions may lead to a domino effect in more complex and dense traffic situations \cite{sunil_modeling_2017}.

To explore the effectiveness of decentralized approaches under uncertainty, this paper investigates the full process of conflict detection and resolution, which forms the core of autonomous separation in high-density airspace. Conflict detection typically involves predicting a loss of separation within a specified look-ahead time based on the current state (position and velocity) of surrounding traffic.

Building upon this foundation, the paper focuses on two representative decentralized resolution algorithms: the Modified Voltage Potential (MVP) method \cite{hoekstra_designing_2002} and the Velocity Obstacle (VO) method \cite{fiorini_motion_1998}. These were selected based on recent studies comparing state-based conflict resolution algorithms under high-density traffic conditions \cite{balasooriyan_multi-aircraft_2017}. Notably, the MVP method has demonstrated superior macroscopic performance in terms of safety, efficiency, and stability. The algorithm even outperforms more complex or jointly optimized algorithms due to its use of implicit coordination and the summation of avoidance vectors in multi-aircraft conflicts \cite{hoekstra_aerial_2021}. The VO approach, on the other hand, provides a geometric and intuitive solution framework widely used in robotics, and has been successfully adapted for air traffic scenarios \cite{van_dam_ecological_2008, ribeiro_velocity_2021}. Together, these methods offer valuable insights into scalable, robust, and interpretable decentralized conflict resolution strategies for future urban air mobility and U-space environments.

\subsection{State-based Conflict Detection}

\begin{figure}[htb]
    \centering
    \includegraphics[width = 0.9\linewidth]{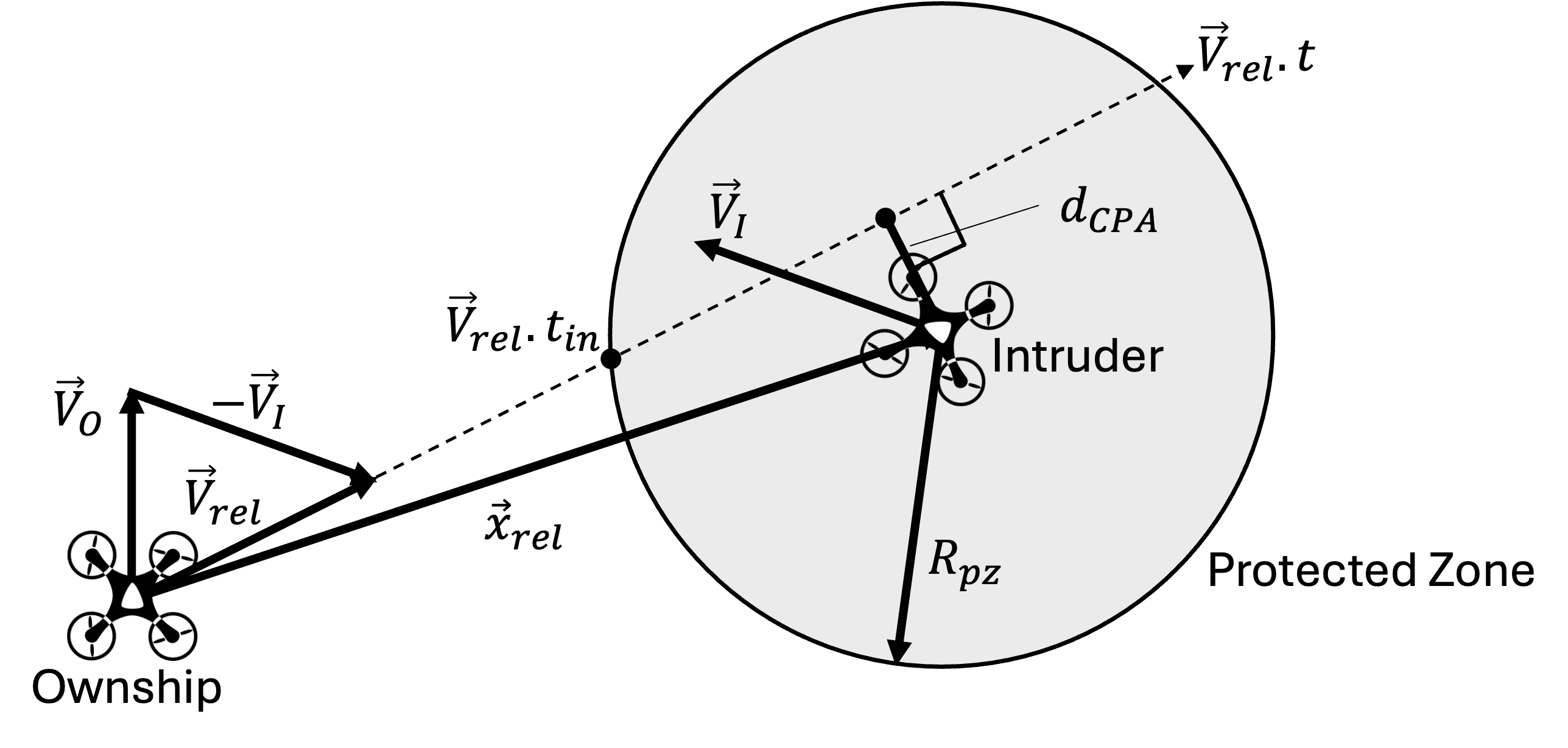}
    \caption{State-based conflict detection by calculating the closest point of approach.}
    \label{fig:conf_detect}
\end{figure}

State-based conflict detection uses the estimated relative trajectory, computed from the current relative position and extrapolated using the relative velocity. A spatial parameter, the radius of the protected zone \( R_{\mathrm{PZ}} \), and a temporal parameter, the look-ahead time \( t_{\text{lookahead}} \), define the condition for conflict. A conflict is said to occur if the magnitude of the projected distance at the closest point of approach, \( \| \mathbf{d}_{\mathrm{CPA}} \| \), is less than \( R_{\mathrm{PZ}} \), and the time to intrusion entry, \( t_{\mathrm{in}} \), is less than \( t_{\text{lookahead}} \), as shown in Eq.~\ref{eq:conf_def}. This situation is illustrated in Fig.~\ref{fig:conf_detect}. The following paragraphs describe the computation of \( \| \mathbf{d}_{\mathrm{CPA}} \| \) and \( t_{\mathrm{in}} \).

\begin{equation}
    \label{eq:conf_def}
    (\|\mathbf{d}_{\mathrm{CPA}}\| < R_{pz}) \land\ (t_{in} < t_{lookahead}) \implies \text{Conflict}
\end{equation}

The ownship aircraft, indexed by \( o \), with its position taken as the origin of the reference frame, moves with velocity \( \mathbf{V}_o \). A potential intruder, indexed by \( i \), located at position \( \mathbf{x}_i \), is flying near the ownship with velocity \( \mathbf{V}_i \). The relative velocity \( \mathbf{V}_{\mathrm{rel}} \) is defined as the difference between the ownship and intruder velocities, \( \mathbf{V}_o - \mathbf{V}_i \). The time to the closest point of approach, \( t_{\mathrm{CPA}} \), is calculated as shown in Eq.~\ref{eq:tcpa}.

\begin{equation}
    \label{eq:tcpa}
    t_{\mathrm{CPA}} = \frac{\mathbf{V}_{\mathrm{rel}} \cdot \mathbf{x}_{\mathrm{rel}}}{\| \mathbf{V}_{\mathrm{rel}} \|^2}
\end{equation}

Once \( t_{\mathrm{CPA}} \) is known, the vector \( \mathbf{d}_{\mathrm{CPA}} \) can be calculated as shown in Eq.~\ref{eq:dcpa}. With the vector distance at the closest point of approach determined, its magnitude \( \| \mathbf{d}_{\mathrm{CPA}} \| \) can be calculated. Then, using vector geometry or the Pythagorean relation, the time to intrusion entry \( t_{\mathrm{in}} \) can be obtained as shown in Eq.~\ref{eq:tin}.

\begin{equation}
    \label{eq:dcpa}
    \mathbf{d}_{\mathrm{CPA}} = \mathbf{x}_{\mathrm{rel}} - \mathbf{V}_{\mathrm{rel}} \cdot t_{\mathrm{CPA}}
\end{equation}

\begin{equation}
    \label{eq:tin}
    t_{\mathrm{in}} = t_{\mathrm{CPA}} - \frac{\sqrt{R_{\mathrm{PZ}}^2 - \| \mathbf{d}_{\mathrm{CPA}} \|^2}}{\| \mathbf{V}_{\mathrm{rel}} \|}
\end{equation}

The advantages of state-based conflict detection lie in its general applicability, owing to its clear and simple definition, and its robustness to deviations from the planned trajectory or cleared flight path. It is flexible and suitable for operations where the future trajectory is unknown, such as surveillance missions.

However, its main limitation is the potential to miss conflicts resulting from planned changes in the velocity vector. One approach to mitigate such false negatives is to apply the conflict detection logic not only to current states, but also to intended future velocity vectors. This enables conflict detection to function as a preventive tool, particularly when an aircraft plans to turn, change altitude, or change speed. This concept can be stated as a traffic rule: when not in conflict, an aircraft shall not change its heading, vertical speed, or velocity vector in a manner that would result in a conflict within the look-ahead time.

\subsection{Conflict Resolution Methods} 

\begin{figure}[htb]
    \centering
    \includegraphics[width = 0.8\linewidth]{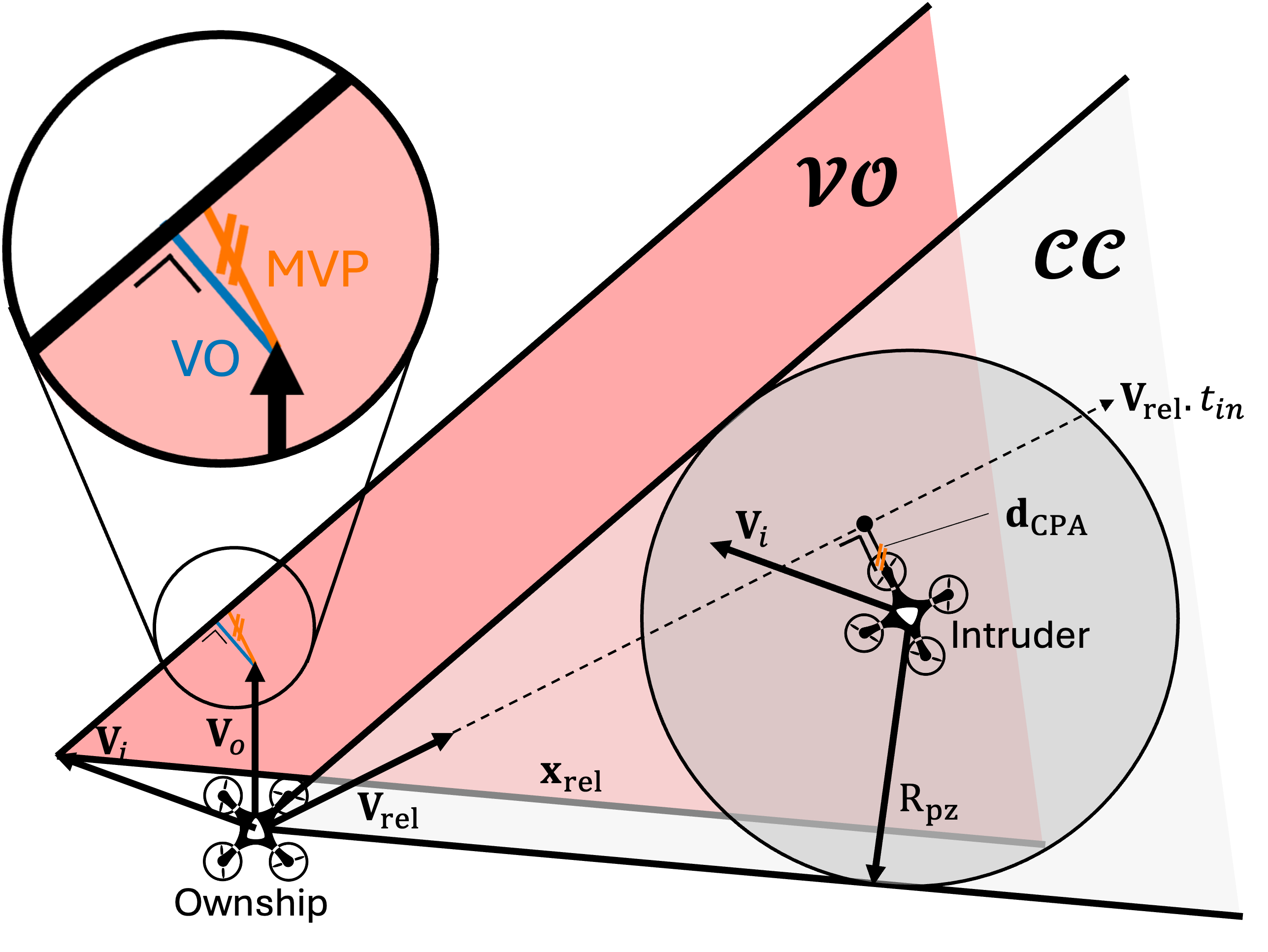}
    \caption{Solution space illustration and highlighted Velocity Obstacle (VO) and Modified Voltage Potential (MVP) choice of velocity for conflict resolution.}
    \label{fig:solution_space_illustration}
\end{figure}

Conflict resolution methods can be understood formally by examining the space of feasible velocity changes available to the ownship. A widely used representation of this solution space is the concept of Velocity Obstacles (\(\mathcal{VO}\)), originally introduced in the context of robotics~\cite{fiorini_motion_1998} and later adapted for aerial conflict resolution. A \(\mathcal{VO}\) is constructed by first defining a collision cone (\(\mathcal{CC}\)), bounded by tangents from the ownship to the protected zone around the intruder, which contains all relative velocity vectors that would lead to a predicted loss of separation. Translating this cone by the intruder’s velocity yields the \(\mathcal{VO}\) in absolute velocity space. The resulting region, illustrated in Fig.~\ref{fig:solution_space_illustration}, represents all ownship velocities that would result in a conflict within a specified time horizon. To ensure safe separation, the ownship must select a resolution velocity vector that lies outside the \(\mathcal{VO}\) set.

Building on this basic concept, several variations of the \(\mathcal{VO}\) framework have been developed to address multi-agent interactions, operational constraints, and traffic management rules. The Selective Velocity Obstacle (SVO)~\cite{jenie_selective_2015} incorporates right-of-way rules on top of the VO concept so that only non-priority aircraft manoeuvrer. The Optimal Reciprocal Collision Avoidance (ORCA)~\cite{berg_optimal_2010} extends the original formulation to cooperative scenarios, reducing oscillatory behaviour by sharing avoidance responsibility between agents. ORCA computes mutually feasible velocities as intersections of permitted half-planes, selecting an option close to the preferred trajectory while ensuring separation. For constant-speed platforms, Constant Speed ORCA (CSORCA)~\cite{durand_constant_2018} modifies the geometry to respect speed constraints and avoid deadlocks. The Dual-Horizon ORCA (DH-ORCA)~\cite{alligier_dual-horizon_2023} extends ORCA with two time horizons. A short horizon for standard ORCA constraints and a longer “cross” horizon for optional CSORCA-like constraints. The optional constraint is applied only when an aircraft’s preferred course is predicted to cross another’s trajectory within the cross horizon, with the decision made independently using only the aircraft’s own intent.

While the aforementioned methods are derived directly from the \(\mathcal{VO}\) set, other strategies can still be interpreted within the same velocity space framework. The Modified Voltage Potential (MVP) can be analysed in terms of the $\mathcal{VO}$ set, since its output corresponds to a point in the admissible region outside all conflict cones. In this study, MVP is compared against the shortest-way-out (SWO) variation—a geometric \(\mathcal{VO}\)-based strategy that selects the minimum-change velocity required to exit the forbidden region. In this paper, this approach is referred to as the VO method. The comparison is motivated by the fact that their resolution velocities are often geometrically close in velocity space, as shown in Figure~\ref{fig:solution_space_illustration}. This paper aims to examines how both methods perform under uncertainty. The following subsections present the mathematical formulation of each conflict resolution strategy.

\subsubsection{Modified Voltage Potential Algorithm}

The Airborne Separation Assurance System (ASAS), developed in 2002 by \cite{hoekstra_designing_2002} is the Modified Voltage Potential (MVP) algorithm. It was originally inspired by the work of Eby and Kelly \cite{eby_self-organizational_1995}. The MVP algorithm has been evaluated in multiple contexts, including both crewed and uncrewed aircraft, and under varying traffic densities \cite{ribeiro_review_2020, sunil_modeling_2017, balasooriyan_multi-aircraft_2017}.

Figure~\ref{fig:solution_space_illustration} illustrates the computation of the resolution vector in the MVP method. The resolution vector is derived using the relative position at the closest point of approach (CPA). The mathematical formulation of the resolution manoeuvre is given in Eq.~\ref{eq:mvp_dv}. An additional parameter, \( \varepsilon \), is introduced to ensure that the resulting resolution velocity avoids grazing the intruder's protected zone.

The resolution manoeuvre in the MVP method is defined by the following expression:

\begin{equation}
    \label{eq:mvp_dv}
    \mathbf{dV} = 
    \frac
    {
        \left( \frac{R_{\mathrm{PZ}}}{\varepsilon} - \| \mathbf{d}_{\mathrm{CPA}} \| \right)
    }
    {
        t_{\mathrm{CPA}} \cdot \| \mathbf{d}_{\mathrm{CPA}} \|
    }
    \cdot \mathbf{d}_{\mathrm{CPA}}
\end{equation}

where \( \varepsilon \) is a geometric buffer parameter that ensures the resolution vector avoids tangency with the protected zone, given by:

\begin{equation}
    \label{eq:mvp_epsilon}
    \varepsilon = \cos\left( 
        \arcsin\left(
            \frac{R_{\mathrm{PZ}}}{\| \mathbf{x}_{\mathrm{rel}} \|}
        \right)
        - 
        \arcsin\left(
            \frac{\| \mathbf{d}_{\mathrm{CPA}} \|}{\| \mathbf{x}_{\mathrm{rel}} \|}
        \right)
    \right)
\end{equation}

The final resolution velocity is obtained by adding the avoidance vector \( \mathbf{dV} \) to the nominal ownship velocity \( \mathbf{V}_{\text{own}} \), yielding:

\begin{equation}
    \label{eq:mvp_vres}
    \mathbf{V}_{\text{res}} = \mathbf{V}_{\text{own}} + \mathbf{dV}
\end{equation}

This formulation ensures that the ownship adjusts its velocity in a direction aligned with the projected distance at the closest point of approach, \( \mathbf{d}_{\mathrm{CPA}} \), scaled appropriately to ensure separation prior to \( t_{\mathrm{in}} \). When both the ownship and the projected closest point lie outside the protected zone, the correction factor \( \varepsilon \) refines the resolution manoeuvre to prevent grazing the protected boundary. Otherwise, a direct linear scaling is applied.

\subsubsection{Velocity Obstacle Algorithm}

The Velocity Obstacle (VO) algorithm resolves conflicts by selecting an avoidance velocity outside the forbidden region \(\mathcal{VO}\), as defined in the previous subsection. In the optimal-change formulation, referred to in this paper as the VO method, the resolution velocity \(\mathbf{V}_{\text{res}}\) is chosen as the point on the boundary of \(\mathcal{VO}\), denoted \(\partial \mathcal{VO}\), that minimizes the deviation from the nominal ownship velocity \(\mathbf{V}_{\text{own}}\). This approach, also known as the shortest-way-out strategy, ensures that the manoeuvre preserves the original trajectory as closely as possible while maintaining separation from the intruder, as illustrated in Fig.~\ref{fig:solution_space_illustration}.

\begin{equation}
    \label{eq:vo_optimal}
    \mathbf{V}_{\text{res}} = \arg \min_{\mathbf{V} \in \partial \mathcal{VO}} \left\| \mathbf{V} - \mathbf{V}_{\text{own}} \right\|
\end{equation}
\section{Uncertainty in UAS Navigation System}
\label{sec:nav_uncertainty}

Uncertainty is inherent in any physical system model. It is typically classified into two categories: aleatory and epistemic. Aleatory uncertainty arises from the inherent randomness of the system, while epistemic uncertainty results from incomplete knowledge, often introduced through assumptions or simplifications in the mathematical model.

Within the CNS system, different sources of uncertainty emerge. In the communication domain, uncertainty manifests as message drops or latency. Although epistemic uncertainty may arise due to model simplification, these phenomena are predominantly aleatory, as they naturally occur in communication channels. In navigation systems, measurement errors in position and velocity are also considered aleatory, as they stem from intrinsic sensor variability and environmental interactions. For surveillance systems, particularly in the context of ADS-L as an integrated CNS component, uncertainty can be attributed to both communication and navigation aspects due to the dependent-surveillance nature. Therefore, this study focuses primarily on quantifying uncertainty arising from aleatory sources.

In the field of Uncertainty Quantification (UQ), propagation analysis is central to understanding how uncertainties in input variables affect output behaviour. In linear systems, the distribution of outputs typically reflects that of the inputs. However, in non-linear systems, the mapping may distort the output distribution by shifting the mean, inducing skewness, or amplifying the tails.

For conflict resolution, input variables such as position and velocity are subject to measurement error, which propagates through the conflict detection and resolution (CD\&R) algorithms described in Section~\ref{sec:autonomous}. This problem setup corresponds to a forward uncertainty propagation, where stochastic inputs are propagated through the conflict detection and resolution logic to evaluate their impact on safety performance. 
Since the CD\&R algorithms form a non-linear mapping from these inputs to outputs such as miss distance or resolution success, the resulting output distributions cannot be computed analytically and must instead be approximated numerically. Monte Carlo simulation is therefore required to accurately capture the complex interactions between the input uncertainties and the resolution outcomes.

Some recent approaches incorporate navigation uncertainty directly into the conflict detection metric by modelling the relative position as a Gaussian random variable. 
Closed-form expressions have been proposed to compute the probability that an aircraft enters a predefined protected volume—such as a cuboid, ellipsoid, or cylinder—under trivariate Gaussian assumptions~\cite{zou_collision_2021}. 
This framework has also been extended to compute minimum separation distances that satisfy a specified Equivalent Level of Safety (ELoS), enabling risk-based separation design for pre-tactical airspace planning~\cite{zhong_demarcation_2024}.

While these probabilistic methods embed uncertainty into the detection logic itself, state-based conflict detection and resolution maintains the deterministic assumption and introduces navigation uncertainty at the evaluation stage through position and velocity perturbations. This approach uses a simple circular protected zone in two dimensions and extendable to spherical or cylindrical shapes in three dimensions \cite{jenie_three-dimensional_2016}. It avoids assumptions about more complex protected-zone geometries or fixed error models. The effect of the navigational error can be mitigated by quantifying it and tuning the spatio-temporal parameters to achieve a desirable CD\&R performance.

This distinction is supported by previous studies on the effects of CNS-related uncertainties, which have shown that both communication and navigation uncertainties degrade autonomous separation performance for UAS~\cite{rahman_effect_2024}. 
In particular,~\cite{rahman_autonomous_2024} provides a detailed analysis of how position uncertainty, as a component of navigation error, impacts the safety performance of conflict resolution. The degradation occurs because the uncertainty perturbs the resolution velocity away from the true value that would otherwise ensure successful conflict avoidance. 

To further examine the role of navigation systems as sources of uncertainty, the following section first describes the types of navigation systems commonly used in UAS, followed by the associated uncertainty modelling.

\subsection{UAS Navigation Systems}

Unmanned Aerial Systems (UAS) typically rely on Global Navigation Satellite Systems (GNSS) and Inertial Measurement Units (IMUs) to navigate in outdoor environments, providing absolute positioning and inertial references essential for flight control. However, GNSS availability is limited in indoor or GNSS-denied environments, restricting its applicability in such scenarios. Furthermore, due to the limited accuracy of GNSS and the drift-prone nature of IMUs, visual-based navigation methods have gained prominence as complementary approaches. Techniques such as Visual Odometry and Visual-Inertial Odometry enable UAS to estimate motion by analysing camera imagery, either independently or in combination with inertial data, thereby enhancing robustness in GNSS-denied settings. Visual Odometry, initially introduced in \cite{nister_visual_2004} and further reviewed in \cite{scaramuzza_visual_2011}, estimates motion solely from visual inputs and has become essential in robotic navigation. Visual-Inertial Odometry extends this capability by fusing visual and inertial measurements to achieve real-time, robust state estimation, as demonstrated in benchmarking studies \cite{delmerico_benchmark_2018}. Despite the development of these alternative methods, current standards for U-space operations and tactical separation remain heavily dependent on GNSS. Specifically, the proposed use of ADS-L for situational awareness, as outlined in the technical specifications~\cite{european_union_aviation_safety_agency_easa_technical_2022}, assumes the continuous availability and reliability of GNSS-derived data. The primary parameters transmitted via ADS-L relevant to tactical separation are the aircraft's position, ground speed, and ground track. These values are obtained from GNSS-based navigation solutions and are encoded in the transmitted payload. The system further defines navigation performance in terms of accuracy bounds, expressed as 95\% confidence intervals. For instance, horizontal position accuracy is categorized into thresholds such as \( <30~\mathrm{m} \), \( <10~\mathrm{m} \), and \( <3~\mathrm{m} \), while velocity accuracy may be defined as \( <10~\mathrm{m/s} \), \( <3~\mathrm{m/s} \), or \( <1~\mathrm{m/s} \), depending on the GNSS quality and system configuration.

Although the ADS-L specification provides accuracy bounds, it does not explicitly define uncertainty using Gaussian distributions or covariance matrices. However, in probabilistic robotics and state estimation, it is common practice to model sensor uncertainty as zero-mean Gaussian noise for analytical and computational tractability \cite{thrun_probabilistic_2002}. Under this assumption, the 95\% bounds specified by ADS-L are interpreted as approximately \( \pm 2\sigma \), allowing the derivation of a diagonal covariance matrix based on the stated horizontal position and velocity accuracy.

For the purposes of this study, which addresses uncertainty quantification and its propagation through state-based conflict detection and resolution (CD\&R) algorithms, the Gaussian assumption is adopted for modelling navigation uncertainty. This approach facilitates the evaluation of how uncertainty in input variables affects CD\&R outcomes and the safety performance of decentralized separation strategies through Monte Carlo simulations. The following subsection presents the mathematical formulation of position and velocity uncertainties, based on the ADS-L specification.

\subsection{Uncertainty Modelling}
\label{subsec:uncertainty_modeling}
To quantify the impact of navigation errors on conflict detection and resolution (CD\&R) algorithms, the position and velocity of each aircraft are modelled as stochastic variables. Specifically, both position and velocity are treated as two-dimensional Gaussian random vectors with independent components. This modelling approach is widely adopted in probabilistic robotics and air traffic management ~\cite{thrun_probabilistic_2002, paielli_conflict_1997, pour_probability_2019} and aligns with the uncertainty descriptors provided in the ADS-L technical specification, which defines accuracy using 95\% confidence bounds. Under the assumption of zero-mean Gaussian noise, these bounds are approximated as \( \pm 2\sigma \), allowing the construction of corresponding covariance matrices. These models form the basis for uncertainty propagation in the CD\&R algorithms using Monte Carlo simulation techniques.

The position vectors of the ownship and intruder aircraft are defined as:

\begin{equation}
\label{eq:stochastic_pos_def}
\mathbf{x}_o \sim \mathcal{N}(\boldsymbol{\mu}_o, \Sigma_o), \quad
\mathbf{x}_i \sim \mathcal{N}(\boldsymbol{\mu}_i, \Sigma_i)
\end{equation}

with nominal positions:

\begin{equation}
\label{eq:mean_pos}
\boldsymbol{\mu}_o = \begin{bmatrix} \bar{x}_o \\ \bar{y}_o \end{bmatrix}, \quad
\boldsymbol{\mu}_i = \begin{bmatrix} \bar{x}_i \\ \bar{y}_i \end{bmatrix}
\end{equation}

and diagonal covariance matrices:

\begin{equation}
\label{eq:std_pos}
\Sigma_o = \begin{bmatrix} \sigma_{x_o}^2 & 0 \\ 0 & \sigma_{y_o}^2 \end{bmatrix}, \quad
\Sigma_i = \begin{bmatrix} \sigma_{x_i}^2 & 0 \\ 0 & \sigma_{y_i}^2 \end{bmatrix}
\end{equation}

The velocity vectors are similarly modelled as:

\begin{equation}
\label{eq:stochastic_vel_def}
\mathbf{V}_o \sim \mathcal{N}(\boldsymbol{\nu}_o, \Sigma_{v_o}), \quad
\mathbf{V}_i \sim \mathcal{N}(\boldsymbol{\nu}_i, \Sigma_{v_i})
\end{equation}

with nominal velocities:

\begin{equation}
\label{eq:mean_vel}
\boldsymbol{\nu}_o = \begin{bmatrix} \bar{v}_{x_o} \\ \bar{v}_{y_o} \end{bmatrix}, \quad
\boldsymbol{\nu}_i = \begin{bmatrix} \bar{v}_{x_i} \\ \bar{v}_{y_i} \end{bmatrix}
\end{equation}

and covariance matrices:

\begin{equation}
\label{eq:std_vel}
\Sigma_{v_o} = \begin{bmatrix} \sigma_{v_{x_o}}^2 & 0 \\ 0 & \sigma_{v_{y_o}}^2 \end{bmatrix}, \quad
\Sigma_{v_i} = \begin{bmatrix} \sigma_{v_{x_i}}^2 & 0 \\ 0 & \sigma_{v_{y_i}}^2 \end{bmatrix}
\end{equation}

The relative position and velocity are defined as:

\begin{equation}
\label{eq:rel_def}
\mathbf{x}_{\mathrm{rel}} = \mathbf{x}_i - \mathbf{x}_o, \quad
\mathbf{V}_{\mathrm{rel}} = \mathbf{V}_o - \mathbf{V}_i
\end{equation}

Since \( \mathbf{x}_i \) and \( \mathbf{x}_o \) are modelled as independent Gaussian variables, their difference is also Gaussian:

\begin{equation}
\label{eq:rel_pos_dist}
\mathbf{x}_{\mathrm{rel}} \sim \mathcal{N}(\boldsymbol{\mu}_{\mathrm{rel}},\ \Sigma_{\mathrm{rel}})
\end{equation}

with

\begin{equation}
\label{eq:rel_pos_mean_cov}
\boldsymbol{\mu}_{\mathrm{rel}} = \boldsymbol{\mu}_i - \boldsymbol{\mu}_o, \quad
\Sigma_{\mathrm{rel}} = \Sigma_i + \Sigma_o
\end{equation}

The relative velocity follows the same structure:

\begin{equation}
\label{eq:rel_vel_dist}
\mathbf{V}_{\mathrm{rel}} \sim \mathcal{N}(\boldsymbol{\nu}_{\mathrm{rel}},\ \Sigma_{V_{\mathrm{rel}}})
\end{equation}

with

\begin{equation}
\label{eq:rel_vel_mean_cov}
\boldsymbol{\nu}_{\mathrm{rel}} = \boldsymbol{\nu}_o - \boldsymbol{\nu}_i, \quad
\Sigma_{V_{\mathrm{rel}}} = \Sigma_{v_o} + \Sigma_{v_i}
\end{equation}

The equations above define the position and velocity of both the ownship and the intruder as two-dimensional Gaussian random variables, consistent with the ADS-L uncertainty descriptors. Specifically, the positions $\mathbf{x}_o$ and $\mathbf{x}_i$ are modelled as normally distributed vectors with means $\boldsymbol{\mu}_o$ and $\boldsymbol{\mu}_i$, respectively, and diagonal covariance matrices $\Sigma_o$ and $\Sigma_i$. These means represent the nominal (expected) positions in Cartesian coordinates, while the covariance matrices quantify the uncertainty in each axis, assuming independence between the $x$ and $y$ components.

Similarly, the velocities $\mathbf{V}_o$ and $\mathbf{V}_i$ are treated as Gaussian random vectors with mean velocities $\boldsymbol{\nu}_o$ and $\boldsymbol{\nu}_i$, and corresponding covariance matrices $\Sigma_{v_o}$ and $\Sigma_{v_i}$. These models reflect the stochastic nature of velocity estimates from sensor noise.

The relative position $\mathbf{x}_{\mathrm{rel}}$ and relative velocity $\mathbf{V}_{\mathrm{rel}}$ are then defined as the differences between the intruder and ownship states. Since the original variables are independent Gaussians, these relative states are also Gaussian distributed. The mean of the relative position is simply the difference between the mean positions, while its covariance is the sum of the individual covariances. The same holds for the relative velocity.

To align the modelling with the ADS-L reporting standard, reported accuracy bounds are interpreted as circular Gaussian confidence regions. For example, a \(30~\mathrm{m}\) horizontal position accuracy at the 95\% confidence level yields standard deviations of \( \sigma_{x} = \sigma_{y} \approx 6.127~\mathrm{m} \), based on the inverse chi-squared quantile, such that 95\% of samples fall within a \(30~\mathrm{m}\) radius.
\section{Methodology}
\label{sec:methodology}

This study investigates how navigation uncertainties affect the performance of state-based conflict detection and resolution (CD\&R) algorithms. Two experimental phases are conducted. The first examines the propagation of navigation uncertainty into the output variables of conflict detection and resolution. The second evaluates the macroscopic safety implications through repeated simulations across multiple conflict geometries and uncertainty types.

All scenarios are designed as pairwise encounters between one ownship and one intruder. This configuration facilitates controlled variation of initial conditions and isolates the influence of navigation uncertainties. Prior research has indicated that most conflicts remain pairwise, even in high-density airspace~\cite{jenie_selective_2015}.

The conflict geometry is defined by three parameters: the intruder's heading relative to the ownship, the horizontal miss distance at closest point of approach (\(d_{\mathrm{CPA}}\)), and the ground speed. The ownship maintains a fixed heading of \(0^\circ\) and a fixed speed of \(20~\mathrm{kts}\), while the intruder’s heading is varied to alter the encounter geometry. The intruder speed is fixed at \(15~\mathrm{kts}\), serving as a representative value for conflict scenarios. Both aircraft fly at constant altitude. For the CD\&R logic, the protected zone radius is set to \(R_{\mathrm{PZ}} = 50~\mathrm{m}\), and the look-ahead time is fixed at \(t_{\mathrm{lookahead}} = 15~\mathrm{s}\), consistent with operational assumptions in recent literatures~\cite{ribeiro_review_2020, jenie_selective_2015, rahman_effect_2024}.

Uncertainties are modelled in two navigation input variables: position and velocity. These are treated as independent Gaussian random variables with parameters selected based on the ADS-L specification~\cite{european_union_aviation_safety_agency_easa_technical_2022}, as discussed in Section~\ref{subsec:uncertainty_modeling}. Horizontal position uncertainty is set with a \(2\sigma\) bound of \(30~\mathrm{m}\), corresponding to a standard deviation of approximately \(6.127~\mathrm{m}\) in each axis. Velocity uncertainty is represented by \(2\sigma\) bound of \(1~\mathrm{m/s}\), corresponding to a standard deviation approximately \(0.204~\mathrm{m/s}\) per axis~\cite{rahman_validation_2025}.

\subsection{Conflict Detection}
In the uncertainty propagation experiment, each source of navigation uncertainty (either in position or velocity) is introduced independently while all other variables remain deterministic. For each run, the encounter geometry is configured such that the deterministic time to intrusion \( t_{\mathrm{in}} \) equals the look-ahead threshold \( t_{\mathrm{lookahead}} \). By fixing this reference point, the simulation evaluates how uncertainty in the input propagates through the conflict detection and resolution pipeline. The initial separation \( \|\mathbf{d}_{\mathrm{CPA}}\| \) and relative heading angle \( \Delta\psi \) are varied systematically to sample a range of conflict scenarios.

For every scenario, \(10^4\) Monte Carlo samples are drawn based on the assumed Gaussian distribution of the uncertain input variable. Each sample is propagated through the conflict detection equations, producing distributions over four output variables: time to closest point of approach \( t_{\mathrm{CPA}} \), distance at closest point of approach \( \|\mathbf{d}_{\mathrm{CPA}}\| \), time to intrusion \( t_{\mathrm{in}} \), and detection probability. While the first three variables are deterministic in nominal conditions, the introduction of randomness causes them to become stochastic. The detection outcome, originally a binary decision, is expressed in probabilistic terms due to the uncertainty of the input.

The detection probability is estimated empirically from the Monte Carlo samples. Letting \( \left( \|\mathbf{d}_{\mathrm{CPA}}^{(i)}\|, t_{\mathrm{in}}^{(i)} \right) \) denote the outputs of sample \( i \), the detection probability is computed as the average number of samples satisfying the conflict condition, as shown in Equation~\eqref{eq:conf_detect_prob}.

\begin{align}
\label{eq:conf_detect_prob}
P_{\text{detect}} \approx \hat{P}_{\text{detect}} 
&= \frac{1}{N} \sum_{i=1}^{N} \mathbf{1} \Big[ 
    \left( \|\mathbf{d}_{\mathrm{CPA}}^{(i)}\| < R_{pz} \right) \nonumber \\
&\quad\quad\quad\quad\quad\quad \land\ \left( t_{\mathrm{in}}^{(i)} < t_{\mathrm{lookahead}} \right) 
\Big]
\end{align}

The analysis of single-sample detection probability provides only an instantaneous view of system performance. In operational settings, however, conflict detection is not limited to a single observation but occurs repeatedly as new surveillance updates arrive. A missed detection at one instant does not necessarily imply a total failure, since the following opportunities may still identify the conflict before intrusion. Conversely, persistent non-detection arises only when every detection attempt during the observation window fails. The final component of the conflict detection analysis therefore extends to this time-sequenced setting, modelling detection as a series of Bernoulli trials with evolving probability of detection.

\subsection{Conflict Resolution}
\label{subsec:cr_method}

Following the identification of a potential conflict, a resolution manoeuvre must be generated to restore separation within the look-ahead time. This section considers two resolution algorithms: the Modified Voltage Potential (MVP) and the Velocity Obstacle (VO) method, both introduced in Section~\ref{sec:autonomous}. These algorithms receive the relative state as input and output a resolution velocity vector \( \mathbf{V}_{\text{res}} \), based on the formulations provided in Eqs.~\eqref{eq:mvp_dv} and \eqref{eq:vo_optimal}, respectively.

To evaluate the effect of navigation uncertainty on conflict resolution, the same Monte Carlo samples generated for the detection phase are propagated through each resolution algorithm. In each case, uncertainty is applied to either position or velocity while holding the other input constant. Under position uncertainty, the ownship and intruder positions are perturbed using the Gaussian models introduced in Section~\ref{subsec:uncertainty_modeling}, while their velocities remain fixed. This affects both \( t_{\mathrm{CPA}} \) and \( \mathbf{d}_{\mathrm{CPA}} \). Conversely, when velocity uncertainty is considered, only the velocity vectors are sampled while positions remain deterministic. These uncertainties will create randomness and alter the resolution vector produced by MVP and VO.

For each sample, the resolution velocity \( \mathbf{V}_{\text{res}}^{(i)} \) is computed and stored. To assess whether the sample leads to successful conflict avoidance, the resolution vector is evaluated in the Velocity Obstacle (VO) frame. If \( \mathbf{V}_{\text{res}}^{(i)} \notin \mathcal{VO} \), the resolution is considered successful.

In addition to that, the effectiveness of the maneuver is further assessed by computing the post-resolution \( \| \mathbf{d}_{\mathrm{CPA}} \| \). Each aircraft is propagated forward by one second using the assigned resolution velocity. Let \( \Delta t = 1~\mathrm{s} \) be the time step. The updated positions are computed as:
\[
\begin{aligned}
\mathbf{x}_{\mathrm{own}}^{+} &= \mathbf{x}_{\mathrm{own}} + \mathbf{V}_{\mathrm{own}}^{\text{res}} \cdot \Delta t, \\
\mathbf{x}_{\mathrm{int}}^{+} &= \mathbf{x}_{\mathrm{int}} + \mathbf{V}_{\mathrm{int}}^{\text{res}} \cdot \Delta t
\end{aligned}
\]
From these positions, the updated relative state is calculated as:
\[
\mathbf{x}_{\mathrm{rel}}^{+} = \mathbf{x}_{\mathrm{int}}^{+} - \mathbf{x}_{\mathrm{own}}^{+}, \quad 
\mathbf{V}_{\mathrm{rel}}^{+} = \mathbf{V}_{\mathrm{own}}^{\text{res}} - \mathbf{V}_{\mathrm{int}}^{\text{res}}.
\]
The post-resolution distance at closest point of approach (\(\|\mathbf{d}_{\mathrm{CPA}}^{+}\|\)) is then evaluated using \( \mathbf{x}_{\mathrm{rel}}^{+} \) and \( \mathbf{V}_{\mathrm{rel}}^{+} \) to the Eq.~\eqref{eq:dcpa}.

This value measures the minimum predicted separation after the resolution maneuver is executed. Together with the VO classification, it provides a quantitative view of conflict resolution effectiveness under uncertainty.

\subsection{BlueSky Simulation}
In the conflict simulation phase, a batch of encounter scenarios is generated over a discretized grid of initial conditions. Each scenario defines a pairwise horizontal conflict geometry using parameters such as relative bearing (\(\Delta\psi\)), intruder speed, and a fixed initial \( \| \mathbf{d}_{\mathrm{CPA}} \| \). The simulation environment is implemented using the BlueSky open-source air traffic simulator~\cite{hoekstra_bluesky_2016}, which has been extended to support stochastic models of navigation uncertainty and to facilitate sample-based evaluation of conflict detection and resolution (CD\&R) outcomes.

Each configuration is defined by a unique combination of uncertainty sources and resolution algorithm. This configuration is then repeated simulations are executed to account for statistical variability. Noise is injected into position and/or velocity based on user-specified uncertainty modes, and is propagated through the full CD\&R pipeline. The conflict detection step evaluates pairwise interactions between an ownship and intruder over a look-ahead horizon. If a conflict is predicted, a resolution command is issued by either the MVP or VO algorithm, depending on the selected strategy. The initial condition of each encounter is designed such that the time to conflict entry satisfies \( t_{\mathrm{in}} = 1.5 \, t_{\mathrm{lookahead}} \), allowing the conflict to be detected sufficiently in advance under navigation uncertainty. Navigation data is updated at a rate of 1~Hz, and a communication uncertainty model is applied such that the navigation state is successfully received with 80\% probability at each update cycle. Each simulation configuration is repeated 50.000 times to achieve statistical significance.

Throughout each simulation run, the relative distance between the aircraft is tracked over time to evaluate the distance at closest point of approach (CPA). This distance is used to determine whether a loss of separation (LOS) has occurred. Two metrics are computed at the end of each batch: the Intrusion Prevention Rate (IPR) defined in Eq.~\eqref{eq:IPR} and the the distribution of CPA distances.

\begin{equation}
\label{eq:IPR}
    \mathrm{IPR} = \frac{n_{\text{conflict}} - n_{\text{LOS}}}{n_{\text{conflict}}}
\end{equation}

In Eq.~\eqref{eq:IPR}, \(n_{\text{conflict}}\) denotes the total number of conflict instances and \(n_{\text{LOS}}\) is the number of encounters in which the horizontal separation falls below the protected zone radius. The IPR metric is bounded between 0.0 and 1.0 and serves as a summary indicator of resolution effectiveness, higher values signify a greater fraction of conflicts resolved without loss of separation.

Another key metric is the distribution of \( \|\mathbf{d}_{\mathrm{CPA}}\| \), which indicates how close the aircraft come during an encounter. This metric helps evaluate how effectively the resolution algorithms increase the initial separation, originally set to 0 meters, to a safe distance exceeding the protected zone radius of 50 meters across different conflict geometries.
\section{Results and Discussion}
\label{sec:results}

This section presents the results of the experiments conducted to assess the effects of navigation uncertainty on conflict detection and resolution (CD\&R) algorithms. The analysis is structured into three subsections. First, the propagation of position and velocity uncertainties through the conflict detection process is examined, highlighting how these input variabilities affect the key detection metrics. Second, the performance of two conflict resolution strategies, Modified Voltage Potential (MVP) and Velocity Obstacle (VO), is evaluated under uncertainty, with a focus on the distribution of the resolution velocities and their ability to increase the \( \| \mathbf{d}_{\mathrm{CPA}} \| \) to a safe value. Finally, large-scale simulations using the BlueSky platform provide an assessment of overall system safety, quantified by the intrusion prevention rate (IPR).

\subsection{Conflict Detection}
\label{subsec:cd}

Under deterministic conditions, conflict detection produces a binary outcome. Either a conflict is detected or it is not. However, under uncertainty, this true or false outcome becomes probabilistic. The presence of noise in input variables such as position and velocity introduces variability into the conflict detection process, resulting in a detection probability that reflects the likelihood of a certain conflict condition being detected as true. To understand how this randomness propagates into the final detection outcome, it is necessary to analyse the key intermediate variables: the time to closest point of approach ($t_{\mathrm{CPA}}$), the distance at closest point of approach vector ($\mathbf{d}_{\mathrm{CPA}}$), and the time to intrusion entry ($t_{\mathrm{in}}$). Deriving the distributions of these variables, where possible, offers insight into the probabilistic behaviour of the conflict detection process. Section \ref{subsubsec:cd_pos_uncertainty} presents analytical solutions and approximations for these variables under position uncertainty, followed by the velocity uncertainty case in Section \ref{subsubsec:cd_vel_uncertainty}. Then, this paper presents the conflict detection probability in \ref{subsubsec:detection_probability}.

\subsubsection{Position Uncertainty}
\label{subsubsec:cd_pos_uncertainty}

To distinguish the effects of different sources of uncertainty, the subscript \(p\) is used to denote variables influenced by position uncertainty. When subjected to position uncertainty, the calculation of \( t_{\mathrm{CPA,p}} \), the time to closest point of approach, becomes a linear function of the random position vectors of both the ownship and the intruder. As a result, the mean and variance of \( t_{\mathrm{CPA,p}} \) can be derived analytically, and the variable follows a Gaussian distribution:

\begin{equation}
\label{eq:tcpa_pos_uncertainty}
    t_{\mathrm{CPA,p}} \sim \mathcal{N}(\mu_{t_{\mathrm{CPA,p}}},\ \sigma_{t_{\mathrm{CPA,p}}}^2)
\end{equation}

with the mean and variance given by

\begin{equation}
\label{eq:mean_tcpa_pos}
\mu_{t_{\mathrm{CPA,p}}} = \frac{\mathbf{V}_{\mathrm{rel}} \cdot \boldsymbol{\mu}_{\mathrm{rel}}}{\|\mathbf{V}_{\mathrm{rel}}\|^2}
\end{equation}

\begin{equation}
\label{eq:var_tcpa_pos}
\sigma_{t_{\mathrm{CPA,p}}}^2 = \frac{\mathbf{V}_{\mathrm{rel}}^T \Sigma_{\mathrm{rel}}\, \mathbf{V}_{\mathrm{rel}}}{{\|\mathbf{V}_{\mathrm{rel}}\|^4}}
\end{equation}

Under position uncertainty, the distribution of the closest point of approach vector, $\mathbf{d}_{\mathrm{CPA}}$, is confined to a one-dimensional subspace that is orthogonal to the relative velocity. This follows from the fact that $\mathbf{d}_{\mathrm{CPA}}$ can be expressed as a linear projection of the relative position vector $\mathbf{x}_{\mathrm{rel}}$ onto the subspace orthogonal to $\mathbf{v}_{\mathrm{rel}}$, as formally proven in ~\ref{app:dcpa_vector_dist}. Specifically, $\mathbf{d}_{\mathrm{CPA}} = P\, \mathbf{x}_{\mathrm{rel}}$, where $P$ is the projection matrix that removes the component along $\mathbf{v}_{\mathrm{rel}}$. As a result, $\mathbf{d}_{\mathrm{CPA}}$ follows a Gaussian distribution with non-zero variance only in the orthogonal direction and zero variance in the direction aligned with $\mathbf{v}_{\mathrm{rel}}$. The distribution lies along a single axis defined by the unit vector $\mathbf{v}_{\perp}$ orthogonal to the relative velocity. The transformed scalar variable $z = \mathbf{v}_{\perp}^\top \mathbf{d}_{\mathrm{CPA}}$ thus follows a univariate Gaussian distribution:

\begin{equation}
\label{eq:dcpa_transformed}
    z \sim \mathcal{N}\left( \mathbf{v}_{\perp}^\top P\, \boldsymbol{\mu}_{\mathrm{rel}},\; \mathbf{v}_{\perp}^\top P\, \Sigma_{\mathrm{rel}}\, P^\top\, \mathbf{v}_{\perp} \right)
\end{equation}

\begin{figure}[!htb]
    \centering
    \includegraphics[width = 0.8\linewidth]{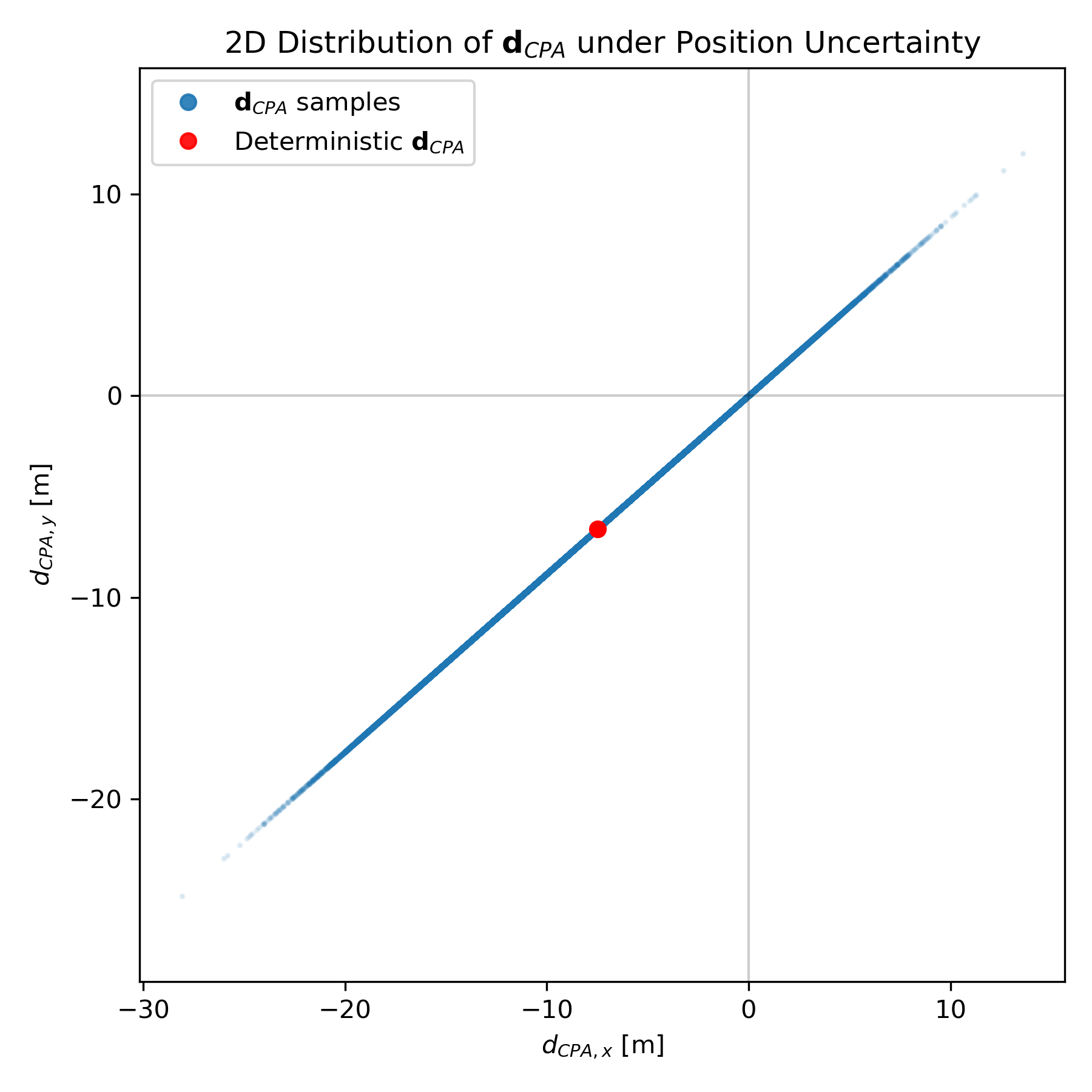}
    \caption{Distribution of $\mathbf{d}_{\mathrm{CPA}}$ under position uncertainty. The spread is constrained to the direction orthogonal to $\mathbf{V}_{\mathrm{rel}}$.}
    \label{fig:dcpa_2d_pos_uncertainty}
\end{figure}

From this, the magnitude of the closest point of approach vector, $\|\mathbf{d}_{\mathrm{CPA,p}}\|$, corresponds to the absolute value of the normally distributed scalar $z$. Therefore, it follows a folded normal distribution:

\begin{equation}
\label{eq:dcpa_pos_uncertainty}
    \|\mathbf{d}_{\mathrm{CPA,p}}\| \sim \left| \mathcal{N}(\mu_{z,\mathrm{p}},\ \sigma_{z,\mathrm{p}}^2) \right|
\end{equation}

where $\mu_{z,\mathrm{p}}$ and $\sigma_{z,\mathrm{p}}^2$ are the mean and variance of the univariate normal distribution defined in Equation~\ref{eq:dcpa_transformed}.

This formulation is consistent with the illustration in Figure~\ref{fig:dcpa_2d_pos_uncertainty}. In the figure, the origin represents the point along the $\mathbf{V}_{\mathrm{rel}} \cdot t_{\mathrm{CPA}}$ line where the magnitude of $\mathbf{d}_{\mathrm{CPA}}$ is zero. Each sample of $\mathbf{d}_{\mathrm{CPA}}$ lies along the line orthogonal to the relative velocity. The magnitude $\|\mathbf{d}_{\mathrm{CPA,p}}\|$ corresponds to the Euclidean distance from the origin to each sample along this line. Since the scalar projection $z$ is normally distributed, the magnitude $\|\mathbf{d}_{\mathrm{CPA,p}}\|$ is simply the absolute value of $z$, which results in a folded normal distribution as expressed in Equation~\ref{eq:dcpa_pos_uncertainty}.

The approximation of the time to intrusion entry, $t_{\mathrm{in}}$, is modelled as a non-linear function of the relative position $\mathbf{x}_{\mathrm{rel}}$. Given that $\mathbf{x}_{\mathrm{rel}} \sim \mathcal{N}(\boldsymbol{\mu}_{\mathrm{rel}}, \Sigma_{\mathrm{rel}})$ and the relative velocity $\mathbf{V}_{\mathrm{rel}}$ is deterministic, the exact distribution of $t_{\mathrm{in}}$ can not be expressed analytically. A first-order approximation can be obtained using the delta method. The non-linear transformation is denoted by $h(\mathbf{x}_{\mathrm{rel}})$, and the approximated mean and variance are given by:

\begin{equation}
\label{eq:mean_tin_pos}
\mu_{t_{\mathrm{in,p}}} \approx h(\boldsymbol{\mu}_{\mathrm{rel}})
\end{equation}

\begin{equation}
\label{eq:var_tin_pos}
\sigma_{t_{\mathrm{in,p}}}^2 \approx \nabla h(\boldsymbol{\mu}_{\mathrm{rel}})^\top \Sigma_{\mathrm{rel}}\, \nabla h(\boldsymbol{\mu}_{\mathrm{rel}})
\end{equation}

The function $h(\cdot)$ is defined in ~\ref{app:tin_pos_uncertainty} as a function of both the time and distance to closest point of approach. The gradient $\nabla h(\boldsymbol{\mu}_{\mathrm{rel}})$ describes how small changes in the relative position affect the estimated time to intrusion. The approximation holds only when the condition $\|\mathbf{d}_{\mathrm{CPA}}(\boldsymbol{\mu}_{\mathrm{rel}})\|^2 < R^2$ is satisfied, ensuring the square root term remains real.

\begin{figure}[!htb]
    \centering
    \includegraphics[width = 0.6\linewidth]{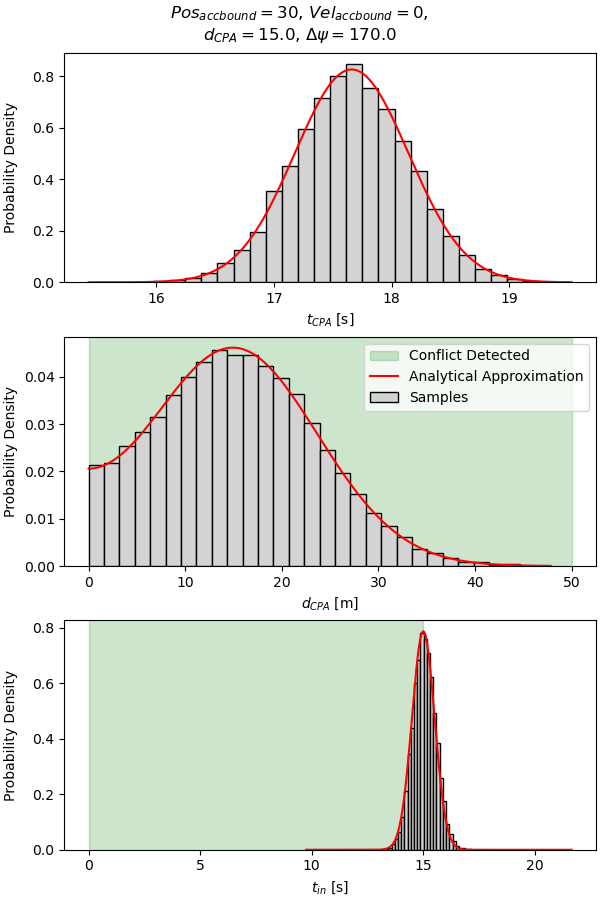}
    \caption{Probability density comparison between numerical simulation and analytical approximation for $t_{\mathrm{CPA}}$, $\|\mathbf{d}_{\mathrm{CPA}}\|$, and $t_{\mathrm{in}}$ when $\mathbf{d}_{\mathrm{CPA}} = 15$ meters. The approximation closely matches the simulation.}
    \label{fig:CD_position_uncertainty_approximation_dcpa15_dpsi_170}
\end{figure}

\begin{figure}[!htb]
    \centering
    \includegraphics[width = 0.6\linewidth]{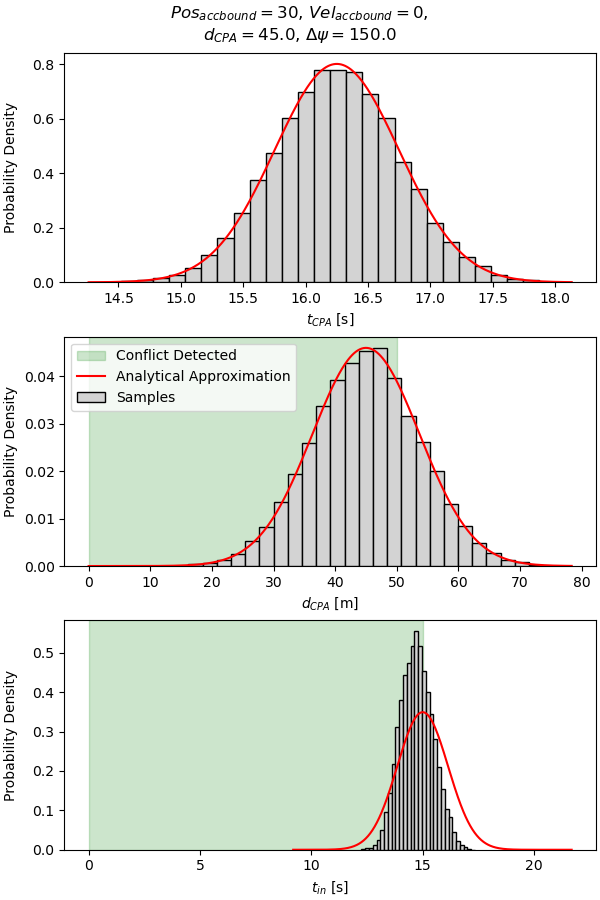}
    \caption{Same comparison when $\|\mathbf{d}_{\mathrm{CPA}}\| = 45$ meters, near the $R_{PZ}$ boundary. Deviation occurs in $t_{\mathrm{in}}$ due to samples in $\mathbf{d}_{\mathrm{CPA}}$ exceeding $R_{PZ}^2$, where $t_{\mathrm{in}}$ becomes undefined.}
    \label{fig:CD_position_uncertainty_approximation_dcpa45_dpsi_150}
\end{figure}

Figures~\ref{fig:CD_position_uncertainty_approximation_dcpa15_dpsi_170} and~\ref{fig:CD_position_uncertainty_approximation_dcpa45_dpsi_150} compare the numerical simulation results with the analytical solutions or approximations for the distributions of $t_{\mathrm{CPA}}$, $\|\mathbf{d}_{\mathrm{CPA}}\|$, and $t_{\mathrm{in}}$. In both cases, the analytical results closely match the numerical distributions. However, a significant deviation occurs in the distribution of $t_{\mathrm{in}}$ when the $\mathbf{d}_{\mathrm{CPA}}$ is \(45~\mathrm{m}\) near the boundary of the $R_{PZ}$, as shown in Figure~\ref{fig:CD_position_uncertainty_approximation_dcpa45_dpsi_150}. This discrepancy arises because a portion of the samples yield values of $\|\mathbf{d}_{\mathrm{CPA}}\|^2$ greater than $R_{PZ}^2$, rendering $t_{\mathrm{in}}$ undefined for those parts.

\subsubsection{Velocity Uncertainty}
\label{subsubsec:cd_vel_uncertainty}

When velocity uncertainty is present, the calculations of the time to closest point of approach \( t_{\mathrm{CPA,v}} \) and the magnitude of the closest point of approach vector \( \| \mathbf{d}_{\mathrm{CPA,v}} \| \) become functions of the relative velocity vector \( \mathbf{V}_{\mathrm{rel}} \), which is modelled as a Gaussian random variable. Since both expressions are non-linear in \( \mathbf{V}_{\mathrm{rel}} \), their exact distributions can not be analytically defined. To approximate their mean and variance, we apply the delta method as in the $t_{\mathrm{in}}$ calculation for position uncertainty.

Under velocity uncertainty, the time to closest point of approach, \( t_{\mathrm{CPA,v}} \), is a non-linear function of the relative velocity vector \( \mathbf{V}_{\mathrm{rel}} \sim \mathcal{N}(\boldsymbol{\mu}_{\mathrm{v}}, \Sigma_v) \). Using the delta method, the distribution of \( t_{\mathrm{CPA,v}} \) can be approximated as:

\begin{equation}
\label{eq:tcpa_vel_uncertainty}
t_{\mathrm{CPA,v}} \sim \mathcal{N}(\mu_{t_{\mathrm{CPA,v}}},\ \sigma^2_{t_{\mathrm{CPA,v}}})
\end{equation}

\begin{equation}
\mu_{t_{\mathrm{CPA,v}}} \approx f(\boldsymbol{\nu}_{\mathrm{rel}}), \quad
\sigma^2_{t_{\mathrm{CPA,v}}} \approx \nabla f(\boldsymbol{\nu}_{\mathrm{rel}})^\top \Sigma_{V_{\mathrm{rel}}}\, \nabla f(\boldsymbol{\nu}_{\mathrm{rel}})
\end{equation}

where \( f(\mathbf{V}_{\mathrm{rel}}) \) is the non-linear function defined in ~\ref{app:velocity_uncertainty_app}, along with its gradient \( \nabla f \).

The distribution of the distance at closest point of approach vector, $\mathbf{d}_{\mathrm{CPA}}$, is strongly influenced by the uncertainty in the relative velocity $\mathbf{V}_{\mathrm{rel}}$. Since $t_{\mathrm{CPA}}$ varies with each realization of $\mathbf{V}_{\mathrm{rel}}$, so does the projected point $\mathbf{x}_{\mathrm{rel}} - t_{\mathrm{CPA}} \cdot \mathbf{V}_{\mathrm{rel}}$, which defines $\mathbf{d}_{\mathrm{CPA}}$. Geometrically, this projection traces out an arc, centred at the midpoint of $\mathbf{x}_{\mathrm{rel}}$, with radius equal to half its magnitude. This result is formalized in ~\ref{app:locus_projection}, which shows that the set of all orthogonal projections of a fixed point onto lines through another point forms a circle.

The radius of this circular distribution increases with higher relative speed. This is because the conflict is assumed to occur at a fixed look-ahead time, $t_{\mathrm{in}} = t_{\mathrm{lookahead}}$, meaning that larger velocities imply a greater extrapolated distance from the current position, and thus a larger spread in the projected points. 
The proof in ~\ref{app:locus_projection} also shows that with the same relative velocity, when the ownship and intruder get closer in space, the radius of the distribution will be reduced.

\begin{figure}[!htb]
    \centering
    \includegraphics[width = 0.7\linewidth]{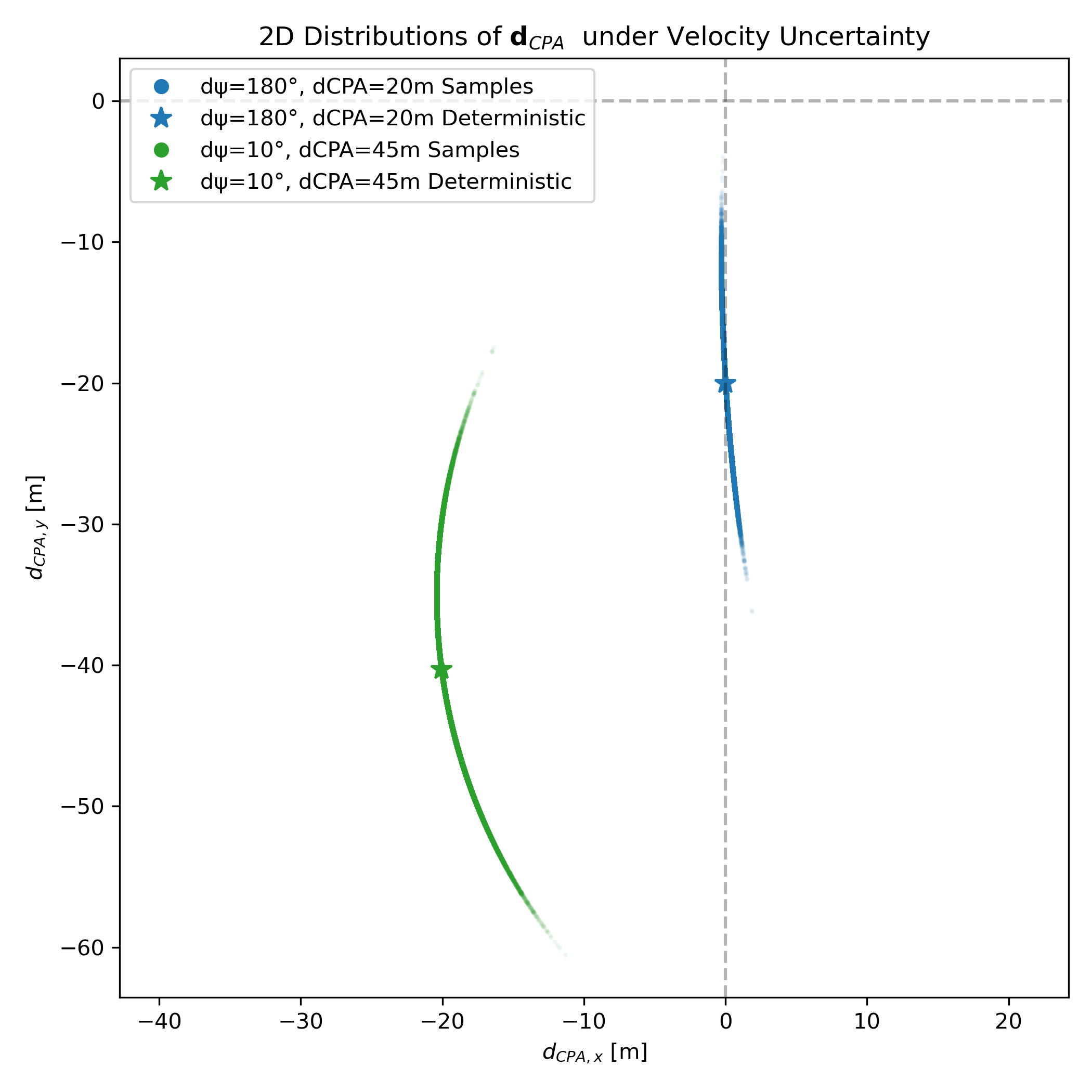}
    \caption{Two-dimensional distribution of $\mathbf{d}_{\mathrm{CPA}}$ under velocity uncertainty, showing arc-shaped uncertainty structures for different conflict geometries.}
    \label{fig:velo_uncertainty_dcpa_dist}
\end{figure}

Figure~\ref{fig:velo_uncertainty_dcpa_dist} illustrates the two-dimensional distribution of the $\mathbf{d}_{\mathrm{CPA}}$ under velocity uncertainty for two different conflict geometries. The blue arc corresponds to a case with an initial offset of \(20~\mathrm{m}\) and a relative heading difference of $180^\circ$, resulting in a high relative speed. The green arc represents a configuration with a larger initial offset of \(45~\mathrm{m}\) but a smaller heading difference of $10^\circ$, leading to a lower relative speed. In both cases, the conflict is defined to occur at a fixed look-ahead time. The higher relative velocity results in a larger relative distance, which increases the radius of the arc traced by the $\mathbf{d}_{\mathrm{CPA}}$ samples. In contrast, the lower-speed case yields a more tightly curved arc due to the shorter extrapolation distance.

Unlike the position uncertainty case, where the distribution of $\mathbf{d}_{\mathrm{CPA}}$ lies along a straight line orthogonal to the relative velocity vector, velocity uncertainty induces a curved projection geometry. As shown in ~\ref{app:locus_projection}, the set of projection points from a fixed position onto varying velocity directions traces out a circular arc, centred at half the relative position with a magnitude of half the relative position.

Each realization of $\mathbf{d}_{\mathrm{CPA}}$ under velocity uncertainty can be naturally expressed in polar form as shown in Eq.~\eqref{eq:dcpa_decomp}. This decomposition allows the expression of the angular component \( \phi \) as a closed-form distribution. ~\ref{app:pn_phi} proves that \( \phi \) follows a projected normal distribution. This result arises from the fact that \( \phi \) is a deterministic rotation of the relative velocity angle \( \theta \), which is itself projected normal. Since the projected normal family is closed under rotational transformation, \( \phi \) inherits this distribution. While the form of the distribution is known, its exact parameters depend on the encounter geometry and are analytically complex to derive.

\begin{equation}
\label{eq:dcpa_decomp}
  \mathbf{d}_{\mathrm{CPA}} = \frac{\mathbf{x}_{\mathrm{rel}}}{2} + \|\frac{\mathbf{x}_{\mathrm{rel}}}{2}\| \begin{bmatrix} \cos\phi \\ \sin\phi \end{bmatrix},
\end{equation}

On the other hand, the magnitude \( \|\mathbf{d}_{\mathrm{CPA}}\| \), which is computed from the origin to each projected point, cannot be expressed in a closed-form distribution. This is due to the non-linear dependence of the trigonometric components shown in Eq.~\eqref{eq:dcpa_decomp}. However, an approximate distribution for the magnitude can still be obtained using a first-order expansion. As outlined in ~\ref{app:velocity_uncertainty_app}, we apply the delta method to linearize \( \|\mathbf{d}_{\mathrm{CPA}}\| \) around the mean. The resulting distribution is approximately folded normal, reflecting the non-negativity of the distance.

\begin{equation}
\label{eq:dcpa_vel_uncertainty}
\|\mathbf{d}_{\mathrm{CPA,v}}\| \sim \left| \mathcal{N}(\mu_{d_{\mathrm{CPA,v}}},\ \sigma_{d_{\mathrm{CPA,v}}}^2) \right|
\end{equation}

\begin{align}
\label{eq:mean_dcpa_vel}
\mu_{d_{\mathrm{CPA,v}}} &\approx \left| \mathbf{x}_{\mathrm{rel}} - \mu_{t_{\mathrm{CPA,v}}} \cdot \boldsymbol{\nu}_{\mathrm{v}} \right| \\
\sigma_{d_{\mathrm{CPA,v}}}^2 &\approx \nabla g(\boldsymbol{\nu}_{\mathrm{v}})^\top \Sigma_{V_{\mathrm{rel}}}\, \nabla g(\boldsymbol{\nu}_{\mathrm{v}})
\end{align}

Here, the function \( g(\mathbf{V}_{\mathrm{rel}}) = | \mathbf{x}_{\mathrm{rel}} - f(\mathbf{V}_{\mathrm{rel}}) \cdot \mathbf{V}_{\mathrm{rel}} | \) maps the velocity input to the CPA distance, and its gradient is evaluated using the chain rule. The full derivation is provided in ~\ref{app:velocity_uncertainty_app}.

The accuracy of the folded normal approximation for $\|\mathbf{d}_{\mathrm{CPA,v}}\|$ under velocity uncertainty relies on the assumption that the non-linear mapping from $\mathbf{V}_{\mathrm{rel}}$ to $\mathbf{d}_{\mathrm{CPA}}$ is approximately linear in the region where most probability mass is concentrated. This assumption becomes more valid when the relative speed $|\mathbf{V}_{\mathrm{rel}}|$ is large, which stabilizes the geometry and flattens the arc traced by $\mathbf{d}_{\mathrm{CPA}}$ over variations in velocity directions.

As shown in Figure~\ref{fig:velo_uncertainty_dcpa_dist}, high relative velocity leads to an arc-shaped distribution that closely resembles a straight line segment. In such cases, the Euclidean norm of $\mathbf{d}_{\mathrm{CPA}}$ becomes nearly a linear function of velocity deviations, making the delta method a reliable approximation. Conversely, when the relative speed is low, the arc becomes more curved, and the projection geometry becomes increasingly non-linear in Cartesian coordinates.

In the case of high velocity uncertainty, another source of approximation error arises. The spread in sampled velocity directions can include realizations that are nearly orthogonal to the mean relative velocity. These lead to $\mathbf{d}_{\mathrm{CPA}}$ vectors that are rotated away from the true direction, including extreme cases where the projections form a near-complete circle. Even if the magnitude of these velocity vectors remains close to the mean, the directional deviation introduces asymmetry in the distribution of $\|\mathbf{d}_{\mathrm{CPA,v}}\|$, violating the conditions under which the folded normal approximation is valid.

Therefore, the approximation degrades when both the relative velocity is low and the uncertainty is high, due to increased curvature and bias in the distribution of the projected $\mathbf{d}_{\mathrm{CPA}}$ vectors. These effects highlight the non-linear nature of the transformation and the limitations of linearization-based methods in such conditions.

\begin{figure}[!htb]
    \centering
    \includegraphics[width = 0.6\linewidth]{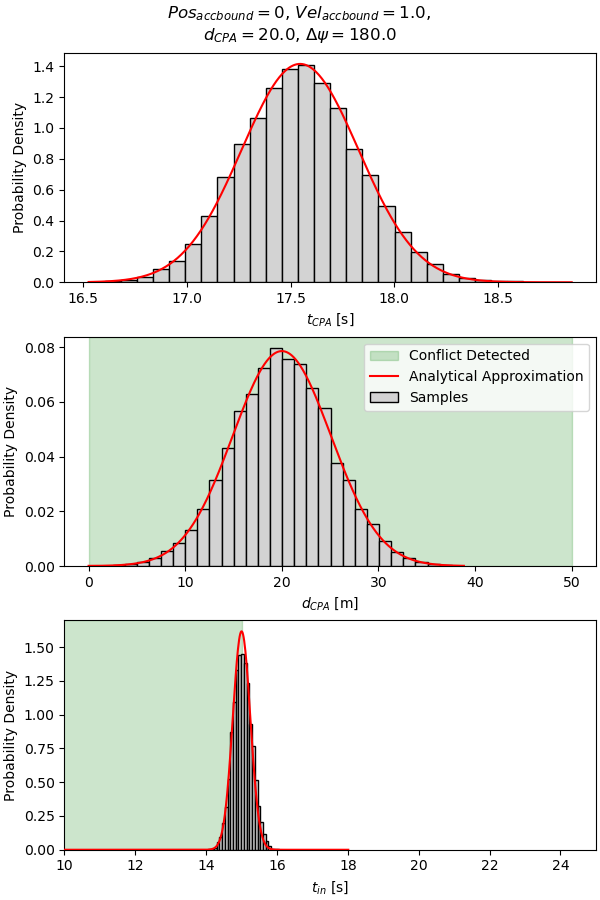}
    \caption{Probability density comparison between numerical simulation and analytical approximation for $t_{\mathrm{CPA}}$, $\|\mathbf{d}_{\mathrm{CPA}}\|$, and $t_{\mathrm{in}}$ under velocity uncertainty. Shown here is the case with high relative velocity: $\mathrm{d}_{\mathrm{CPA}} = 20$~m and $\Delta\psi = 180^\circ$. The approximation closely matches the simulation.}
    \label{fig:CD_vel_uncertainty_highrelvelo}
\end{figure}

\begin{figure}[!htb]
    \centering
    \includegraphics[width = 0.6\linewidth]{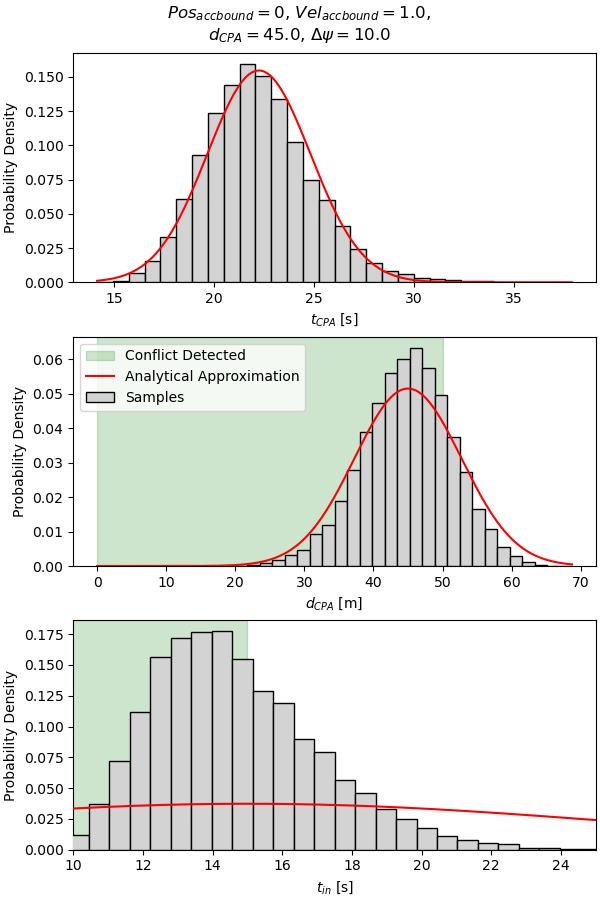}
    \caption{Same comparison under lower relative velocity: $\mathrm{d}_{\mathrm{CPA}} = 45$~m and $\Delta\psi = 10^\circ$. Deviation appears in all metrics due to increased curvature and directional bias in the $\mathbf{d}_{\mathrm{CPA}}$ distribution, reducing the validity of the linear approximation.}
    \label{fig:CD_vel_uncertainty_lowrelvelo}
\end{figure}

To complete the analysis under velocity uncertainty, the intrusion entry time \( t_{\mathrm{in}} \) is also approximated using the delta method. As defined in Equation~\eqref{eq:tin_function}, \( t_{\mathrm{in}} \) depends non-linearly on both \( t_{\mathrm{CPA}} \) and \( \|\mathbf{d}_{\mathrm{CPA}}\| \), which are themselves functions of the random vector \( \mathbf{V}_{\mathrm{rel}} \). The composite function is defined as:

\begin{equation}
\label{eq:h_vel_uncertainty}
h(\mathbf{V}_{\mathrm{rel}}) = f(\mathbf{V}_{\mathrm{rel}}) - \frac{\sqrt{R^2 - g(\mathbf{V}_{\mathrm{rel}})^2}}{|\mathbf{V}_{\mathrm{rel}}|}
\end{equation}

where \( f \) and \( g \) represent the CPA time and distance functions, respectively. The mean and variance of \( t_{\mathrm{in}} \) are then approximated by evaluating \( h \) and its gradient at the mean velocity:

\begin{equation}
\label{eq:mean_tin_vel}
\mu_{t_{\mathrm{in, v}}} \approx h(\boldsymbol{\nu}_{{v}})
\end{equation}

\begin{equation}
\label{eq:var_tin_vel}
\sigma^2_{t_{\mathrm{in, v}}} \approx \nabla h(\boldsymbol{\nu}_{{v}})^\top \Sigma_{V_{\mathrm{rel}}} \nabla h(\boldsymbol{\nu}_{{v}})
\end{equation}

The full derivation of \( \nabla h(\boldsymbol{\nu}_{\mathrm{v}}) \), which combines the gradients of \( f \), \( g \), and \( |\mathbf{v}_{\mathrm{rel}}| \), is provided in ~\ref{app:velocity_uncertainty_app}.

Figures~\ref{fig:CD_vel_uncertainty_highrelvelo} and~\ref{fig:CD_vel_uncertainty_lowrelvelo} compare the numerical simulation results with the analytical approximations for the distributions of $t_{\mathrm{CPA}}$, $\|\mathbf{d}_{\mathrm{CPA}}\|$, and $t_{\mathrm{in}}$ under velocity uncertainty. In the high relative velocity case (Figure~\ref{fig:CD_vel_uncertainty_highrelvelo}), the approximation closely follows the simulated distributions across all three metrics. This is valid since the the arc-shaped $\mathbf{d}_{\mathrm{CPA}}$ distribution is approximately linear. In contrast, Figure~\ref{fig:CD_vel_uncertainty_lowrelvelo} shows a low relative velocity case, where the curvature and directional spread of $\mathbf{d}_{\mathrm{CPA}}$ increase. Moreover, the presence of samples with $\|\mathbf{d}_{\mathrm{CPA}}\|^2 > R_{PZ}^2$, for which $t_{\mathrm{in}}$ is undefined, further amplifies the difference between the analytical approximation and the simulation. These results highlight the limitations of linear approximations in low relative speed, high velocity uncertainty situations, where non-linear effects become more prevalent.

An analytical approximation for the combined position and velocity uncertainty case is not provided due to the significant increase in complexity. Moreover, as shown in the velocity uncertainty case, the accuracy of the linear approximation remains limited to specific conditions. Extending this approach to account for both sources of uncertainty would likely result in highly unreliable estimates, thereby undermining the purpose of the approximation.

\subsubsection{Detection Probability}
\label{subsubsec:detection_probability}

After deriving analytical solutions and approximations for \( t_{\mathrm{CPA}} \), \( \|\mathbf{d}_{\mathrm{CPA}}\| \), and \( t_{\mathrm{in}} \), the next variable of interest is the detection probability. Recall that the conflict detection algorithm provides a binary output: True (conflict) or False (no conflict). In the presence of uncertainty, this binary outcome becomes probabilistic. The resulting detection probability reflects the likelihood of a conflict being detected under uncertain input conditions.

To evaluate this probability, an experiment was conducted across various initial values of \( \mathrm{d}_{\mathrm{CPA}} \) and \( \Delta\psi \), under the condition \( t_{\mathrm{in}} = t_{\mathrm{lookahead}} \), which places the conflict geometry exactly at the detection threshold. The detection probability is then approximated using Eq.~\eqref{eq:conf_detect_prob}

\begin{figure}[!htb]
    \centering
    \includegraphics[width = 0.7\linewidth]{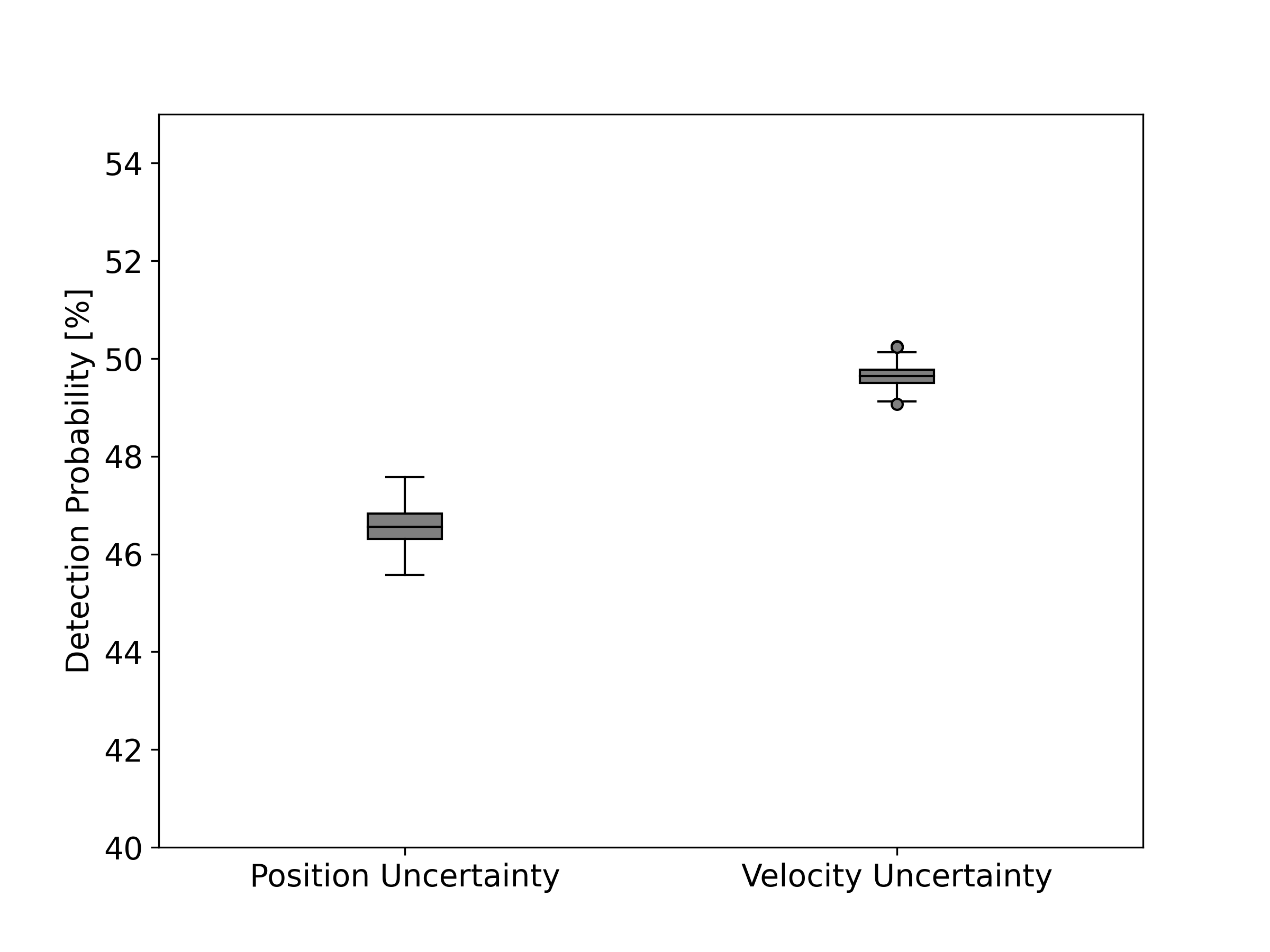}
    \caption{Detection probability under position and velocity uncertainty}
    \label{fig:CD_detect_prob}
\end{figure}

Figure~\ref{fig:CD_detect_prob} presents the distribution of detection probabilities under position and velocity uncertainty, when \( t_{\mathrm{in}} = t_{\mathrm{lookahead}} \), assuming a position standard deviation of 30 meters and a velocity standard deviation of 1~m/s. Under position uncertainty, the mean detection probability is 46.55\%, with a maximum of 47.58\% at \( \Delta\psi = 140^\circ \), \( \|\mathbf{d}_{\mathrm{CPA}}\| = 0 ~\mathrm{m}\), and a minimum of 45.58\% at \( \Delta\psi = 70^\circ \), \( \mathrm{d}_{\mathrm{CPA}} = 15~\mathrm{m}\). In comparison, under velocity uncertainty, the mean detection probability is slightly higher at 48.25\%, with a maximum of 49.61\% at \( \Delta\psi = 150^\circ \), \( \|\mathbf{d}_{\mathrm{CPA}}\| = 25~\mathrm{m}\), and a minimum of 46.82\% at \( \Delta\psi = 90^\circ \), \( \|\mathbf{d}_{\mathrm{CPA}}\| = 15~\mathrm{m}\). For both types of uncertainty, no specific trend was observed in detection probability with respect to relative heading or initial offset. Note that these results are specific to the chosen uncertainty magnitudes and may vary with different configurations.

While the detection probability is under 50\% for \( t_{\mathrm{in}} = t_{\mathrm{lookahead}} \), it is equally interesting to observe the detection probability variation in time. Tracing back to the  previous variables, the \( \mathbf{d}_{\mathrm{CPA}}\) distribution under position uncertainty remains unchanged because it is restricted to the orthogonal projection of the relative velocity vector that remains unchanged. This implies that the variance of \( t_{\mathrm{in}} \) distribution remains the same and only the mean is shifted in time. Depending on the uncertainty level, it is likely that some conflicts are detected earlier in time (when estimated \( t_{\mathrm{in}} > t_{\mathrm{lookahead}} \)) due to the spread of the distribution. Then, as the ownship and intruder get closer, the remaining time to intrusion decreases while the look-ahead threshold remains fixed, causing the detection probability to rise accordingly. Visually, this means that in Figure \ref{fig:CD_position_uncertainty_approximation_dcpa15_dpsi_170}, \ref{fig:CD_position_uncertainty_approximation_dcpa45_dpsi_150}, \ref{fig:CD_vel_uncertainty_highrelvelo}, and \ref{fig:CD_vel_uncertainty_lowrelvelo}, the distribution of time to intrusion shifts to the left into the "green" conflict detected area.

\begin{figure}[!htb]
    \centering
    \includegraphics[width = 1.0\linewidth]{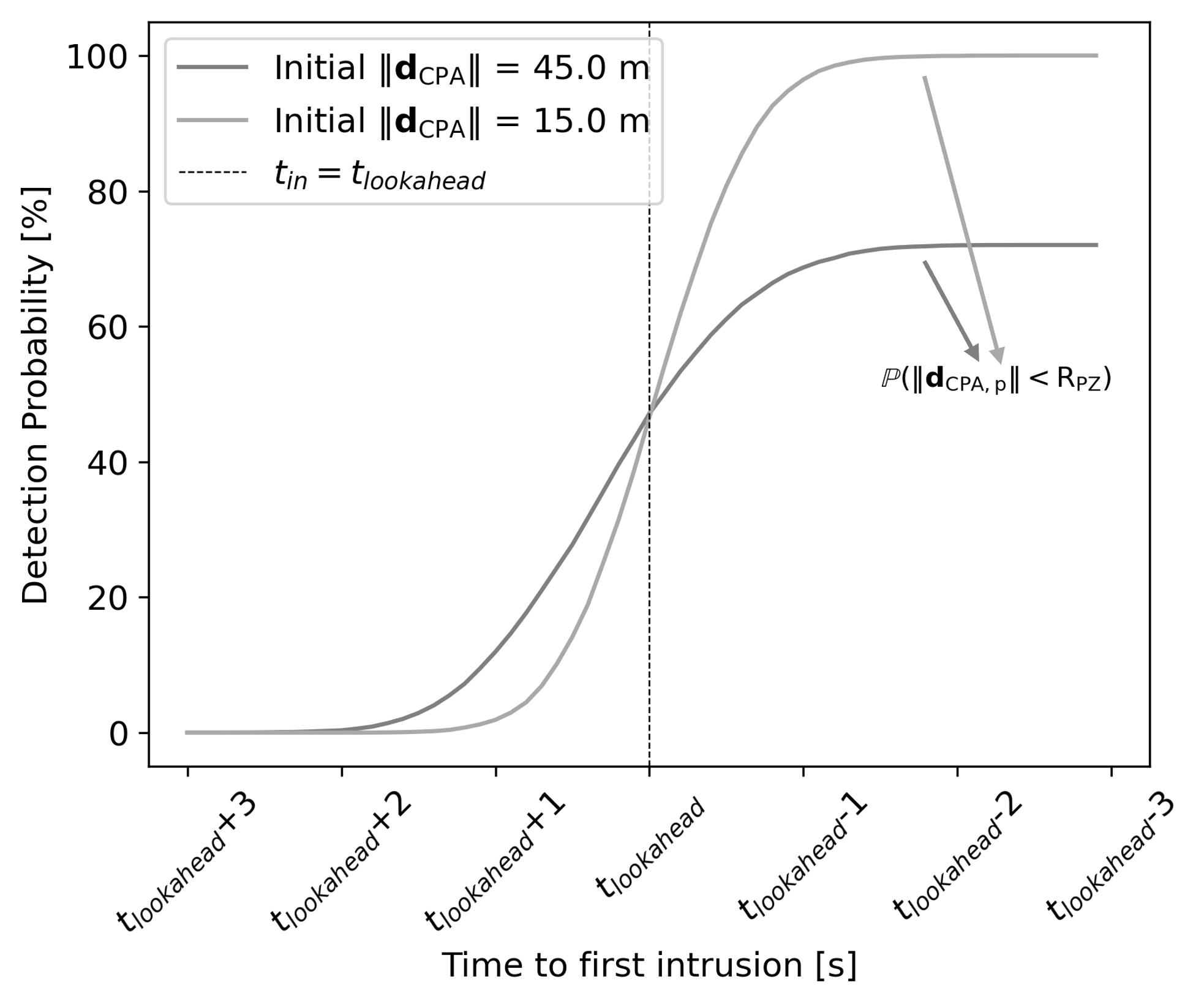}
    \caption{Variation of the time to first intrusion ($t_{\mathrm{in}}$) relative to the look-ahead threshold ($t_{\mathrm{lookahead}}$) under different initial values of $\|\mathbf{d}_{\mathrm{CPA}}\|$. This analysis applies specifically to position uncertainty, where the variance of $\|\mathbf{d}_{\mathrm{CPA}}\|$, and consequently $t_{\mathrm{in}}$, remains constant due to its invariance with respect to aircraft separation distance.
}
    \label{fig:CD_detect_prob_dcpa_tin}
\end{figure}

Mathematically, the detection probability can be decomposed using the chain rule in Equation~\eqref{eq:p_detect_chain}. This formulation shows that detection probability is governed by the joint probability of spatial and temporal conditions. Under position uncertainty, the spatial term is analytically tractable due to the Gaussian assumption on position error, whereas the temporal term remains intractable because of its nonlinearity and dependence on $\|\mathbf{d}_{\mathrm{CPA}}\|$ and the relative geometry. Notably, the approximation of $t_{\mathrm{in}}$ becomes unreliable when $\|\mathbf{d}_{\mathrm{CPA}}\|$ approaches the protected zone radius $R_{\mathrm{PZ}}$, as shown in Figures~\ref{fig:CD_position_uncertainty_approximation_dcpa15_dpsi_170} and~\ref{fig:CD_position_uncertainty_approximation_dcpa45_dpsi_150}.

\begin{align}
P_{\mathrm{detect}} &=
\mathbb{P}\left(\lVert \mathbf{d}_{\mathrm{CPA}} \rVert < R_{\mathrm{PZ}}\right) \notag \\
&\quad \cdot
\mathbb{P}\left(t_{\mathrm{in}} < t_{\mathrm{lookahead}} \,\middle|\,
            \lVert \mathbf{d}_{\mathrm{CPA}} \rVert < R_{\mathrm{PZ}}\right)
\label{eq:p_detect_chain}
\end{align}

Figure~\ref{fig:CD_detect_prob_dcpa_tin} illustrates how the detection probability varies as $t_{\mathrm{in}}$ changes relative to the look-ahead threshold $t_{\mathrm{lookahead}}$ (i.e. as time progresses). The different curves illustrate different initial values of $\|\mathbf{d}_{\mathrm{CPA}}\|$, given that there is no conflict resolution applied. A key observation is that both the slope and the maximum achievable detection probability are defined by the nominal closest point of approach, due to the fixed radius of the protected zone. This means that grazing conflict can remain undetected, while for more severe conflicts, only the initial time of detection is affected by the uncertainty. Even though the nominal value of $\|\mathbf{d}_{\mathrm{CPA}}\|$ may fall below $R_{\mathrm{PZ}}$, random perturbations can push sampled values beyond this boundary. As a result, the maximum detection probability is given by Equation~\eqref{eq:p_detect_max}. Due to the temporal conditional probability in Equation~\eqref{eq:p_detect_chain}, the likelihood of conflict detection becomes 100\% since the $t_{\mathrm{in}}$ falls well below $t_{\mathrm{lookahead}}$ as the intrusion becomes imminent.

\begin{equation}
    P_\mathrm{detect, max} = \mathbb{P}\left( \|\mathbf{d}_{\mathrm{CPA}}\| < \mathrm{R_{PZ}} \right)
    \label{eq:p_detect_max}
\end{equation}

An additional inference can be drawn when \( P_\mathrm{detect, max} = 1.0 \). In this case, the detection probability curve effectively corresponds to the cumulative distribution function (CDF) of the time to first intrusion random variable \( t_{\mathrm{in}} \). As a result, the slope of the detection probability curve reflects the spread of the underlying probability density function (PDF) of \( t_{\mathrm{in}} \), which is governed by the level of position uncertainty. Higher uncertainty leads to a wider distribution of \( t_{\mathrm{in}} \), producing a shallower slope, whereas lower uncertainty results in a sharper transition around the look-ahead time threshold. This relationship highlights that not only the maximum detection probability, but also the rate at which it increases with decreasing \( t_{\mathrm{in}} \), is sensitive to the level of navigational uncertainty.

Under velocity uncertainty, the distribution of \( \mathbf{d}_{\mathrm{CPA}} \) varies with the separation between aircraft. As shown in \ref{fig:velo_uncertainty_dcpa_dist}, the radius of the \( \mathbf{d}_{\mathrm{CPA}} \) increases with distance and contracts as the aircraft move closer. Consequently, the variance of \( t_{\mathrm{in}} \) decreases with reduced separation, leading to a sharper rise in detection probability as \( t_{\mathrm{in}} \) decreases. Similar to the position uncertainty case, early detections are still possible when \( t_{\mathrm{in}} > t_{\mathrm{lookahead}} \), and detection probability continues to rise as the aircraft converge.

To conclude, the detection probability depends on the conflict situation. When the spatial and temporal parameters are decoupled, a more severe \( \mathbf{d}_{\mathrm{CPA}} \) leads to a higher detection probability than a grazing conflict. Then, as time progresses and the time to intrusion decreases, the detection probability increases. The detection classification is further discussed in the next subsection.

\subsubsection{Operational Implications}

The previous subsection analysed instantaneous detection probability under different encounter geometries. In practice, multiple observations occur between the first possible detection and the onset of intrusion. A persistently undetected conflict arises only if every detection opportunity is missed throughout this interval.

When surveillance messages are exchanged at discrete times, each opportunity to detect a conflict constitutes an independent Bernoulli trial with a time varying success probability \(p_t\), obtained from Equation \eqref{eq:p_detect_chain}. The probability of observing no detection prior to intrusion is the product of the complementary probabilities, as shown in Equation \eqref{eq:pr_no_detect_discrete}. In the equation, \(\mathcal{T} = \{t_0, t_1, \dots, t_n\}\) enumerates the sampling times from the first possible detection \(t_0\) up to the intrusion entry time \(t_{\mathrm{in}}\).

\begin{equation}
\Pr(\text{no detect}) \;=\; \prod_{t \in \mathcal{T}} \bigl(1 - p_t\bigr)
\label{eq:pr_no_detect_discrete}
\end{equation}

Figure~\ref{fig:CD_detect_prob_dcpa_tin} shows that for small nominal \(\|\mathbf{d}_{\mathrm{CPA}}\|\) and \(P_{\mathrm{detect,max}}=1\), the term \(p_t\) becomes \(100\%\) and the \(\Pr(\text{no detect})\) is effectively \(0\). For grazing encounters (i.e., those with large nominal projected \(\|\mathbf{d}_{\mathrm{CPA}}\|\)), \(P_{\mathrm{detect,max}}<1\) constrains the ceiling of \(p_t\), and the resulting \(\Pr(\text{no detect})\) can be non-negligible.

As an example, consider a case with nominal projected \(\|\mathbf{d}_{\mathrm{CPA}}\| = 45~\mathrm{m}\) (with 50 meters separation standard) and horizontal position accuracy \(30~\mathrm{m}\). With a look-ahead time of \(5~\mathrm{s}\), the probability of no detection prior to intrusion is \(0.319\%\). Increasing the look-ahead horizon to \(6~\mathrm{s}\) and \(7~\mathrm{s}\) reduces the no-detection probability to \(0.089\%\) and \(0.025\%\), respectively. Extending the horizon introduces additional observations with positive \(p_t\), which multiplies additional factors \((1-p_t)\) in \eqref{eq:pr_no_detect_discrete} and thereby lowers the overall non-detection probability.

Introducing conflict resolution modifies the evolution of \(p_t\). Once a detection occurs, a manoeuvre is initiated. If the modified geometry still results in a conflict, successive potentially positive detections occur under updated \(p_t\) values that reflect the new conflict state. The probability of having an intrusion after a first manoeuvre therefore differs from the one before the manoeuvre, and depends on the effectiveness of the conflict resolution algorithm in resolving the conflict. This is further discussed in the next subsection.
\subsection{Conflict Resolution}
\label{subsec:cr}

Conflict resolution is analysed in the velocity-space domain, where the set of admissible commands can be reprsented by the inverse of the velocity obstacle set (\(\mathcal{VO}\) ). This section examines how navigation uncertainty propagates from the input state to the resolution velocity generated by the Modified Voltage Potential (MVP) and Velocity Obstacle (VO) algorithms. The analysis begins with position uncertainty, followed by velocity uncertainty.

\subsubsection{Position uncertainty}

\begin{figure}[!htb]
    \centering
    \includegraphics[width = 0.7\linewidth]{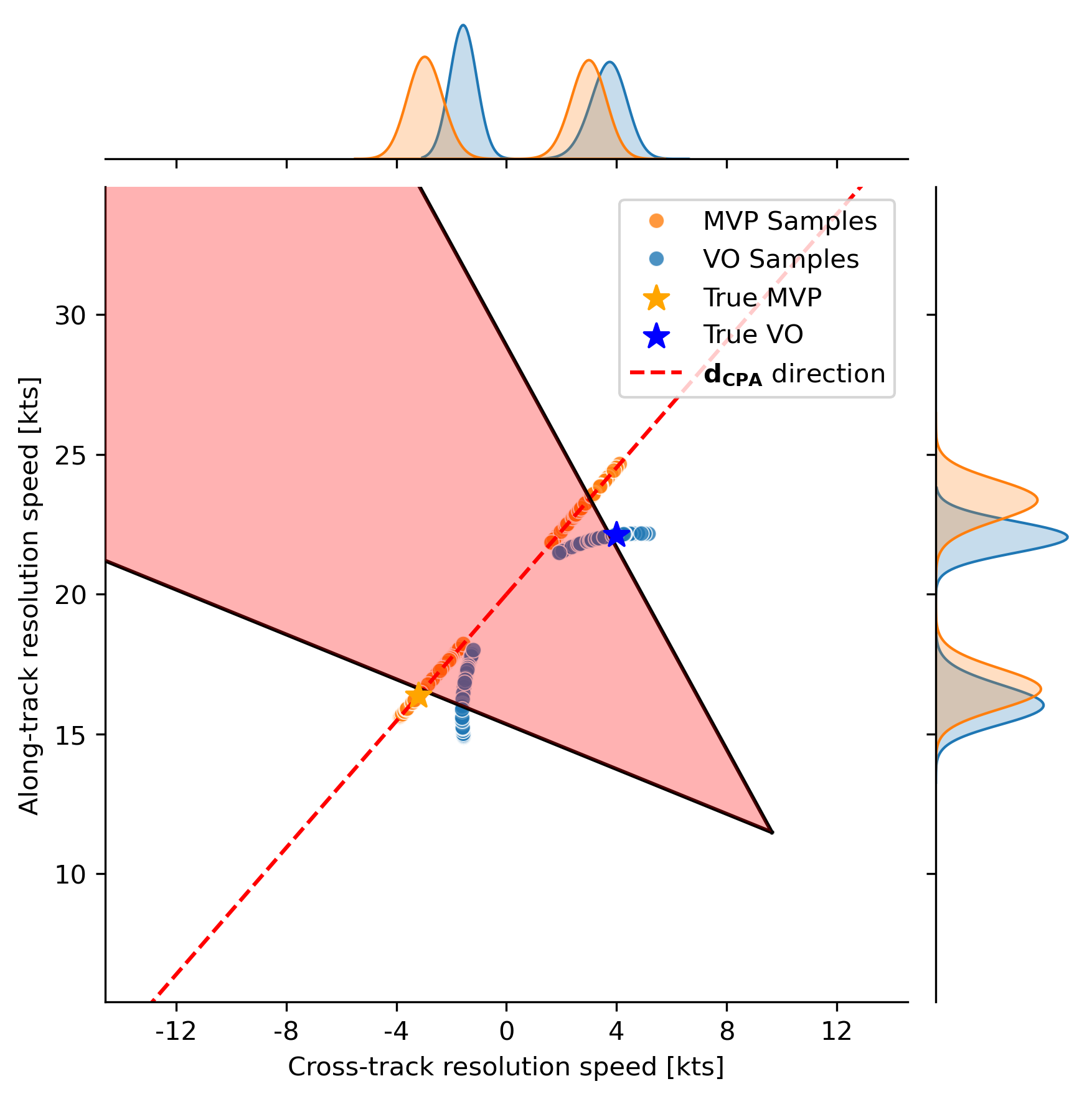}
    \caption{Resolution velocity samples under position uncertainty for \(\Delta\psi = 40^\circ\), \(\|\mathbf{d}_{\mathrm{CPA}}\| = 0~\mathrm{m}\).}
    \label{fig:CR_highrelvelo_dcpa0}
\end{figure}

Figure~\ref{fig:CR_highrelvelo_dcpa0} presents Monte Carlo samples of the resolution velocity under position uncertainty for a conflict geometry with \(\Delta\psi = 40^{\circ}\) and \(\|\mathbf{d}_{\mathrm{CPA}}\| = 0~\mathrm{m}\). The MVP-provided resolution samples are distributed along a straight line, whereas the VO-provided resolutions form an arc. This behaviour is a direct consequence of the formulation in Eq.~\eqref{eq:mvp_dv}, where the resolution velocity is constructed from a scalar term and the vector \(\mathbf{d}_{\mathrm{CPA}}\). The scalar term reflects the variability introduced by the uncertainty in \(t_{\mathrm{CPA}}\) and \(\|\mathbf{d}_{\mathrm{CPA}}\|\), while the vector component retains a Gaussian distribution. As a result, the MVP resolution velocity samples maintain a linear structure that is independent of the relative velocity.

\begin{figure}[!htb]
    \centering
    \includegraphics[width = 0.7\linewidth]{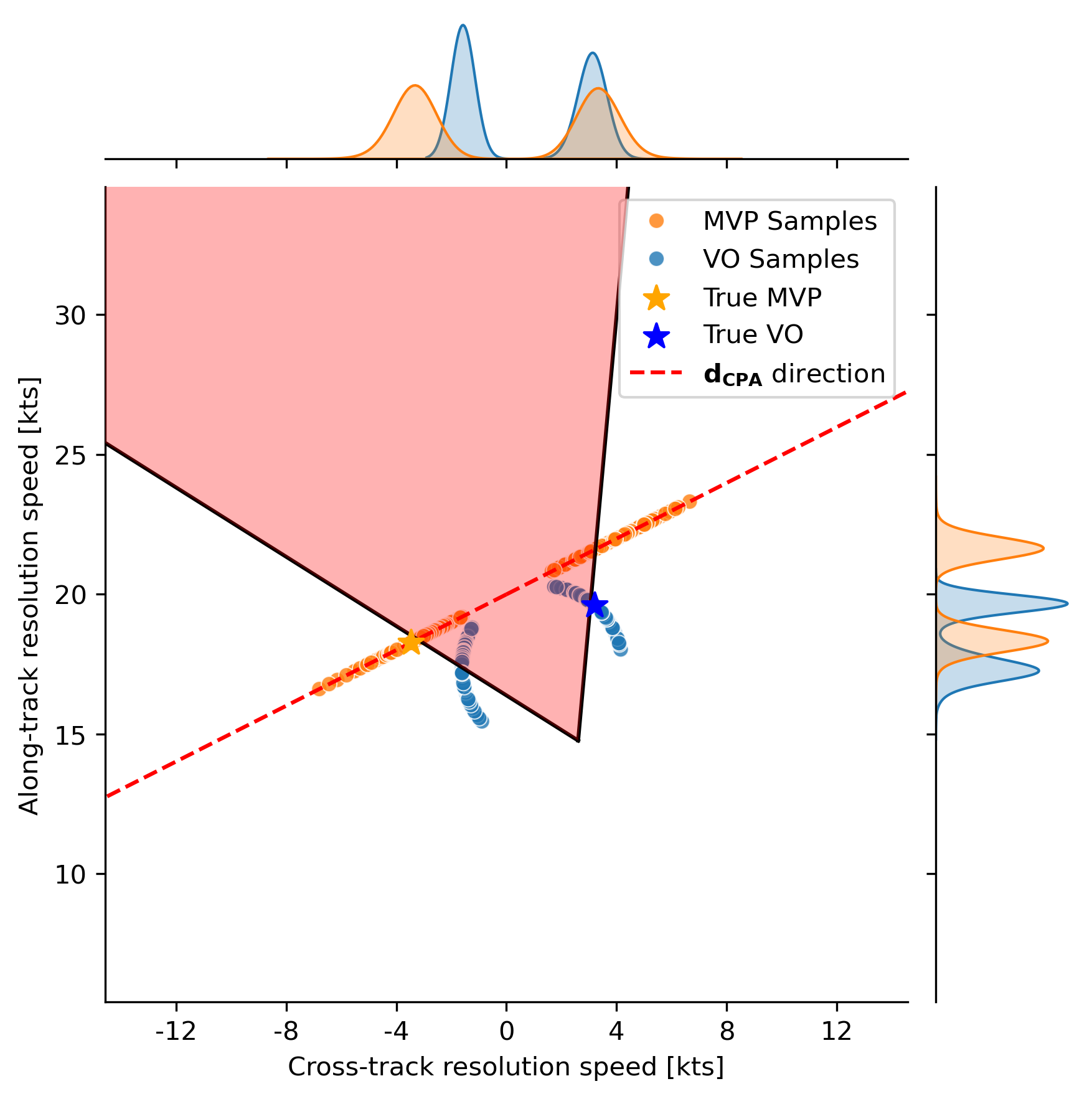}
    \caption{Resolution velocity samples under position uncertainty for \(\Delta\psi = 10^\circ\), \(\|\mathbf{d}_{\mathrm{CPA}}\| = 0~\mathrm{m}\).}
    \label{fig:CR_lowrelvelo_dcpa0}
\end{figure}

In contrast, the VO resolution velocities follow a circular arc. It can be mathematically shown, this arc is centred around the midpoint between the ownship and intruder velocities, with a radius equal to half the magnitude of the relative velocity (the derivation is shown in ~\ref{app:locus_projection}). Consequently, the shape and size of the VO distribution are dependent on the conflict geometry. This effect is shown in Figure~\ref{fig:CR_lowrelvelo_dcpa0}, where a smaller heading angle \(\Delta\psi = 10^{\circ}\) leads to a lower relative velocity and a smaller arc radius, while the MVP distribution retains its linear form. In the limiting case, where speeds are equal and heading differences are small, the arc becomes smaller.

\begin{figure}[!htb]
    \centering
    \includegraphics[width = 0.7\linewidth]{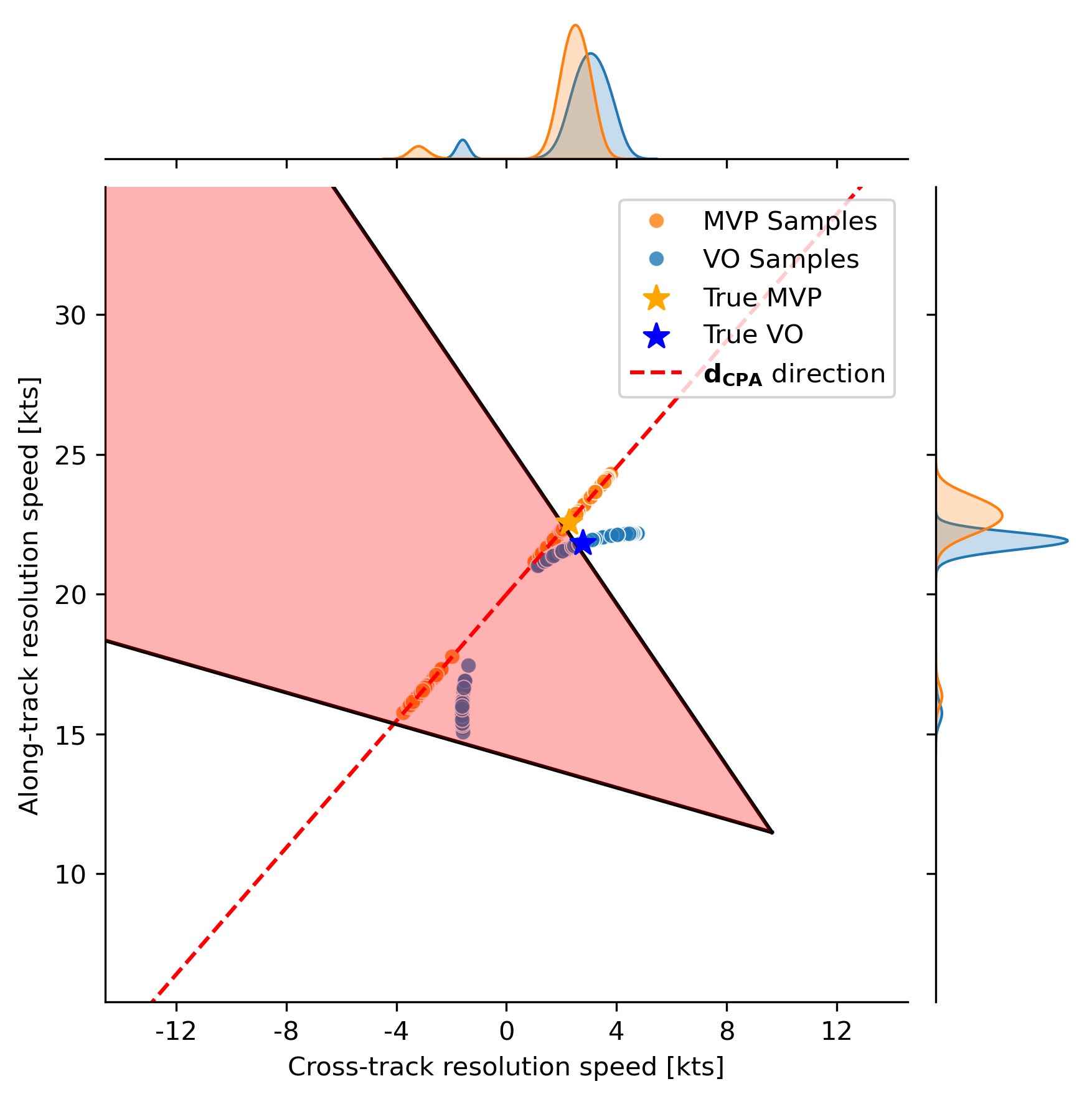}
    \caption{Resolution velocity samples under position uncertainty for \(\Delta\psi = 40^\circ\), \(\|\mathbf{d}_{\mathrm{CPA}}\| = 15~\mathrm{m}\).}
    \label{fig:CR_highrelvelo_dcpa15}
\end{figure}

Another aspect of the resolution velocity distribution is the division of samples across the two legs of the \(\mathcal{VO}\). Figure~\ref{fig:CR_highrelvelo_dcpa15} compares cases with \(\|\mathbf{d}_{\mathrm{CPA}}\| = 0~\mathrm{m}\) and \(15~\mathrm{m}\), both at \(\Delta\psi = 40^\circ\). When the mean of the random vector \(\mathbf{d}_{\mathrm{CPA}}\) is zero, the distribution is symmetric, and the samples are equally divided between both \(\mathcal{VO}\) legs. As the mean increases, the probability mass shifts, concentrating the samples on one side. This bias can be quantified using the transformed variable in Eq.~\eqref{eq:dcpa_transformed}, which depends on the conflict geometry and the relative position covariance \(\Sigma_{\mathrm{rel}}\).  Mathematically, the probability that a sample lies on the dominant \(\mathcal{VO}\) leg corresponds to the probability that the scalar random variable \(z\), defined in Eq.~\eqref{eq:dcpa_transformed}, is positive. This is expressed in Eq.\eqref{eq:dominant_leg_prob}.

\begin{equation}
\label{eq:dominant_leg_prob}
    P_{\text{dominant}} = \mathbb{P}(z > 0)
\end{equation}

This has practical implications. When \(\|\mathbf{d}_{\mathrm{CPA}}\| = 0~\mathrm{m}\), the distribution is symmetric and \(P_{\text{dominant}} \approx 0.5\), indicating that the resolution velocity is equally likely to fall on either leg of the \(\mathcal{VO}\). Under uncertainty, if \(\|\mathbf{d}_{\mathrm{CPA}}\|\) remains close to zero after the first resolution attempt, this may cause reversals in resolution direction. Once a resolution step shifts the mean of \(\|\mathbf{d}_{\mathrm{CPA}}\|\) away from zero, the probability mass concentrates on a single dominant \(\mathcal{VO}\) leg (i.e., \(P_{\text{dominant}} > 0.5\)), and subsequent iterations reinforce this choice, creating a self-reinforcing dynamic that reduces reversals. Conversely, staying near zero sustains oscillatory behaviour. Therefore, increasing \(\|\mathbf{d}_{\mathrm{CPA}}\|\) is essential not only to ensure conflict avoidance, but also to improve confidence in the selected resolution trajectory.

\begin{figure*}[!htbp]
    \centering
    \includegraphics[width=1.0\linewidth]{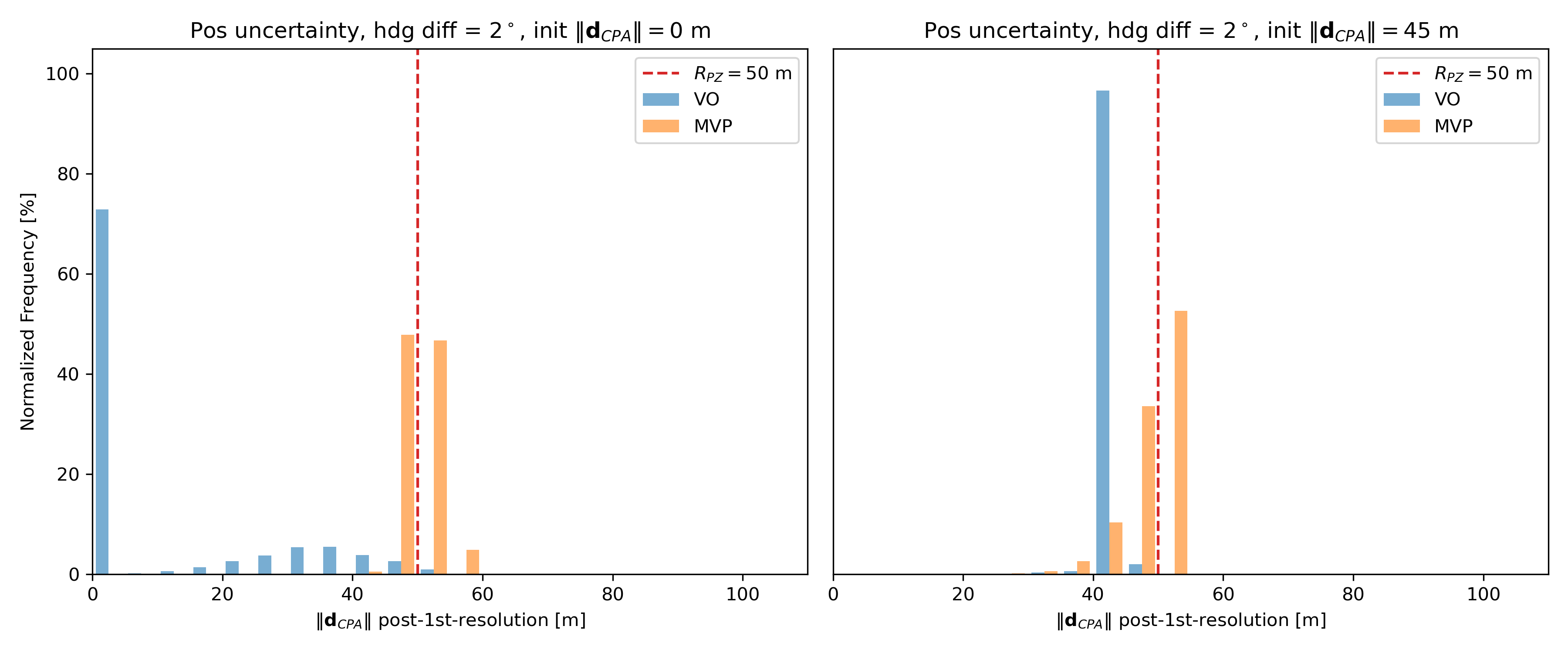}
    \caption{Distribution of post-resolution projected $\|\mathbf{d}_{CPA}\|$ for $\Delta_{\psi} = 2^\circ$ and initial $\|\mathbf{d}_{CPA}\|$ of 0 and 45 meters for position uncertainty}
    \label{fig:pos_uncertainty_2deg_dcpa}
\end{figure*}

\begin{figure*}[!htbp]
    \centering
    \includegraphics[width=1.0\linewidth]{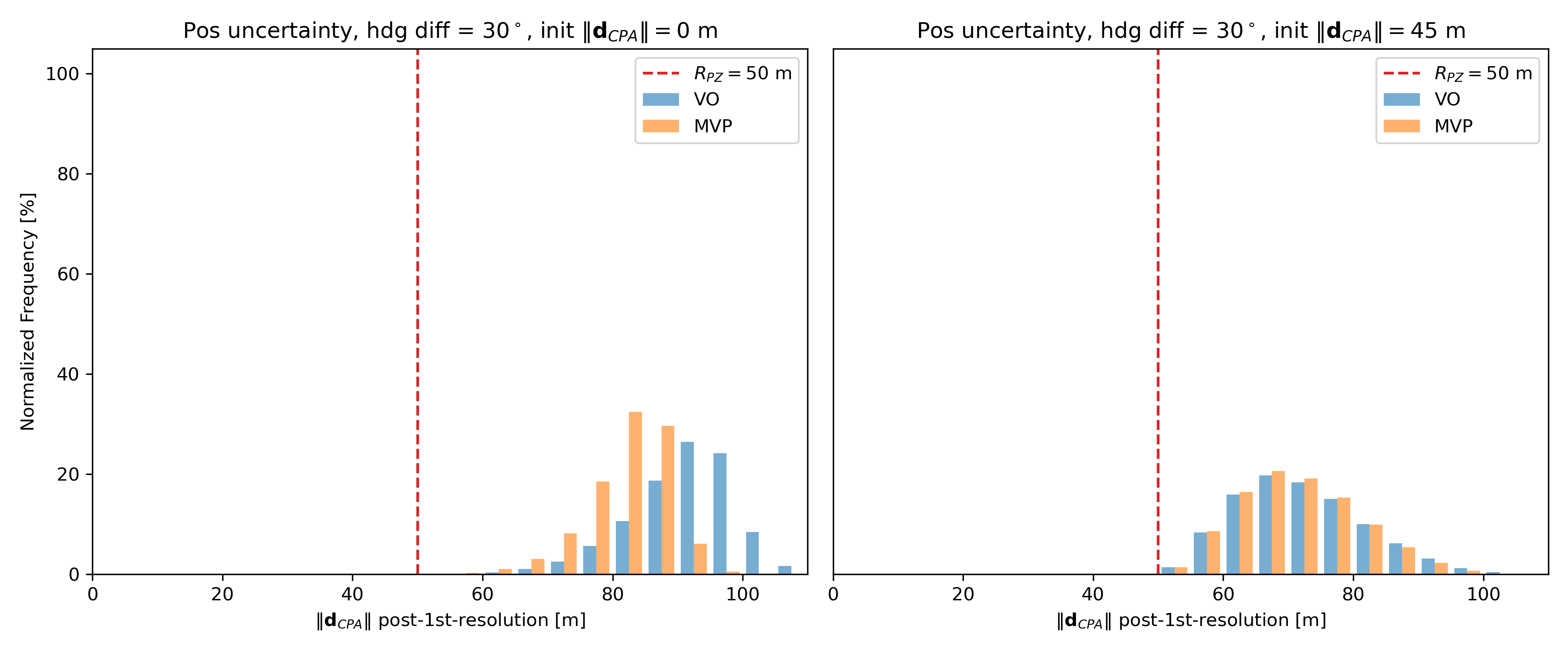}
    \caption{Distribution of post-resolution projected $\|\mathbf{d}_{CPA}\|$ for $\Delta_{\psi} = 30^\circ$ and initial $\|\mathbf{d}_{CPA}\|$ of 0 and 45 meters for position uncertainty}
    \label{fig:pos_uncertainty_30deg_dcpa}
\end{figure*}

To quantify how effectively each algorithm enlarges the \(
\|\mathbf{d}_{\mathrm{CPA}}\|
\), the post-resolution miss distance
\(
\|\mathbf{d}_{\mathrm{CPA}}^{+}\|
\)
is evaluated one second after the manoeuvre has been applied. Figure~\ref{fig:pos_uncertainty_2deg_dcpa} shows the resulting distributions for a heading difference of \(2^{\circ}\) and two initial offsets, \(\|\mathbf{d}_{\mathrm{CPA}}\| = 0~\mathrm{m}\) and \(45~\mathrm{m}\), with both aircraft travelling at \(20~\text{kt}\).
When \(\|\mathbf{d}_{\mathrm{CPA}}\| = 0~\mathrm{m}\), the VO algorithm yields projected \(
\|\mathbf{d}_{\mathrm{CPA}}^{+}\|
\) that remain below \(5~\mathrm{m}\) with high probability, whereas the MVP algorithm produces a better result.  In this shallow-angle case, \(99.0\,\%\) of the VO samples fall below \(50~\mathrm{m}\), compared with \(48.4\,\%\) for MVP.  
For an initial offset of \(45~\mathrm{m}\), the VO distribution is entirely confined below \(50~\mathrm{m}\), while the corresponding proportion for MVP is \(47.4\,\%\).  Although MVP does not always clear the protected zone in a single step, the subsequent detection-and-resolution cycles can correct any residual conflict.

A higher heading difference (\(\Delta\psi = 30^{\circ}\)) is depicted in Figure~\ref{fig:pos_uncertainty_30deg_dcpa}.  Under this geometry all post-resolution distances exceed \(50~\mathrm{m}\) for both algorithms, irrespective of the initial offset, indicating successful conflict avoidance in every realisation.

The contrast between the \(2^{\circ}\) and \(30^{\circ}\) cases stems from the relative speed: shallow angles generate low relative velocities.  Although the deterministic solutions of MVP and VO both lie outside the velocity-obstacle set, their optimisation criteria differ.  MVP explicitly seeks to maximize \(\|\mathbf{d}_{\mathrm{CPA}}\|\) with minimal velocity change, whereas VO minimizes the change in velocity subject only to leaving the obstacle set.  This distinction makes the MVP algorithm intrinsically more robust under position uncertainty, especially in low-speed encounters where the VO arc gets tighter and the admissible region narrows.

The following section applies the same analysis framework to velocity uncertainty, comparing the resulting resolution commands and the corresponding changes in projected \(\|\mathbf{d}_{\mathrm{CPA}}\|\).

\subsubsection{Velocity Uncertainty}
\begin{figure}[!htb]
    \centering
    \includegraphics[width = 0.7\linewidth]{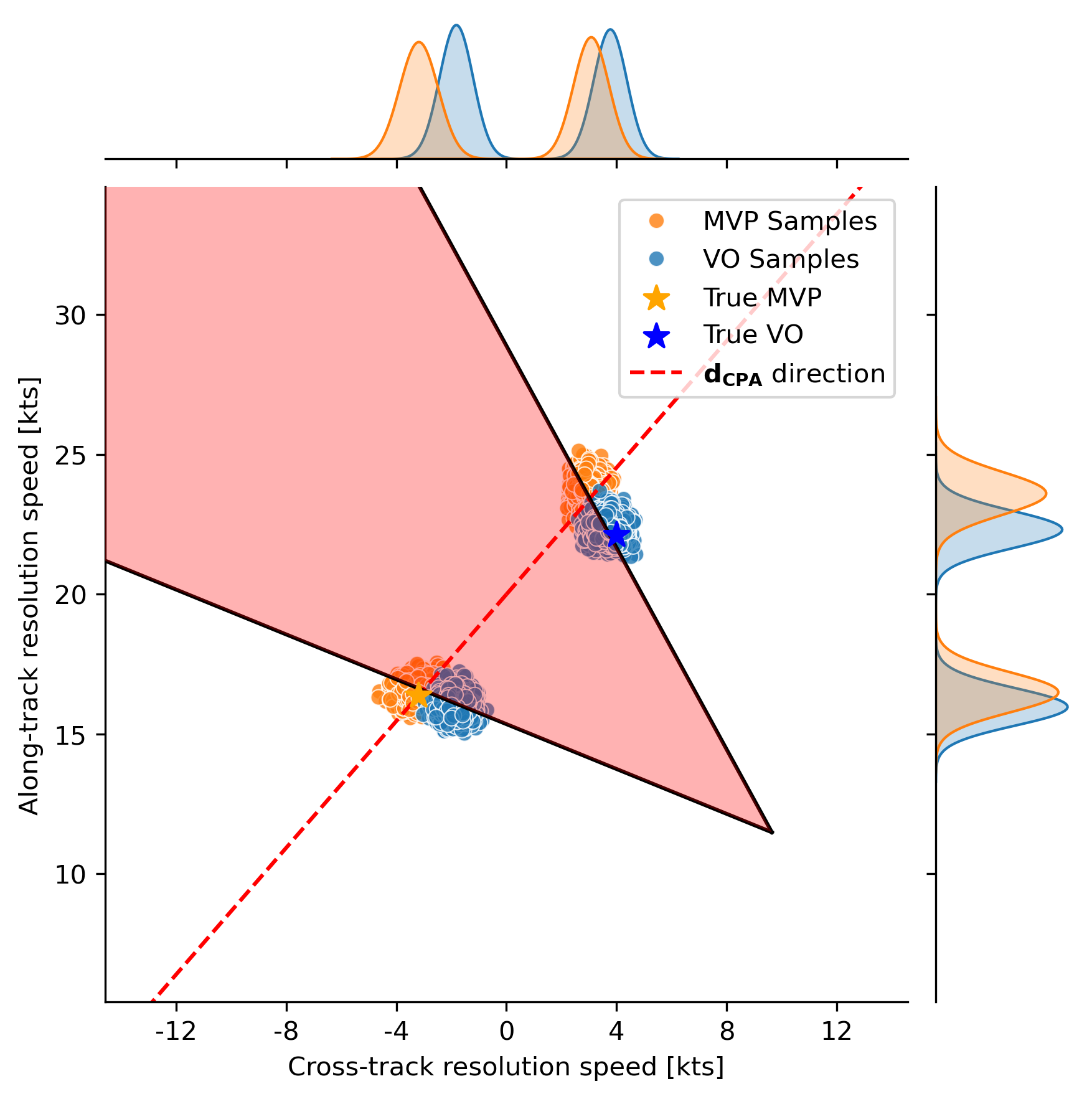}
    \caption{Resolution velocity samples under position uncertainty for \(\Delta\psi = 40^\circ\), \(\|\mathbf{d}_{\mathrm{CPA}}\| = 0~\mathrm{m}\).}
    \label{fig:velo_highangle_VO}
\end{figure}

Figure~\ref{fig:velo_highangle_VO} depicts Monte-Carlo samples of the resolution velocity under velocity uncertainty for \(\Delta\psi = 40^{\circ}\) and \(\|\mathbf{d}_{\mathrm{CPA}}\| = 0~\mathrm{m}\).  Because the random perturbation acts directly on the velocity vector, the resulting distribution is no longer confined to a closed-form formulation.  Even so, the samples remain symmetrically split between the two $\mathcal{VO}$ legs, a consequence of the zero mean in \(\mathbf{d}_{\mathrm{CPA}}\).

\begin{figure}[!htb]
    \centering
    \includegraphics[width = 0.7\linewidth]{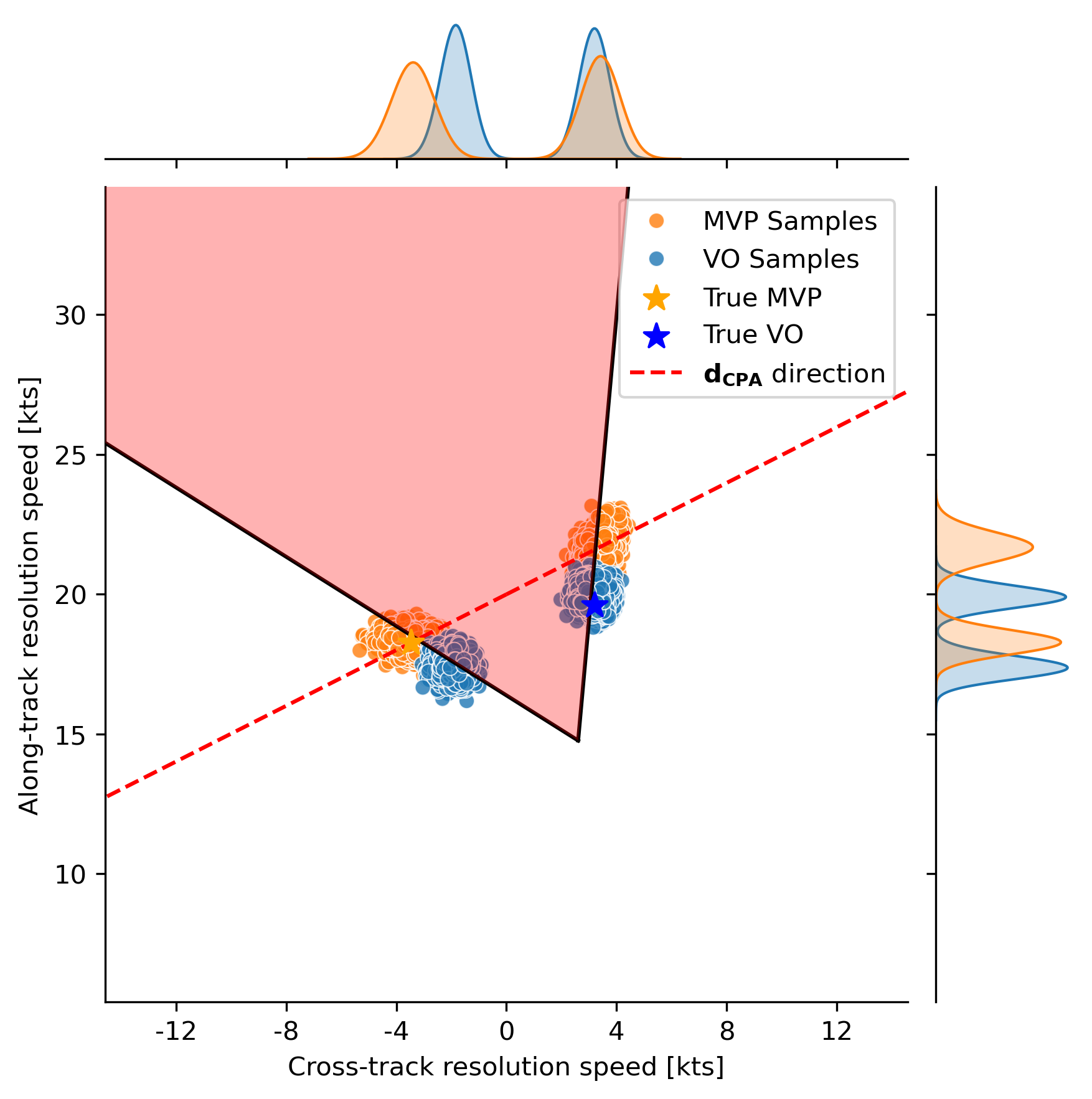}
    \caption{Resolution velocity samples under position uncertainty for \(\Delta\psi = 10^\circ\), \(\|\mathbf{d}_{\mathrm{CPA}}\| = 0~\mathrm{m}\).}
    \label{fig:velo_smallangle_VO}
\end{figure}

A reduced heading difference (\(\Delta\psi = 10^{\circ}\)) is shown in Figure~\ref{fig:velo_smallangle_VO}.  In both algorithms the sample point moves toward the VO intersection point, which corresponds to the intruder velocity in the velocity space.  This shift becomes most pronounced when the relative speed approaches zero, i.e., equal speed combined with a shallow heading angle difference.

\begin{figure*}[!htbp]
    \centering
    \includegraphics[width=1.0\linewidth]{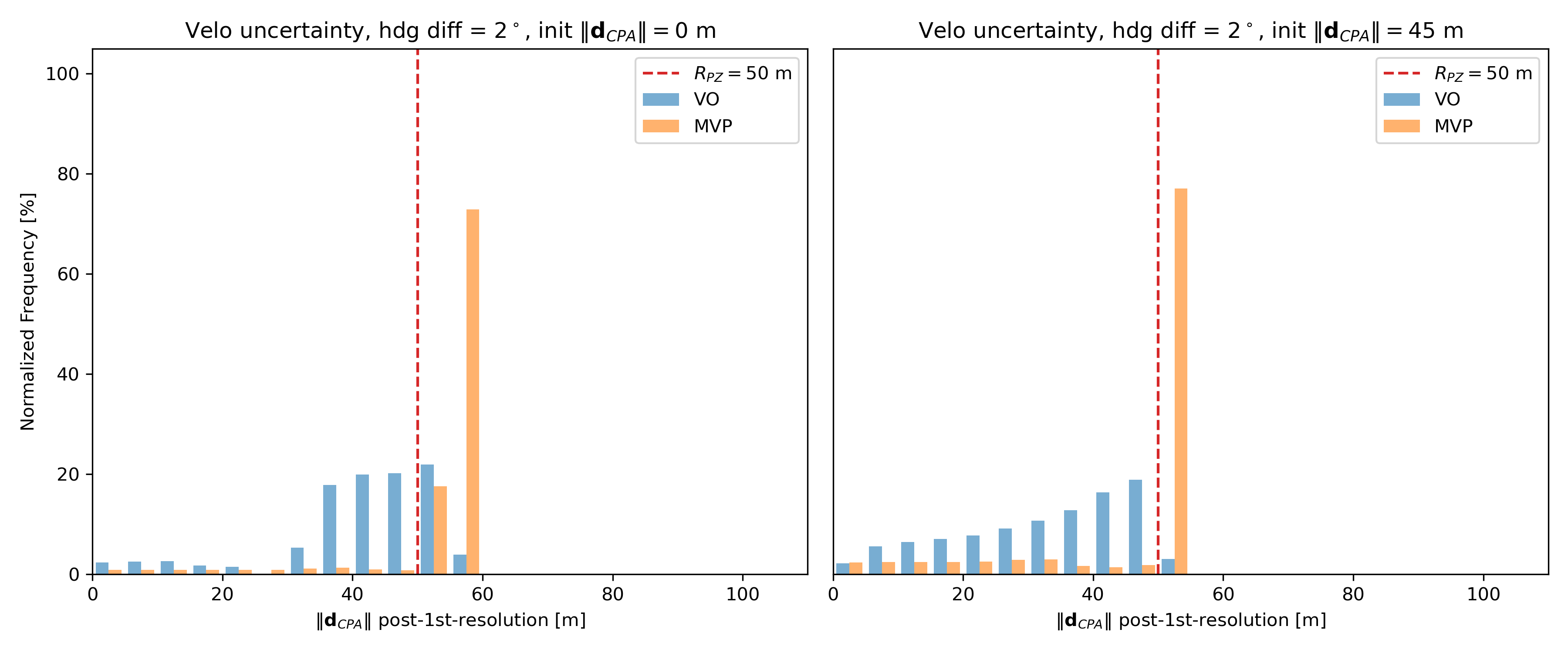}
    \caption{Distribution of post- projected $|\mathbf{d}_{CPA}|$ for $\Delta_{\psi} = 2^\circ$ and initial $|\mathbf{d}_{CPA}|$ of 0 and 45 meters for velocity uncertainty}
    \label{fig:velo_uncertainty_2deg_dcpa}
\end{figure*}

\begin{figure*}[!htbp]
    \centering
    \includegraphics[width=1.0\linewidth]{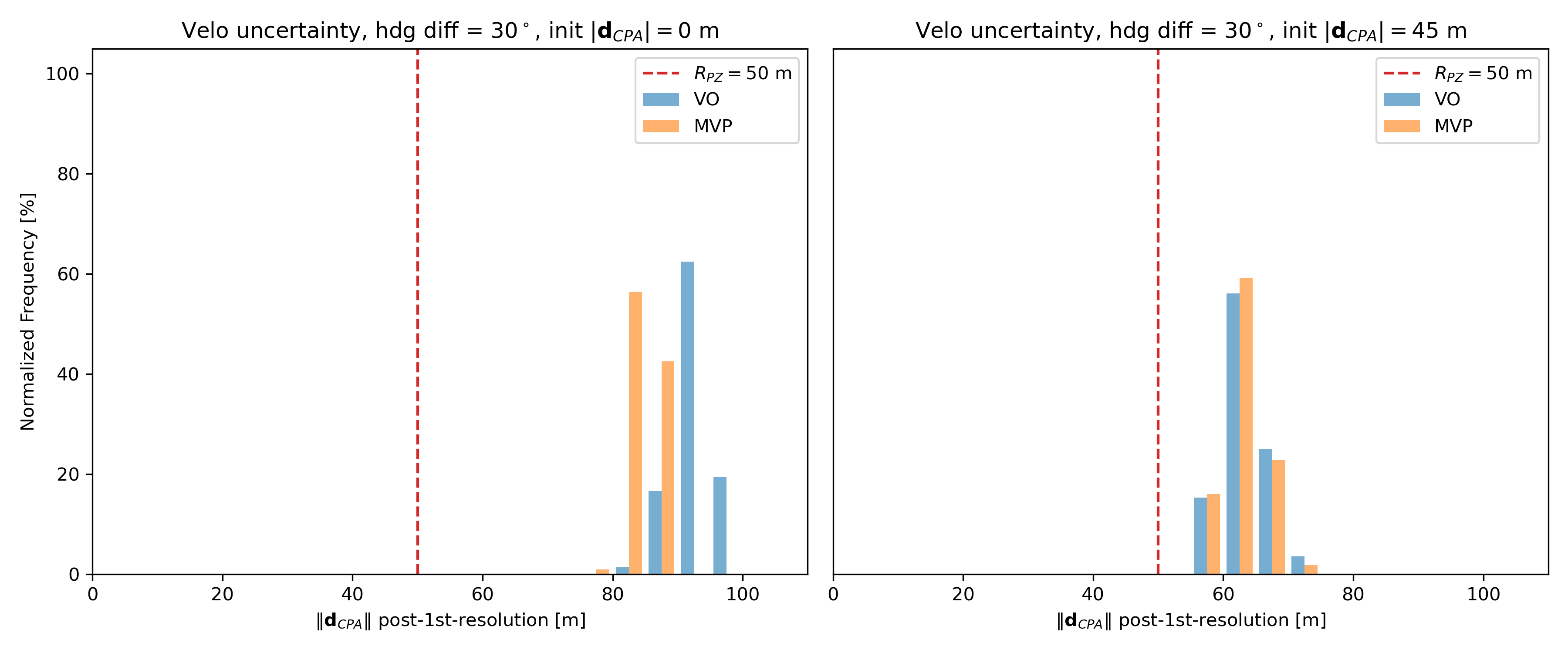}
    \caption{Distribution of post-resolution $|\mathbf{d}_{CPA}|$ for $\Delta_{\psi} = 30^\circ$ and initial $|\mathbf{d}_{CPA}|$ of 0 and 45 meters for velocity uncertainty}
    \label{fig:velo_uncertainty_30deg_dcpa}
\end{figure*}

To illustrate this edge case, Figures~\ref{fig:velo_uncertainty_2deg_dcpa} and \ref{fig:velo_uncertainty_30deg_dcpa} compare \(
\|\mathbf{d}_{\mathrm{CPA}}^{+}\|
\) distributions for heading differences of \(2^{\circ}\) and \(30^{\circ}\).  For \(\Delta\psi = 2^{\circ}\) and an initial \(\|\mathbf{d}_{\mathrm{CPA}}\| = 0~\mathrm{m}\), the VO algorithm leaves \(\|\mathbf{d}_{\mathrm{CPA}}^{+}\| < 50~\mathrm{m}\) in \(74.1\,\%\) of the samples; with an initial offset of \(45~\mathrm{m}\) this fraction rises to \(96.9\,\%\), and some samples even decrease the distance at closest point of approach, indicating a sub-optimal command.  Under the same conditions, MVP limits the proportion of unresolved cases to \(9.5\,\%\) and \(22.9\,\%\), respectively.  At \(\Delta\psi = 30^{\circ}\) all samples for both algorithms exceed the protected-zone radius after one second, confirming successful avoidance.

These observations emphasize the earlier conclusion that, under uncertainty, a conflict-resolution rule is most robust when it explicitly maximizes \(\|\mathbf{d}_{\mathrm{CPA}}\|\). The MVP formulation satisfies this criterion, whereas the VO algorithm seeks the smallest change required to exit the obstacle set.

\subsubsection{Conflict Resolution Dynamics}
The evolution of the \(\|\mathbf{d}_{\mathrm{CPA}}\|\) over successive resolution cycles is governed both by conflict geometry and uncertainty level.  Suppose that the initial offset satisfies \(|\mathbf{d}_{\mathrm{CPA}}| = 0~\mathrm{m}\) and \(\Delta\psi = 2^{\circ}\), the first-iteration outcomes are illustrated in the left panel of Figure~\ref{fig:pos_uncertainty_2deg_dcpa}. Because the detection probability is below \(50\%\) at \(t_{\mathrm{in}} = t_{\mathrm{lookahead}}\), the likelihood of resolving the conflict in a single step is correspondingly limited, especially for the VO algorithm under low relative speed.  The MVP algorithm attains a markedly higher success probability in the same condition.  Under high relative speed the conflict is typically removed after the first manoeuvre, although the event remains probabilistic due to input uncertainty.

For shallow-angle encounters the resolution process exhibits a multi-step character.  Starting from \(|\mathbf{d}_{\mathrm{CPA}}| = 0~\mathrm{m}\), the distribution of the next miss distance is given by Figure~\ref{fig:pos_uncertainty_2deg_dcpa} for two specific conditions. Each range of \(\|\mathbf{d}_{\mathrm{CPA}}\|\) produced in that figure carries its own conditional probability of triggering a successful resolution at the following iteration. A key observation from the figure is that MVP directly shifts the \(\|\mathbf{d}_{\mathrm{CPA}}\|\) away from \(0~\mathrm{m}\). Thus, in the next iteration, the probability of the aircraft performing a counter-productive manoeuvre (such as turning to the opposite direction or the less dominant leg) is much smaller. This is not the case for VO, after the first resolution high probability of the \(\|\mathbf{d}_{\mathrm{CPA}}\|\) still lies close to zero. The sequence continues until \(\|\mathbf{d}_{\mathrm{CPA}}\|\) exceeds the protected-zone radius or until a defined amount of time is reached.

A comprehensive assessment of this iterative behaviour is obtained through large-scale Monte-Carlo simulation in BlueSky.  The resulting safety metrics are presented in the next subsection, providing a macroscopic view of conflict-resolution performance under the combined influence of conflict geometry and uncertainty and also the final \(\|\mathbf{d}_{\mathrm{CPA}}\|\).
\subsection{Safety metric}
\label{subsec:safety}

\begin{figure}[!htb]
    \centering
    \includegraphics[width=0.7\linewidth]{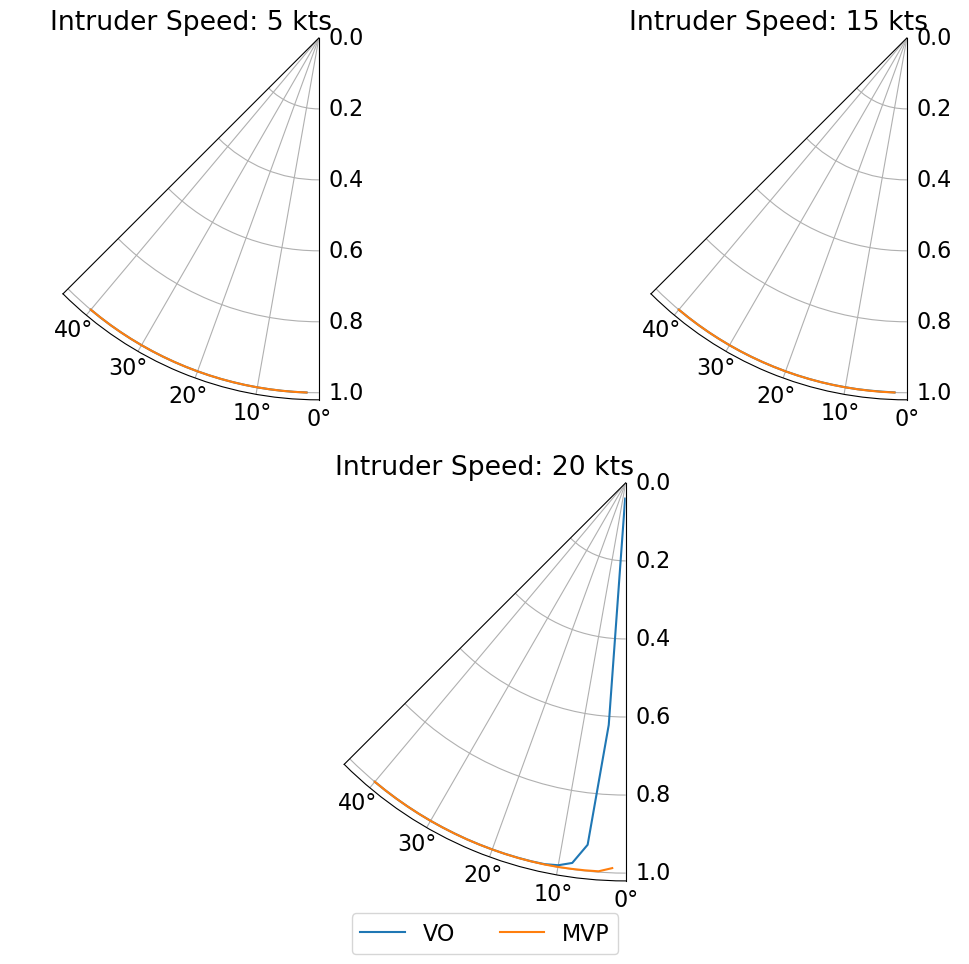}
    \caption{Intrusion prevention rate under position uncertainty for various intruder speeds.}
    \label{fig:ipr_position_uncertainty}
\end{figure}

Figure~\ref{fig:ipr_position_uncertainty} compares the intrusion prevention rate (IPR) of the VO and MVP algorithms under position uncertainty for three intruder speeds.  At \(5~\text{kts}\) and \(15~\text{kts}\) the two methods exhibit similarly low IPR values across all heading differences.  When the intruder speed equals the own-ship speed (\(20~\text{kts}\)), the results diverge: MVP maintains a high IPR for every \(\Delta\psi\), whereas VO shows a notable decrease of performance.  The degradation arises at shallow angles, where the relative speed is small and the VO algorithm fails to enlarge \(\|\mathbf{d}_{\mathrm{CPA}}\|\) under position uncertainty (see Section~\ref{subsec:cr}).

\begin{figure}[!htb]
    \centering
    \includegraphics[width=0.8\linewidth]{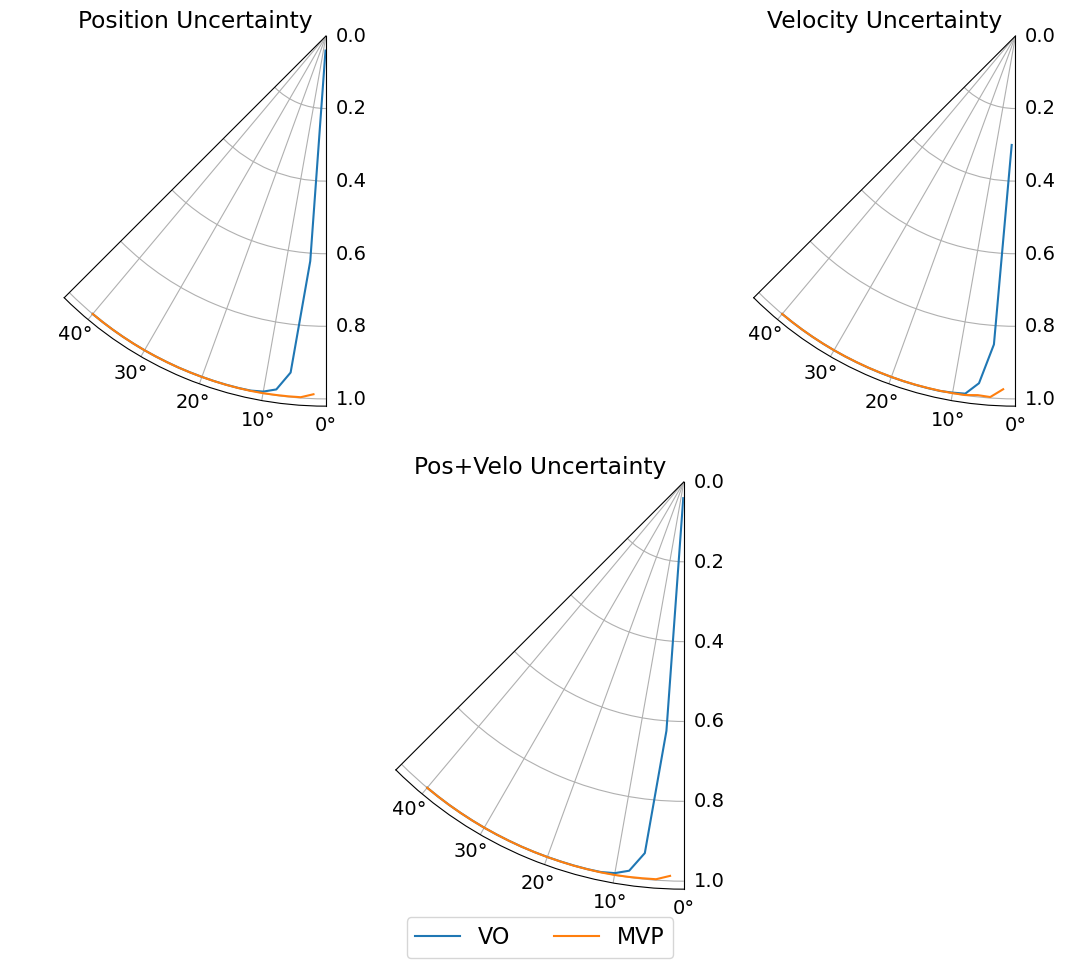}
    \caption{Intrusion prevention rate under position, velocity and combined uncertainty at \(20~\text{kts}\).}
    \label{fig:ipr_combined_uncertainty}
\end{figure}

Figure~\ref{fig:ipr_combined_uncertainty} extends the comparison to three uncertainty models namely; position, velocity and combined (position + velocity), at \(20~\text{kts}\).  Across the full heading range, MVP retains a high IPR, with only a modest reduction below \(\Delta\psi = 10^{\circ}\).  The VO algorithm performs significantly worse, particularly at shallow angles, and reaches its lowest IPR under combined uncertainty.  The contrast reflects the algorithmic objectives: MVP maximizes \(\|\mathbf{d}_{\mathrm{CPA}}\|\), whereas VO minimizes the velocity change subject to exiting the VO set, a strategy that is less effective when uncertainty perturbs the resolution velocity.

A previous study introduced a scaled–speed variant of the VO algorithm that permitted velocity changes up to \(15\,\%\) larger than the nominal command.  Despite this additional control authority, the IPR decreased, indicating that magnifying \(\|\Delta\mathbf{V}\|\) for the resolution velocity does not compensate for VO’s structural limitations \cite{rahman_autonomous_2024}.  The decisive element is the optimisation criterion: MVP selects the resolution velocity that maximizes \(\|\mathbf{d}_{\mathrm{CPA}}\|\); VO merely chooses the smallest deviation that exits the set \(\mathcal{VO}\).

\begin{figure}[!htb]
    \centering
    \includegraphics[width=0.8\linewidth]{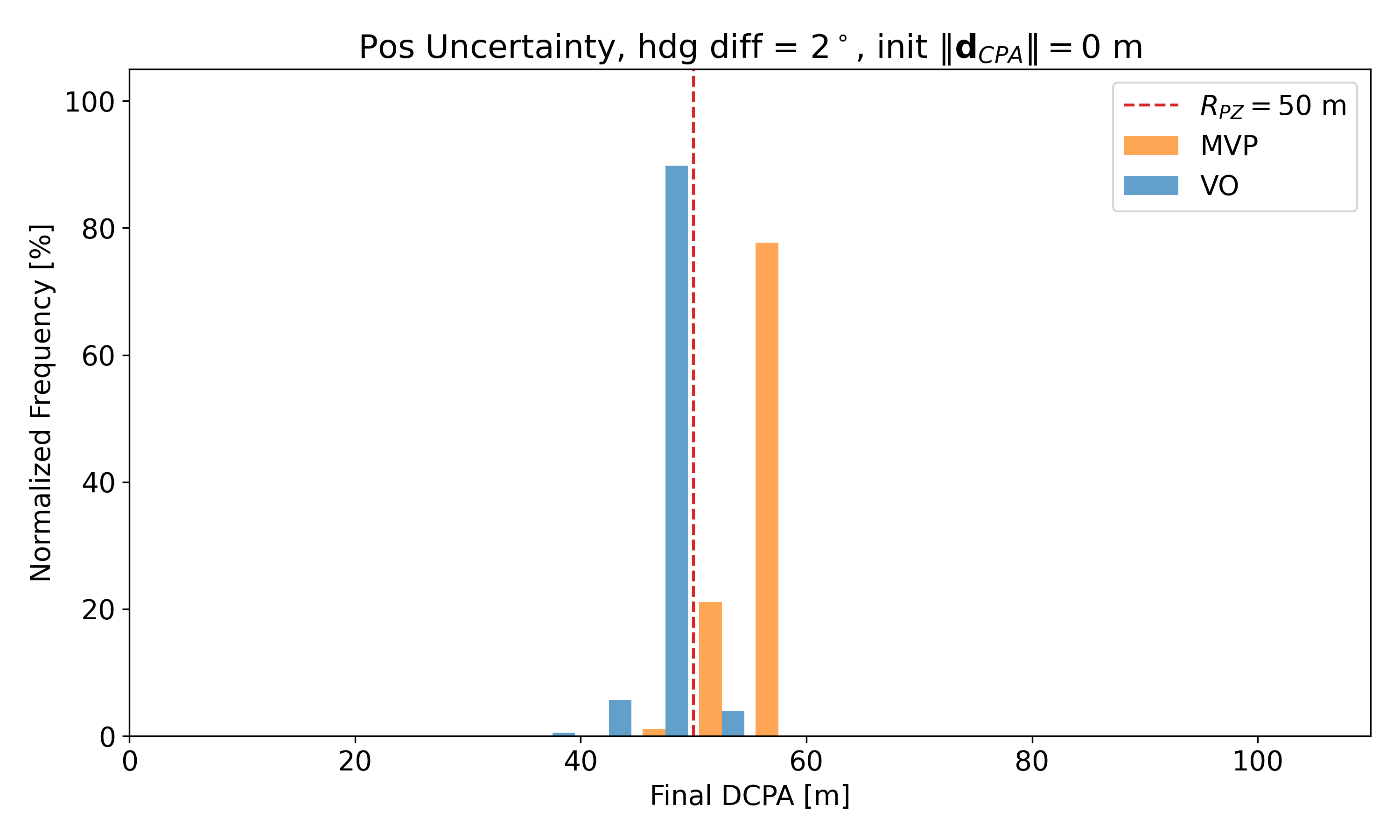}
    \caption{Distribution of \(\|\mathbf{d}_{\mathrm{CPA}}\|\) under position uncertainty}
    \label{fig:final_dcpa_position_uncertainty}
\end{figure}

\begin{figure}[!htb]
    \centering
    \includegraphics[width=0.8\linewidth]{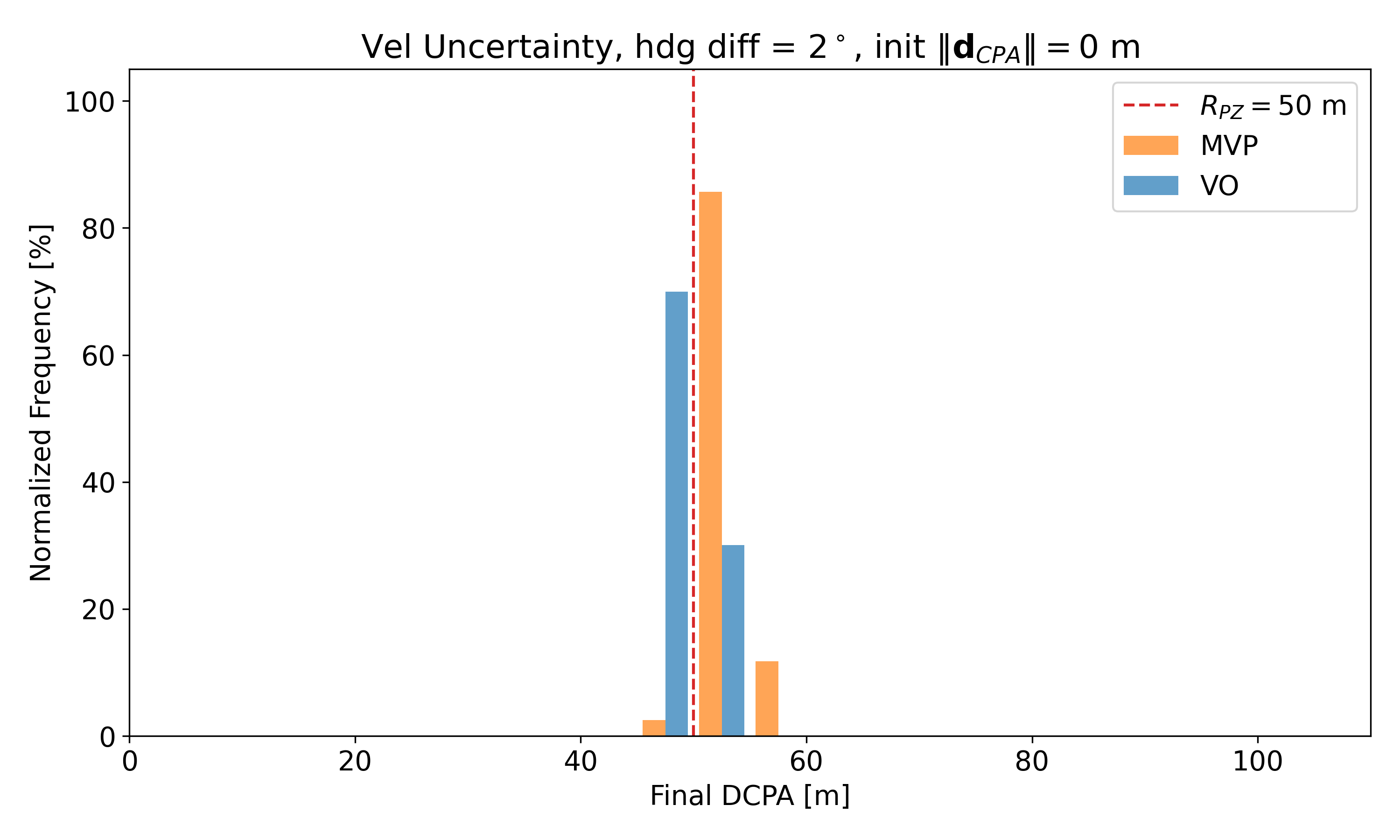}
    \caption{Distribution of \(\|\mathbf{d}_{\mathrm{CPA}}\|\) under velocity uncertainty}
    \label{fig:final_dcpa_velocity_uncertainty}
\end{figure}

Figures~\ref{fig:final_dcpa_position_uncertainty} and~\ref{fig:final_dcpa_velocity_uncertainty} illustrate the distribution of the distance at closest point of approach ($\lVert \mathbf{d}_{\text{CPA}} \rVert$) after executing a resolution manoeuvre under position and velocity uncertainty, respectively. Both figures consider a shallow encounter geometry with a heading difference of $2^{\circ}$ and an initial $\lVert \mathbf{d}_{\text{CPA}} \rVert = 0$~m. These conditions represent a challenging case for conflict resolution, where the relative motion is at its minimum.

In Figure~\ref{fig:final_dcpa_position_uncertainty}, under position uncertainty, the MVP algorithm produces a cluster of post-resolution distances above the protected zone threshold of $50$~m, indicated by the red dashed line. More than $98\%$ of the samples falls above $50$~m, suggesting that MVP reliably generates separation even under uncertainty. While some of the MVP samples fall below $50$~m, this can be solved by increasing the look-ahead time allowing more time for the algorithm to resolve the conflict. In contrast, the Velocity Obstacle (VO) algorithm has close to $96\%$ of the samples falling below the $50$~m threshold. As in MVP, this can be solved by adding more lookahead time but since the fraction that falls below the threshold is large, it will take significantly more time to resolve the conflict. This reflects VO's vulnerability, that is applying minimal velocity changes while avoiding conflict does not always result in a safe separation under position uncertainty.

Figure~\ref{fig:final_dcpa_velocity_uncertainty} shows a similar pattern under velocity uncertainty. MVP continues to demonstrate strong performance, with a tightly grouped distribution above the protected zone radius. VO again produces close to $70\%$ of the samples fall below the threshold. This reinforces the earlier observation that VO’s resolution velocities is not the most reliable under uncertainty in navigation systems, both for position and velocity uncertainty.

\begin{figure*}[htbp]
    \centering
    \includegraphics[width=1.0\linewidth]{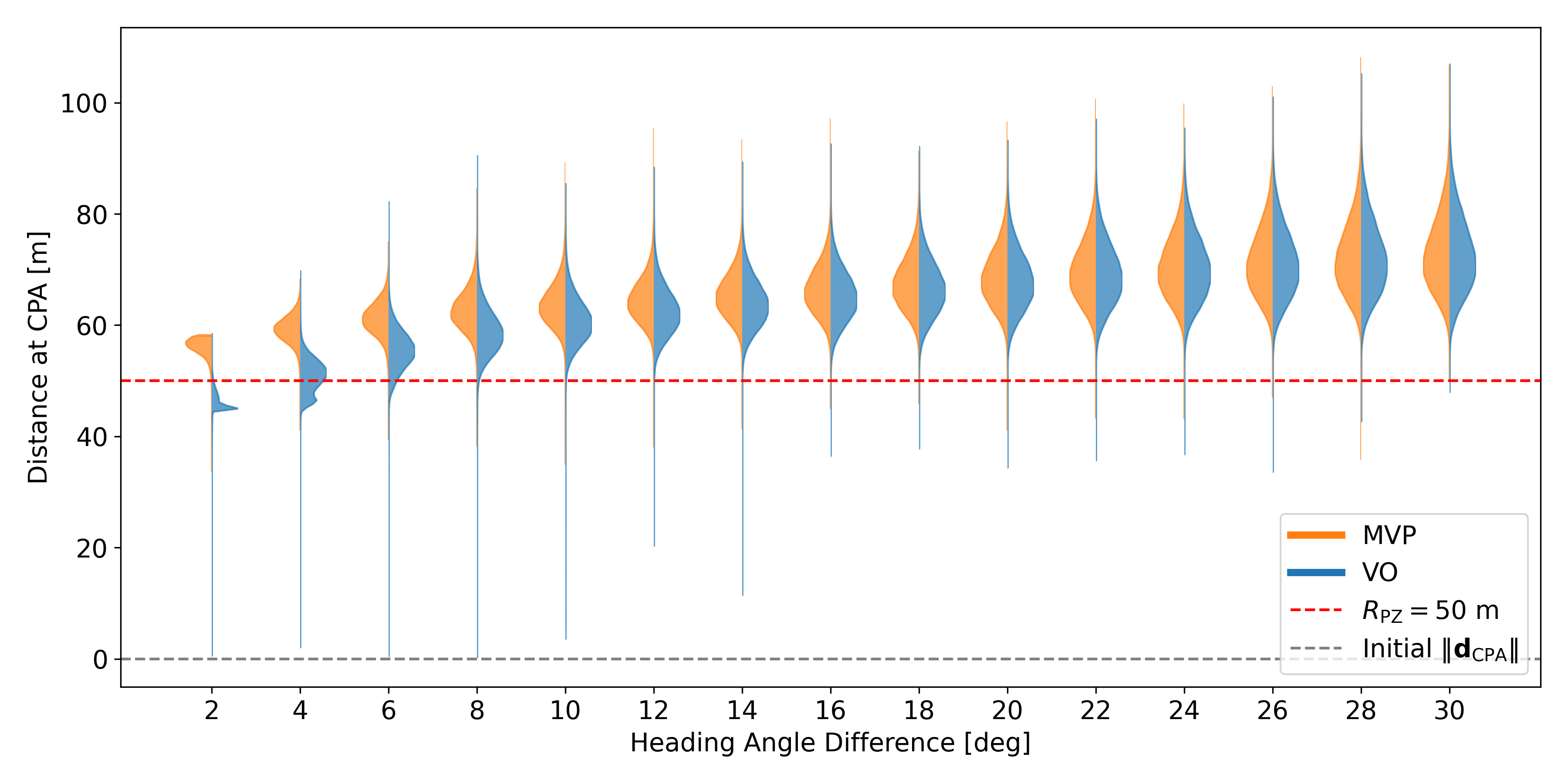}
    \caption{Comparison of $\lVert \mathbf{d}_{\text{CPA}} \rVert$ distribution for MVP and VO at different conflict angles under position uncertainty.}
    \label{fig:violin plot}
\end{figure*}

Figure~\ref{fig:violin plot} extends the analysis by comparing the distribution of $\lVert \mathbf{d}{\text{CPA}} \rVert$ across a range of heading angle differences, from $2^{\circ}$ to $30^{\circ}$. Note that some outlier points show $\lVert \mathbf{d}{\text{CPA}} \rVert$ close to $0$~m, indicating possibility of mid-air collision. Each violin plot summarizes the distribution of $\lVert \mathbf{d}{\text{CPA}} \rVert$ under position uncertainty for both the MVP and VO algorithms. As the conflict angle increases, the relative velocity between the aircraft also increases, and both algorithms tend to generate greater separation with more similar distributions. This occurs because the feasible resolution velocity produced by the VO algorithm, constrained to a circular arc, aligns more closely with the direction that maximizes $\lVert \mathbf{d}_{\text{CPA}} \rVert$. Since the VO algorithm produces resolution velocities along an arc centred at half the relative velocity vector, with a radius equal to half its magnitude, this arc increasingly aligns with the MVP resolution as the relative velocity increases.

Another remark from Figure~\ref{fig:violin plot} is the final miss-distance $\|d_{\mathrm{CPA}}\|$ lies well above the protected-zone radius $R_{PZ}$ at high heading-angle difference. These results indicate a consistent extra distance relative to $R_{PZ}$. Consequently, part of this margin can be exchanged for smaller deviations from the nominal trajectory by modestly relaxing the targeted post-resolution separation (i.e., via tuning $R_{PZ}$) while still ensuring a preferred intrusion prevention rate.

These findings support the following conjecture: the most robust conflict resolution strategy under navigation uncertainty is the one that explicitly maximizes $\lVert \mathbf{d}_{\text{CPA}} \rVert$, especially in geometries with small relative velocity. By doing so, the MVP algorithm consistently maintains safe separation distances and reduces the likelihood of counterproductive resolution manoeuvres. In contrast, VO’s minimal velocity change strategy often proves ineffective under these conditions. These results align with macroscopic trends observed in the intrusion prevention rate (IPR) from the BlueSky simulation campaign.

\section{Conclusion and Future Work}
\label{sec:conclusion}
This study assessed the impact of position and velocity uncertainty on the performance of state-based conflict detection and resolution (CD\&R) algorithms in U-Space. Using Monte Carlo simulation and analytical approximations, the propagation of uncertainty was quantified for key variables in both conflict detection and resolution phases. High-level safety implications were then evaluated using BlueSky simulations.

Monte-Carlo propagation shows that once Gaussian noise is injected into position or velocity, the deterministic variables of state-based detection, time to closest point of approach ($t_{\text{CPA}}$), distance at CPA ($\lVert \mathbf{d}_{\text{CPA}}\rVert$) and time to intrusion ($t_{\text{in}}$), become full probability distributions whose shape depends on conflict geometry and the source of noise. For position uncertainty these distributions remain (folded) Gaussian, but under velocity uncertainty the non-linearity of the projection degrades first-order approximations, especially in low–relative-speed encounters where curved ($\lVert \mathbf{d}_{\text{CPA}}\rVert$) samples appear.  
Crucially, the binary conflict/no-conflict output turns probabilistic: when $t_{\text{in}}$ coincides with the look-ahead threshold, the probability of declaring a conflict falls below $50\,\%$ and rises only as $t_{\text{in}}$ moves further inside the threshold. This sensitivity highlights the need to tune look-ahead time as a function of the navigation noise.

Uncertainty distorts not only the perceived conflict geometry but also the feasible resolution velocities. Under position uncertainty, the resolution velocity from the Modified Voltage Potential (MVP) algorithm align along a straight line defined by $\mathbf{d}_{\text{CPA}}$, while the Velocity Obstacle (VO) algorithm produces samples distributed along a conflict-geometry-dependent arc.  Independently, under velocity uncertainty, the resolution velocity does not show a closed-form solution. Nonetheless, as the relative velocity decreases, the resolution velocity distribution for VO collapses toward the intruder’s velocity, reducing the geometrical space of the resolution. A shallow angle conflict $2^{\circ}$, $20\,\text{kt}$ scenario illustrates the effect: after one second, VO fails to maintain separation in $74\,\%$ of trials, while MVP reduces this to just $9.5\,\%$.  
By explicitly maximizing $\lVert \mathbf{d}_{\text{CPA}} \rVert$, MVP maintains safe separation even under navigational uncertainties, whereas VO’s minimal-change strategy becomes ineffective when relative motion is limited.

Large-scale Monte-Carlo campaigns in the BlueSky simulator translate those geometric differences into macroscopic safety metrics. Under position uncertainty, MVP keeps the intrusion prevention rate (IPR) close to $1.0$ across all heading differences, even when the intruder matches the ownship speed. In contrast, VO safety performance significantly reduced for shallow angle. The same issue exists under velocity and combined uncertainty at  $20\,\text{kts}$, MVP's safety performance remains high with a slight dip at the shallowest angle while VO reaches its lowest IPR under combined uncertainty.

In principle, MVP steers the resolution manoeuvre outward from the predicted closest point of approach. VO, in contrast, seeks the smallest possible change in velocity and therefore performs reliably only when the encounter geometry and the relative speed are desirable. Across analytical derivation, simulation, and full-mission trials, a single principle emerges: robust separation requires an algorithm that maximizes projected distance at closest point of approach, not one that merely escapes the intrusion.

Future work should focus on several critical directions to enhance the reliability and safety of conflict detection and resolution (CD\&R) under navigation uncertainty. First, spatial and temporal parameters such as look-ahead time and protected zone radius should be dynamically tuned based on factors like aircraft speed, navigation uncertainty, and communication reliability, rather than being fixed. Defining these parameters as functions of the conflict context can help maintain a target level of safety, for example ensuring that no more than one in a million conflicts leads to aircraft coming within fifty meters of each other. Additionally, future models should incorporate a broader range of uncertainties beyond basic position and velocity estimates, including environmental effects like wind, assumptions about intruder intent, and gradual sensor degradation. This wider scope will improve the realism and robustness of CD\&R strategies. Lastly, future research must also address uncertainty quantification for various CD\&R algorithms under navigation uncertainty and rigorously challenge the conjecture that the best conflict resolution algorithm is the one that explicitly maximizes the distance at closest point of approach ($\lVert \mathbf{d}_{\text{CPA}} \rVert$). This conjecture should be tested to determine whether it holds universally, and its validity, or lack thereof, should be emphasized as a foundational consideration in the design of future CD\&R systems.

\newpage

\appendix
\section{Proof: Coordinate Transformation of $\mathbf{d}_{\mathrm{CPA}}$ and Its Distribution under Position Uncertainty}
\label{app:dcpa_vector_dist}

Let the relative position be a random vector:
\begin{equation}
\mathbf{x}_{\mathrm{rel}} \sim \mathcal{N}(\boldsymbol{\mu}_{\mathrm{rel}}, \Sigma_{\mathrm{rel}}) \in \mathbb{R}^2
\end{equation}
and the relative velocity be a fixed vector 
\begin{equation}
\mathbf{V}_{\mathrm{rel}} \in \mathbb{R}^2.
\end{equation}
The time to closest point of approach (CPA) is defined as:
\begin{equation}
t_{\mathrm{CPA}} = \frac{\mathbf{V}_{\mathrm{rel}}^\top \mathbf{x}_{\mathrm{rel}}}{\|\mathbf{V}_{\mathrm{rel}}\|^2}
\end{equation}

Then, the vector to CPA is:
\begin{equation}
\mathbf{d}_{\mathrm{CPA}} := \mathbf{x}_{\mathrm{rel}} - t_{\mathrm{CPA}}\, \mathbf{V}_{\mathrm{rel}}
\end{equation}

Define the orthogonal projection matrix:
\begin{equation}
P := I - \frac{\mathbf{V}_{\mathrm{rel}} \mathbf{V}_{\mathrm{rel}}^\top}{\|\mathbf{V}_{\mathrm{rel}}\|^2},
\quad \text{so that} \quad \mathbf{d}_{\mathrm{CPA}} = P\, \mathbf{x}_{\mathrm{rel}}
\end{equation}

Since \( P \) is a linear transformation and \( \mathbf{x}_{\mathrm{rel}} \) is Gaussian, it follows that:
\begin{equation}
\mathbf{d}_{\mathrm{CPA}} \sim \mathcal{N}(P\, \boldsymbol{\mu}_{\mathrm{rel}},\, P\, \Sigma_{\mathrm{rel}}\, P^\top)
\end{equation}

\subsection{Coordinate Transformation}

Define the unit vector in the direction of motion:
\begin{equation}
\mathbf{v}_{\parallel} := \frac{\mathbf{V}_{\mathrm{rel}}}{\|\mathbf{V}_{\mathrm{rel}}\|}
\end{equation}
and the unit vector orthogonal to it (counterclockwise rotation by 90 degrees):
\begin{equation}
\mathbf{v}_{\perp} := 
\begin{bmatrix}
- v_{\parallel, y} \\
\hphantom{-} v_{\parallel, x}
\end{bmatrix}
\end{equation}

These vectors form an orthonormal basis:
\begin{equation}
T := \begin{bmatrix}
\mathbf{v}_{\perp}^\top \\
\mathbf{v}_{\parallel}^\top
\end{bmatrix} \in \mathbb{R}^{2 \times 2}
\end{equation}

Now define the transformed vector:
\begin{equation}
\mathbf{z} := T\, \mathbf{d}_{\mathrm{CPA}}
\end{equation}

\subsection{Mean of Transformed Vector}

Using the linearity of expectation:
\begin{equation}
\mathbb{E}[\mathbf{z}] = T\, \mathbb{E}[\mathbf{d}_{\mathrm{CPA}}] = T\, P\, \boldsymbol{\mu}_{\mathrm{rel}}
\end{equation}

Breaking this into components:
\begin{equation}
\mathbb{E}[\mathbf{z}] = 
\begin{bmatrix}
\mathbf{v}_{\perp}^\top P\, \boldsymbol{\mu}_{\mathrm{rel}} \\
\mathbf{v}_{\parallel}^\top P\, \boldsymbol{\mu}_{\mathrm{rel}}
\end{bmatrix}
\end{equation}

Since \( \mathbf{d}_{\mathrm{CPA}} = P\, \mathbf{x}_{\mathrm{rel}} \) lies in the space orthogonal to \( \mathbf{V}_{\mathrm{rel}} \), and \( \mathbf{v}_{\parallel} \parallel \mathbf{V}_{\mathrm{rel}} \), it follows that:
\begin{equation}
\mathbf{v}_{\parallel}^\top P\, \boldsymbol{\mu}_{\mathrm{rel}} = 0
\end{equation}

Thus, the mean of the transformed variable is:
\begin{equation}
\mathbb{E}[\mathbf{z}] =
\begin{bmatrix}
\mathbf{v}_{\perp}^\top P\, \boldsymbol{\mu}_{\mathrm{rel}} \\
0
\end{bmatrix}
\end{equation}

\subsection{Covariance of Transformed Vector}

The covariance of the transformed variable is given by:
\begin{equation}
\operatorname{Cov}[\mathbf{z}] = T\, P\, \Sigma_{\mathrm{rel}}\, P^\top\, T^\top
\end{equation}

Expanding this using the orthonormal basis:
\begin{equation}
\operatorname{Cov}[\mathbf{z}] =
\begin{bmatrix}
\mathbf{v}_{\perp}^\top P\, \Sigma_{\mathrm{rel}}\, P^\top \mathbf{v}_{\perp} &
\mathbf{v}_{\perp}^\top P\, \Sigma_{\mathrm{rel}}\, P^\top \mathbf{v}_{\parallel} \\
\mathbf{v}_{\parallel}^\top P\, \Sigma_{\mathrm{rel}}\, P^\top \mathbf{v}_{\perp} &
\mathbf{v}_{\parallel}^\top P\, \Sigma_{\mathrm{rel}}\, P^\top \mathbf{v}_{\parallel}
\end{bmatrix}
\end{equation}

Since \( P\, \mathbf{V}_{\mathrm{rel}} = 0 \), and \( \mathbf{v}_{\parallel} \parallel \mathbf{V}_{\mathrm{rel}} \), it follows that:
\begin{equation}
P^\top\, \mathbf{v}_{\parallel} = 0 \quad \Rightarrow \quad \text{All terms involving } \mathbf{v}_{\parallel} \text{ vanish}
\end{equation}

Thus:
\begin{equation}
\operatorname{Cov}[\mathbf{z}] =
\begin{bmatrix}
\sigma_z^2 & 0 \\
0 & 0
\end{bmatrix}
\end{equation}

where:
\begin{equation}
\sigma_z^2 := \mathbf{v}_{\perp}^\top P\, \Sigma_{\mathrm{rel}}\, P^\top\, \mathbf{v}_{\perp}
\end{equation}

\subsection{Conclusion}

The transformed vector \( \mathbf{z} = T\, \mathbf{d}_{\mathrm{CPA}} \) lies entirely along the x axis in the transformed frame. It is a univariate Gaussian:
\begin{equation}
z \sim \mathcal{N}\left(\mathbf{v}_{\perp}^\top P\, \boldsymbol{\mu}_{\mathrm{rel}},\; \mathbf{v}_{\perp}^\top P\, \Sigma_{\mathrm{rel}}\, P^\top\, \mathbf{v}_{\perp} \right)
\end{equation}

\section{Approximation of $t_{\mathrm{in}}$ under Position Uncertainty}
\label{app:tin_pos_uncertainty}

Assume the relative position is normally distributed, $\mathbf{x}_{\mathrm{rel}} \sim \mathcal{N}(\boldsymbol{\mu}_{\mathrm{rel}}, \Sigma_{\mathrm{rel}})$, and the relative velocity $\mathbf{V}_{\mathrm{rel}}$ is deterministic. The time to intrusion entry is modeled as a nonlinear function of $\mathbf{x}_{\mathrm{rel}}$:

\begin{equation}
\label{eq:tin_function}
h(\mathbf{x}_{\mathrm{rel}}) = \frac{\mathbf{V}_{\mathrm{rel}}^\top \mathbf{x}_{\mathrm{rel}}}{\|\mathbf{V}_{\mathrm{rel}}\|^2} - \frac{1}{\|\mathbf{V}_{\mathrm{rel}}\|} \cdot \sqrt{R^2 - \left(\mathbf{v}_\perp^\top P\, \mathbf{x}_{\mathrm{rel}}\right)^2}
\end{equation}

Here, $P = I - \frac{\mathbf{V}_{\mathrm{rel}} \mathbf{V}_{\mathrm{rel}}^\top}{\|\mathbf{V}_{\mathrm{rel}}\|^2}$ is the projection matrix onto the subspace orthogonal to the relative velocity, and $\mathbf{v}_\perp$ is the corresponding unit vector:

\begin{equation}
\mathbf{v}_\perp = \frac{1}{\|\mathbf{V}_{\mathrm{rel}}\|} 
\begin{bmatrix}
- V_{\mathrm{rel},y} \\
\ \, V_{\mathrm{rel},x}
\end{bmatrix}
\end{equation}

Using the delta method, the mean and variance of $t_{\mathrm{in}}$ are approximated as:

\begin{equation}
\mu_{t_{\mathrm{in}}} \approx h(\boldsymbol{\mu}_{\mathrm{rel}}), \quad
\sigma^2_{t_{\mathrm{in}}} \approx \nabla h(\boldsymbol{\mu}_{\mathrm{rel}})^\top \Sigma_{\mathrm{rel}}\, \nabla h(\boldsymbol{\mu}_{\mathrm{rel}})
\end{equation}

The gradient of $h$ at $\boldsymbol{\mu}_{\mathrm{rel}}$ is given by:

\begin{equation}
\nabla h(\boldsymbol{\mu}_{\mathrm{rel}}) = \nabla f(\boldsymbol{\mu}_{\mathrm{rel}}) + \frac{g(\boldsymbol{\mu}_{\mathrm{rel}})}{\|\mathbf{V}_{\mathrm{rel}}\| \sqrt{R^2 - g(\boldsymbol{\mu}_{\mathrm{rel}})^2}} \cdot \nabla g(\boldsymbol{\mu}_{\mathrm{rel}})
\end{equation}

with the component functions defined as:

\begin{align}
f(\mathbf{x}_{\mathrm{rel}}) &= \frac{\mathbf{V}_{\mathrm{rel}}^\top \mathbf{x}_{\mathrm{rel}}}{\|\mathbf{V}_{\mathrm{rel}}\|^2}, \quad
&\nabla f(\boldsymbol{\mu}_{\mathrm{rel}}) = \frac{\mathbf{V}_{\mathrm{rel}}}{\|\mathbf{V}_{\mathrm{rel}}\|^2} \\
g(\mathbf{x}_{\mathrm{rel}}) &= \mathbf{v}_\perp^\top P\, \mathbf{x}_{\mathrm{rel}}, \quad
&\nabla g(\boldsymbol{\mu}_{\mathrm{rel}}) = \mathbf{v}_\perp^\top P
\end{align}

This approximation is valid only when $|\mathbf{d}_{\mathrm{CPA}}(\boldsymbol{\mu}_{\mathrm{rel}})|^2 < R^2$, ensuring the square root remains real.

\section{Derivation of Delta Method Approximations for Velocity Uncertainty}
\label{app:velocity_uncertainty_app}

\subsection{Time to Closest Point of Approach (CPA)}

Let the relative position vector be \( \mathbf{x}_{\text{rel}} \) and the relative velocity vector 
\begin{equation}
\mathbf{V}_{\text{rel}} \sim \mathcal{N}(\mathbf{\mu}_v, \Sigma_v).
\end{equation}

The time to CPA is defined by the nonlinear function:
\begin{equation}
f(\mathbf{V}_{\text{rel}}) = \frac{\mathbf{V}_{\text{rel}} \cdot \mathbf{x}_{\text{rel}}}{\mathbf{V}_{\text{rel}} \cdot \mathbf{V}_{\text{rel}}}
\end{equation}

Using the delta method, the approximate mean and variance of \( t_{\mathrm{CPA}} \) are:
\begin{equation}
\mu_{t_{\mathrm{CPA}}} \approx f(\mathbf{\mu}_v)
\end{equation}

\begin{equation}
\sigma^2_{t_{\mathrm{CPA}}} \approx \nabla f(\mathbf{\mu}_v)^\top \Sigma_v \nabla f(\mathbf{\mu}_v)
\end{equation}

where the gradient is:
\begin{equation}
\nabla f(\mathbf{\mu}_v) = \frac{\mathbf{x}_{\text{rel}}}{\|\mathbf{\mu}_v\|^2} - \frac{2 (\mathbf{\mu}_v \cdot \mathbf{x}_{\text{rel}})}{\|\mathbf{\mu}_v\|^4} \mathbf{\mu}_v
\end{equation}

\subsection{Distance at CPA}

Define the nonlinear function:
\begin{equation}
g(\mathbf{V}_{\text{rel}}) = \left\| \mathbf{x}_{\text{rel}} - f(\mathbf{V}_{\text{rel}}) \cdot \mathbf{V}_{\text{rel}} \right\|
\end{equation}

Let
\begin{equation}
\mathbf{d}_{\mathrm{CPA}} = \mathbf{x}_{\text{rel}} - t_{\mathrm{CPA}} \cdot \mathbf{V}_{\text{rel}}.
\end{equation}
Then the folded normal distribution of \( \|\mathbf{d}_{\mathrm{CPA}}\| \) has approximate mean and variance:
\begin{equation}
\mathbf{\mu}_{d_{\mathrm{CPA}}} \approx g(\mathbf{\mu}_v), \quad
\sigma^2_{d_{\mathrm{CPA}}} \approx \nabla g(\mathbf{\mu}_v)^\top \Sigma_v \nabla g(\mathbf{\mu}_v)
\end{equation}

To compute \( \nabla g(\mathbf{\mu}_v) \), we use the chain rule. Define
\begin{equation}
\mathbf{d} = \mathbf{x}_{\text{rel}} - f(\mathbf{V}_{\text{rel}}) \cdot \mathbf{V}_{\text{rel}}
\end{equation}

Then
\begin{equation}
\nabla g(\mathbf{\mu}_v) = \frac{\mathbf{d}}{\|\mathbf{d}\|} \left( - \nabla f(\mathbf{\mu}_v) \cdot \mathbf{\mu}_v - f(\mathbf{\mu}_v) \cdot I \right)
\end{equation}

\subsection{Intrusion Entry Time \( t_{\text{in}} \)}

Let
\begin{equation}
h(\mathbf{V}_{\text{rel}}) = f(\mathbf{V}_{\text{rel}}) - \frac{\sqrt{R^2 - g(\mathbf{V}_{\text{rel}})^2}}{\|\mathbf{V}_{\text{rel}}\|}
\end{equation}

Then the mean and variance of \( t_{\text{in}} \) are approximated as:
\begin{equation}
\mu_{t_{\text{in}}} \approx h(\mathbf{\mu}_v)
\end{equation}

\begin{equation}
\sigma^2_{t_{\text{in}}} \approx \nabla h(\mathbf{\mu}_v)^\top \Sigma_v \nabla h(\mathbf{\mu}_v)
\end{equation}

The gradient \( \nabla h(\mathbf{\mu}_v) \) involves the gradients of \( f(\mathbf{V}_{\text{rel}}) \), \( g(\mathbf{V}_{\text{rel}}) \), and \( \|\mathbf{V}_{\text{rel}}\| \), as follows:
\begin{equation}
\nabla h(\mathbf{\mu}_v) = \nabla f(\mathbf{\mu}_v) + \frac{g(\mathbf{\mu}_v) \nabla g(\mathbf{\mu}_v)}{\|\mathbf{\mu}_v\| \sqrt{R^2 - g(\mathbf{\mu}_v)^2}} + \frac{\sqrt{R^2 - g(\mathbf{\mu}_v)^2}}{\|\mathbf{\mu}_v\|^3} \mathbf{\mu}_v
\end{equation}

This final expression highlights the nonlinearity of \( t_{\text{in}} \), and the dependence on both the CPA time and distance. It assumes \( \|\mathbf{d}_{\mathrm{CPA}}\| < R \) to ensure the square root is real.

\section{Locus of Projections of a Point onto Lines Through Another Point}
\label{app:locus_projection}

Let \(A\) and \(B\) be two distinct points in the plane.  
We show that the set of all points obtained by projecting \(A\) perpendicularly onto lines that pass through \(B\) forms a circle.  
This circle is centered at the midpoint of segment \(\overline{AB}\) and has radius equal to half the distance between \(A\) and \(B\).

\subsection{Coordinate System Setup}
Without loss of generality, place:
\begin{equation}
B = (0, 0), \qquad A = (2d, 0), \qquad \text{where } d > 0
\end{equation}

Then the midpoint of \(\overline{AB}\) is
\begin{equation}
M = (d, 0)
\end{equation}

and the distance between \(A\) and \(B\) is
\begin{equation}
|AB| = 2d
\end{equation}
so the radius of the claimed circle is
\begin{equation}
r = d
\end{equation}

\subsection{Lines Passing Through \(B\)}

\begin{itemize}
  \item Non-vertical lines:  
  A line through \(B\) with slope \(m\) has the form \(y = mx\), which can also be written as:
  \begin{equation}
  mx - y = 0
  \end{equation}

  \item Vertical line:  
  The vertical line through \(B\) has equation
  \begin{equation}
  x = 0
  \end{equation}
  We handle this special case separately in Section \ref{sec:vertical_case}.
\end{itemize}

\subsection{Projection of \(A\) onto a Non-Vertical Line}

To find the perpendicular projection of a point \((x_0, y_0)\) onto a line given in standard form \(ax + by + c = 0\), the formula is:
\begin{equation}
\left( \frac{b(bx_0 - ay_0) - ac}{a^2 + b^2}, \frac{a(-bx_0 + ay_0) - bc}{a^2 + b^2} \right)
\end{equation}

For our line \(mx - y = 0\), we have \(a = m\), \(b = -1\), \(c = 0\).  
Applying the formula to point \(A = (2d, 0)\), we get:
\begin{equation}
P_m = \left( \frac{2d}{m^2 + 1}, \frac{2dm}{m^2 + 1} \right)
\end{equation}
which is the projection point of \(A\) onto the line \(y = mx\).

\subsection{Distance from the Projection Point to the Midpoint}

We now compute the distance from the projection point \(P_m\) to the midpoint \(M = (d, 0)\):
\begin{equation}
\begin{aligned}
\|P_m M\|^2
&= \left( \frac{2d}{m^2 + 1} - d \right)^2 + \left( \frac{2dm}{m^2 + 1} \right)^2 \\
&= \frac{d^2(1 - m^2)^2 + 4d^2m^2}{(m^2 + 1)^2} \\
&= \frac{d^2(m^4 + 2m^2 + 1)}{(m^2 + 1)^2} \\
&= d^2
\end{aligned}
\end{equation}

So
\begin{equation}
\|P_m M\| = d
\end{equation}
independent of \(m\). This confirms that every projection point lies on the circle centered at \(M = (d, 0)\) with radius \(d\).

\subsection{Vertical Line Case}
\label{sec:vertical_case}

For the vertical line \(x = 0\), the perpendicular projection of \(A = (2d, 0)\) is simply the point
\begin{equation}
B = (0, 0)
\end{equation}

Its distance to the midpoint \(M = (d, 0)\) is also:
\begin{equation}
\|BM\| =  d
\end{equation}
so \(B\) also lies on the same circle.

\subsection{Conclusion}

We have shown that as we vary the line through \(B\), the perpendicular projection of \(A\) always lies at a constant distance \(d\) from the midpoint of segment \(\overline{AB}\).  
Thus, the complete set of these projection points forms a circle centered at \(M\) with radius 
\begin{equation}
d = \tfrac{1}{2}\|AB\|
\end{equation}

\section{Proof that the Angle $\phi$ of $\mathbf d_{\mathrm{CPA}}$ Follows a Projected Normal Distribution}
\label{app:pn_phi}

Let the relative position be the fixed vector $\mathbf{x}_{\mathrm{rel}} \in \mathbb{R}^2$ and let the relative velocity be the random vector
\begin{equation}
  \mathbf{V}_{\mathrm{rel}} \sim \mathcal{N}(\boldsymbol\nu,\, \Sigma_{\mathbf{V}_{\mathrm{rel}}}), \qquad
  \mathbf{V}_{\mathrm{rel}} = \begin{bmatrix} v_x \\ v_y \end{bmatrix}.
\end{equation}
Define $\theta = \operatorname{atan2}(v_y, v_x)$ so that (by definition) $\theta$ is distributed as a projected normal~\cite{mardia_directional_2000}. We prove that the angle
\begin{equation}
  \phi = \operatorname{atan2}(d_{\mathrm{CPA},y}, d_{\mathrm{CPA},x})
\end{equation}
associated with the closest–point-of-approach (CPA) vector $\mathbf{d}_{\mathrm{CPA}}$ is also projected normal.

\subsection{Rotate the coordinate frame}
Without loss of generality, rotate the coordinate frame so that
\begin{equation}
  \mathbf{x}_{\mathrm{rel}} = (2d, 0), \qquad 2d = \| \mathbf{x}_{\mathrm{rel}} \| > 0.
\end{equation}
A deterministic rotation preserves the class of projected normal distributions.

\subsection{Express $\mathbf{V}_{\mathrm{rel}}$ in polar form}
We represent the relative velocity as
\begin{equation}
  \mathbf{V}_{\mathrm{rel}} = v \begin{bmatrix} \cos\theta \\ \sin\theta \end{bmatrix}, \qquad v > 0,
\end{equation}
where $v$ and $\theta$ are the magnitude and direction of the Gaussian distributed velocity vector.

\subsection{Compute the CPA vector}
Using the standard CPA time formula,
\begin{equation}\label{eq:tCPA}
  t_{\mathrm{CPA}} = \frac{\mathbf{V}_{\mathrm{rel}}^{\!\top} \mathbf{x}_{\mathrm{rel}}}{\| \mathbf{V}_{\mathrm{rel}} \|^2} = \frac{2d \cos\theta}{v},
\end{equation}
we find the CPA vector:
\begin{align}
  \mathbf{d}_{\mathrm{CPA}} &= \mathbf{x}_{\mathrm{rel}} - t_{\mathrm{CPA}} \mathbf{V}_{\mathrm{rel}} \\
  &= 2d \begin{bmatrix} 1 - \cos^2\theta \\ -\cos\theta \sin\theta \end{bmatrix}
   = 2d \begin{bmatrix} \sin^2\theta \\ -\cos\theta \sin\theta \end{bmatrix}.
\end{align}
As shown in Appendix~D, $\mathbf{d}_{\mathrm{CPA}}$ under velocity uncertainty traces a circular arc of radius $d$ centered at $(d, 0)$.

\subsection{Direction of $\mathbf{d}_{\mathrm{CPA}}$}
We identify the angle $\phi$ of $\mathbf{d}_{\mathrm{CPA}}$ by observing that it points in the same direction as
\begin{equation}
  \begin{bmatrix} \sin\theta \\ -\cos\theta \end{bmatrix},
\end{equation}
which is a counterclockwise rotation of $\begin{bmatrix} \cos\theta \\ \sin\theta \end{bmatrix}$ by $90^\circ$:
\begin{equation}
  \begin{bmatrix} \cos(\theta - \tfrac{\pi}{2}) \\ \sin(\theta - \tfrac{\pi}{2}) \end{bmatrix}
  = \begin{bmatrix} \sin\theta \\ -\cos\theta \end{bmatrix}.
\end{equation}
Therefore, the angle of $\mathbf{d}_{\mathrm{CPA}}$ is
\begin{equation}\label{eq:phi_shift}
  \phi = \theta - \tfrac{\pi}{2} \quad (\bmod\; 2\pi).
\end{equation}

\subsection{Distribution of $\phi$}

From Equation~\eqref{eq:phi_shift}, we have \(\phi = \theta - \tfrac{\pi}{2}\), meaning the angle \(\phi\) is obtained by rotating \(\theta\) clockwise by \(90^\circ\). Since \(\theta\) is distributed as a projected normal~\cite{mardia_directional_2000}, and the projected normal family is closed under deterministic rotations, it follows that \(\phi\) is also projected normal.

If the relative position vector \(\mathbf{x}_{\mathrm{rel}}\) is not aligned with the \(x\)-axis, the entire coordinate system undergoes an additional constant rotation by \(\arg(\mathbf{x}_{\mathrm{rel}})\). This results in an extra shift in \(\phi\), but again, the projected normal family is preserved under such transformations.

Thus, \(\phi\) follows a projected normal distribution. The precise parameters of this distribution depend on the geometry of the problem and the original Gaussian used to define \(\theta\). Deriving them in closed form is nontrivial and beyond the scope of this paper.

\newpage

\section{Figures}
\label{app:figures}

\begin{figure}[!h]
    \centering
    \begin{minipage}[b]{0.45\linewidth}
        \centering
        \includegraphics[width=\linewidth]{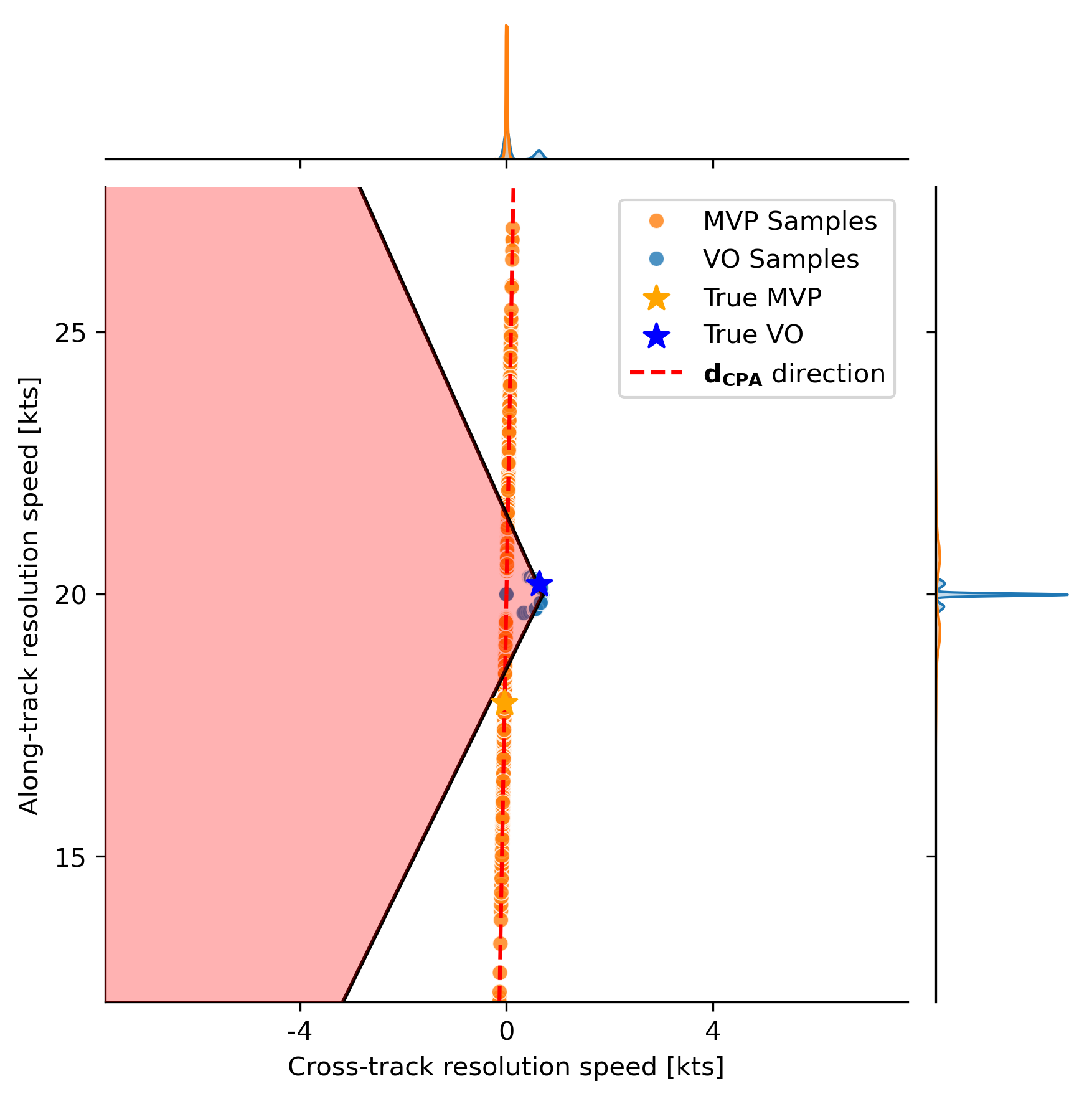}
        \caption{Resolution velocity distribution under position uncertainty 
        ($\sigma_{x_{o}} = \sigma_{x_{i}} = 6.127$), $\Delta_{\psi} = 2^\circ$, 
        $|\mathbf{d}_{CPA}| = 0$ m}
        \label{fig:pos_2deg_dcpa0m}
    \end{minipage}
    \hfill
    \begin{minipage}[b]{0.45\linewidth}
        \centering
        \includegraphics[width=\linewidth]{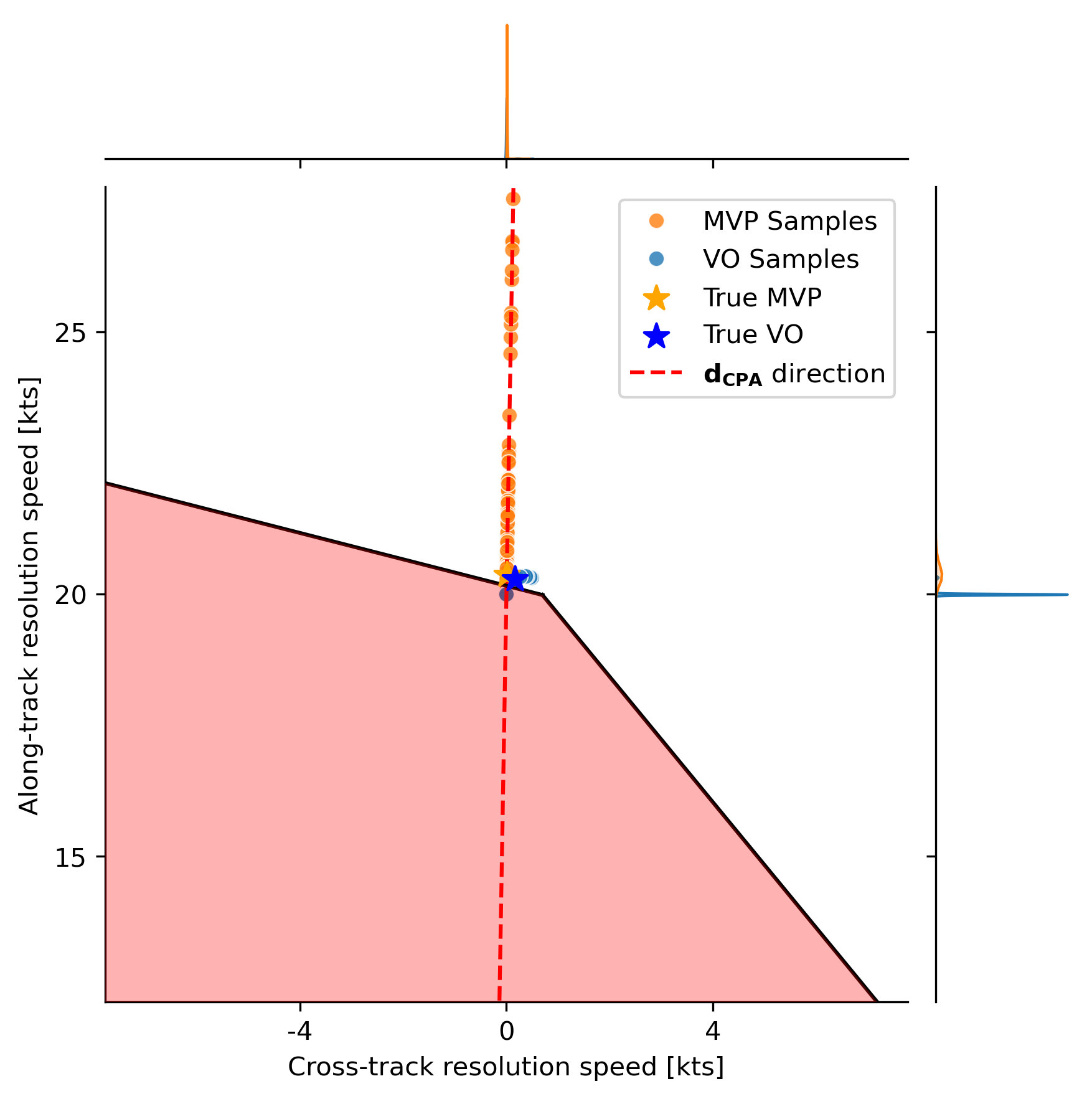}
        \caption{Resolution velocity distribution under position uncertainty 
        ($\sigma_{x_{o}} = \sigma_{x_{i}} = 6.127$), $\Delta_{\psi} = 2^\circ$, 
        $|\mathbf{d}_{CPA}| = 45$ m}
        \label{fig:pos_2deg_dcpa45m}
    \end{minipage}
\end{figure}

\begin{figure}[!h]
    \centering
    \begin{minipage}[b]{0.45\linewidth}
        \centering
        \includegraphics[width=\linewidth]{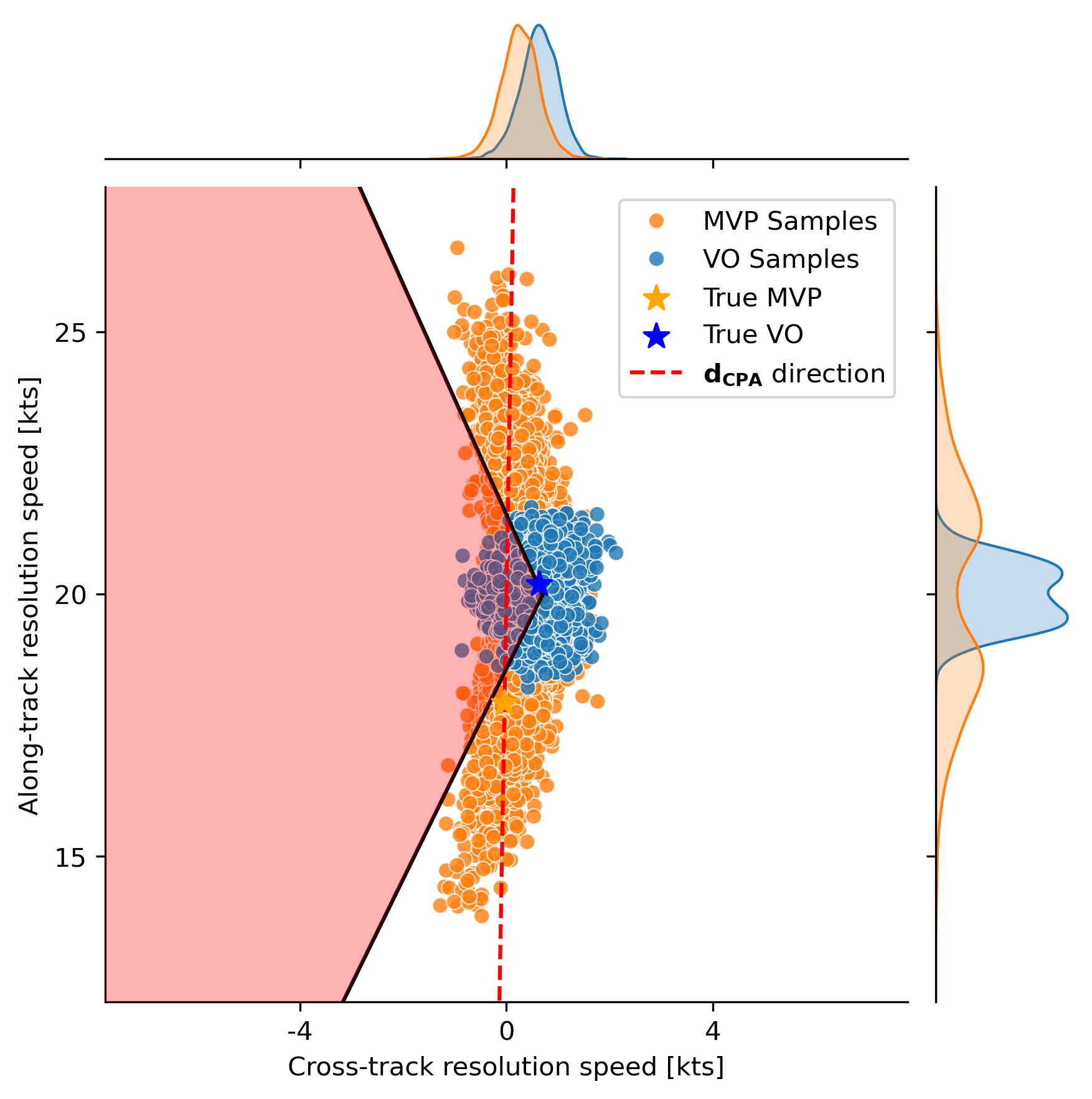}
        \caption{Resolution velocity distribution under velocity uncertainty 
        ($\sigma_{v_{x_o}} = \sigma_{v_{x_i}} = 0.204$), $\Delta_{\psi} = 2^\circ$, 
        $|\mathbf{d}_{CPA}| = 0$ m}
        \label{fig:velo_2deg_dcpa0m}
    \end{minipage}
    \hfill
    \begin{minipage}[b]{0.45\linewidth}
        \centering
        \includegraphics[width=\linewidth]{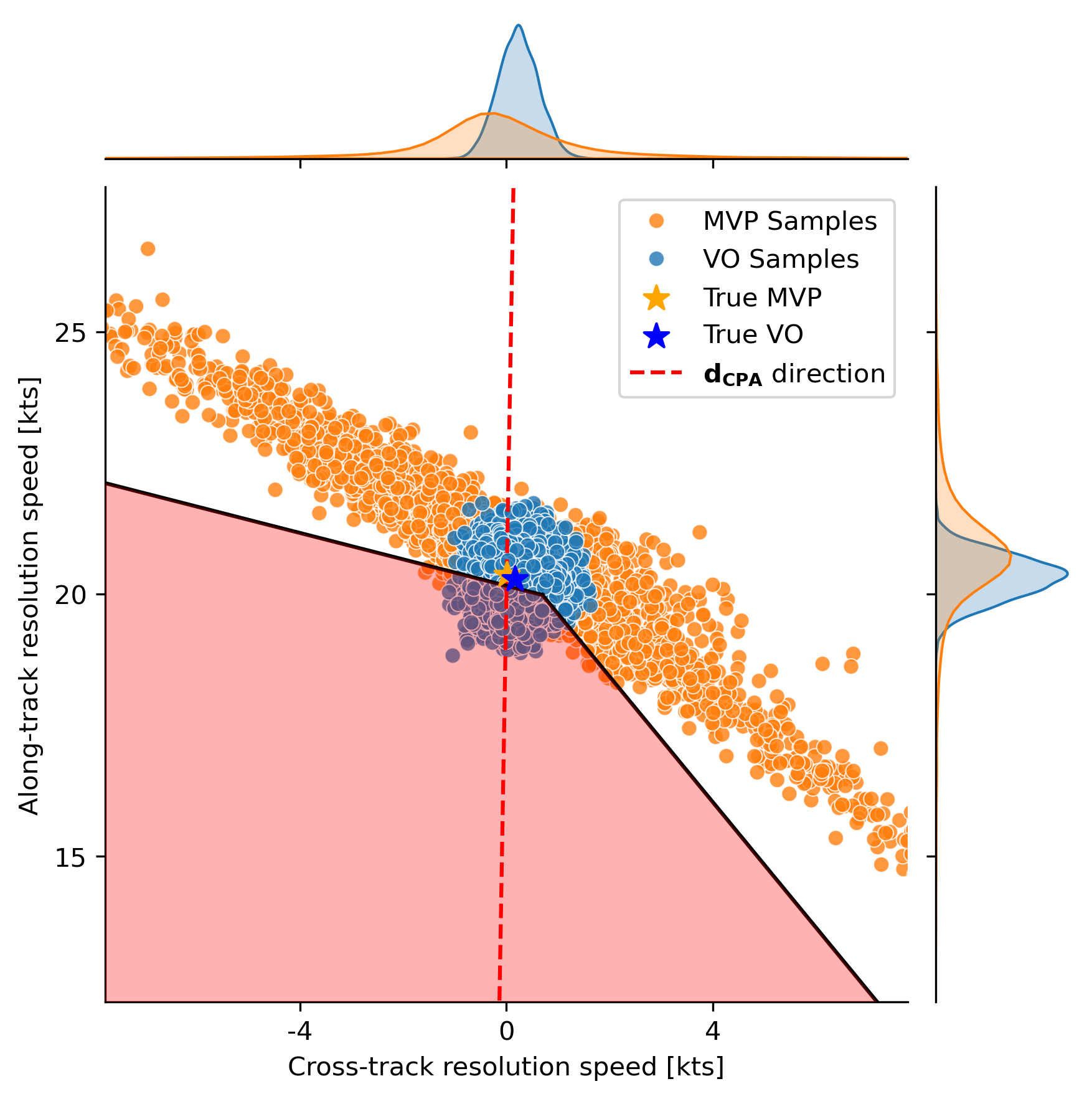}
        \caption{Resolution velocity distribution under velocity uncertainty 
        ($\sigma_{v_{x_o}} = \sigma_{v_{x_i}} = 0.204$), $\Delta_{\psi} = 2^\circ$, 
        $|\mathbf{d}_{CPA}| = 45$ m}
        \label{fig:velo_2deg_dcpa45m}
    \end{minipage}
\end{figure}

\section*{Declaration of generative AI and AI-assisted technologies in the writing process.
}

\noindent
During the preparation of this work, the author(s) used DeepSeek and ChatGPT to assist in analysing formal proofs and mathematical derivations included in the manuscript. Additionally, ChatGPT was used to check spelling and ensure consistency in verb usage. These tools also helped improve the flow of text in selected sections. The author(s) reviewed and edited all AI-generated content and take full responsibility for the final version of the publication.

\bibliographystyle{elsarticle-num}
\bibliography{reference}

\end{document}